\documentclass[11pt,a4paper]{article}
\usepackage[T1]{fontenc}
\usepackage[utf8]{inputenc}
\usepackage{lmodern}
\usepackage[a4paper,margin=25mm]{geometry}
\usepackage{amsmath,amssymb}
\usepackage{graphicx,booktabs,tabularx,array,longtable}
\usepackage{ragged2e,microtype,setspace}
\usepackage[numbers,square,sort&compress]{natbib}
\usepackage{caption}
\usepackage{placeins}
\usepackage{hyperref}
\hypersetup{hidelinks,pdftitle={GraphPathology enables geometry-aware reconstruction of spatial transcriptomics from sparse molecular measurements},pdfauthor={Truong-Son Hy}}
\renewcommand{\arraystretch}{1.16}
\RaggedRight

\newcommand{\um}{\ensuremath{\mu\mathrm{m}}}
\newcolumntype{Y}{>{\RaggedRight\arraybackslash}X}
\newcommand{\PaperTitle}{GraphPathology enables geometry-aware reconstruction of spatial transcriptomics from sparse molecular measurements}
\newcommand{\AuthorNames}{Truong-Son Hy\textsuperscript{1,*}}
\newcommand{\AuthorAffiliations}{\textsuperscript{1}Department of Computer Science, University of Alabama at Birmingham, Birmingham, Alabama, United States}
\newcommand{\Correspondence}{\textsuperscript{*}Correspondence: \href{mailto:thy@uab.edu}{thy@uab.edu}}

\newcommand{\CodeRepository}{https://github.com/HySonLab/GraphPathology}

\usepackage{needspace}

\begin{document}

\begin{center}
{\Large\bfseries \PaperTitle\par}
\vspace{12 pt}
\AuthorNames\par
\vspace{5 pt}
\AuthorAffiliations\par
\vspace{5 pt}
\Correspondence\par
\end{center}
\vspace{8 pt}
\section*{Abstract}
Reconstructing spatial transcriptomics from limited molecular measurements requires accounting for tissue context and measurement placement. We introduce GraphPathology, a geometry-aware framework separating histological representations from reconstruction with observed molecular anchors. We investigated different forms of graph neural networks (GNNs) through matched graph and non-graph visual comparisons. In two breast-cancer cohorts comprising 25 evaluation patients, reconstruction from 20\% distributed molecular observations achieved median patient Pearson correlations of 0.359 and 0.416. Controlled comparisons in 22 previously evaluated patients held hidden targets, estimator and anchor count fixed. Excluding anchors from a central region reduced median paired correlation by 0.1541 and produced a median paired relative increase in mean squared error of 45.0\%; all patients deteriorated. A graph-based predictor improved anchor-assisted correlation and errors over its matched non-graph counterpart in all 22 patients with both visual backbones, but did not outperform molecular interpolation overall. These findings identify measurement placement as a determinant of reconstruction accuracy and distinguish contextual gains from the value of measured expression. Measuring one fifth of locations motivates potential location-dependent savings; total assay savings require prospective validation. Code: \url{https://github.com/HySonLab/GraphPathology}.

\noindent\textbf{Keywords:} spatial transcriptomics; sparse reconstruction; geometry-aware graph learning; measurement geometry; histopathology; breast cancer

\section{Introduction}
\label{sec:introduction}

Spatial transcriptomics (ST) connects molecular measurements to the organization of intact tissue, making it possible to study gene-expression patterns in their anatomical context~\cite{stahl2016}. This spatial perspective is particularly important in cancer, where malignant, stromal and immune populations occupy heterogeneous neighborhoods rather than forming a well-mixed system. Breast-cancer studies have used spatially resolved measurements to characterize regional transcriptional programs, tumor-associated cell populations and tissue architectures that are obscured when spatial relationships are lost~\cite{andersson2021,wu2021,bassiouni2023}. Reconstructing molecular profiles across tissue is therefore more than a prediction problem: it concerns which aspects of biological organization can be recovered from the measurements available at particular locations.

Despite these opportunities, practical ST studies must balance molecular coverage, spatial resolution, tissue coverage and experimental resources. Histological images provide a complementary source of information, motivating computational methods that predict expression from morphology or integrate images with measured transcriptomes~\cite{he2020,monjo2022,bergenstrahle2022,istar2024}. Paired resources such as HEST-1k further enable systematic investigation of expression--morphology relationships across samples and technologies~\cite{jaume2024}. These developments raise an important question for resource-constrained studies: how much information can be recovered when molecular profiles are measured at only a subset of tissue locations, and when does histological context add useful information beyond those measurements?

We address this question through a specific within-section reconstruction task. A predefined gene panel is observed at selected locations, termed \emph{molecular anchors}, and the objective is to estimate the same panel at the remaining locations. Coordinates and histology are available throughout the retained tissue, but the complete target-expression vector is withheld at every unobserved site. Sparsity here refers to the fraction of locations with available molecular measurements, not to zeros within a measured expression vector. The task consequently differs from denoising measured counts, predicting an entirely unassayed set of genes, or increasing the resolution of a molecular map. It also differs from image-only prediction because same-section molecular observations are explicitly available at inference.

Measurement coverage alone does not specify the difficulty of this task. Two observation sets with the same number of anchors can differ substantially in their proximity to the locations being reconstructed. Distributed measurements may provide nearby support throughout a section, whereas exclusion of measurements from a contiguous region forces reconstruction to rely on more distant anchors. Spatially structured evaluation is therefore essential when assessing predictions intended for spatially structured missingness~\cite{roberts2017}. However, a comparison between distributed and regional masks can also change the evaluated tissue and the reconstruction operator. To attribute an observed difference to anchor placement, the measurement count, hidden targets and estimator must instead be held fixed.

Graph representation learning provides a natural language for expressing these relationships. Tissue locations can be represented as nodes, while edges encode spatial neighborhoods or other relationships among their features. Established graph neural networks (GNNs) learn contextual representations through neighborhood aggregation and message passing~\cite{kipf2017,hamilton2017,gilmer2017}. Related ideas have been applied to general visual recognition, hierarchical histopathology analysis and spatial-transcriptomics analysis~\cite{han2022vig,pati2022hact,hu2021spagcn}. For sparse reconstruction, however, two uses of a graph must be distinguished: a graph that contextualizes histological features and a graph that transfers measured expression from available anchors. Their edges describe different information pathways, and a gain from access to molecular observations should not automatically be attributed to the image representation or graph architecture.

This distinction also determines the appropriate comparison. An image-only predictor must infer expression from relationships learned in other samples, whereas a molecular interpolator receives patient-specific expression from the section under evaluation. Comparing them assesses the value of those observations, but does not isolate the benefit of a particular learning algorithm. The incremental value of morphology is better assessed by supplying the same anchors to molecular-only and image--molecular predictors. Cross-study benchmarking has further shown the importance of evaluating generalization, data characteristics and downstream relevance rather than relying on a single prediction score~\cite{wang2025}. Accordingly, architecture, information access and measurement placement should be varied separately wherever possible.

Here, we introduce \emph{GraphPathology}, a geometry-aware reconstruction and evaluation framework that separates these factors. A coordinate-defined \emph{tissue-context graph} supports the representation of histological neighborhoods, while an observation-dependent \emph{molecular-support graph} connects unobserved sites to measured anchors. Molecular reconstruction, image prediction, fixed fusion and observed-residual interpolation are evaluated as distinct routes. The framework accommodates different forms of GNNs without requiring that learned image context improve every reconstruction setting. Its methodological contribution is the explicit separation of representation learning, permitted molecular information and observation geometry, together with controlled comparisons that identify which components account for predictive performance.

We evaluate a development-selected 50-gene panel in two independent breast-cancer studies comprising 25 original evaluation patients. Additional estimator, placement and neural comparisons reuse these previously evaluated cohorts and are interpreted as exploratory follow-up analyses. The evaluation combines distributed coverage, regional missingness, patient-resolved effects and checks that hidden expression cannot influence predictions. This design connects the practical motivation for sparse acquisition to a testable question about reconstruction reliability, without equating fewer retrospectively revealed locations with demonstrated savings in laboratory expenditure.

\noindent\textbf{Contributions and principal findings.}
The study makes four complementary contributions:
\begin{itemize}
\item \textbf{An explicit two-graph reconstruction framework.} GraphPathology distinguishes histological context from observed molecular support and evaluates molecular-only, image-only and combined predictors under defined information constraints. This formulation enables comparisons that separate the contribution of measured anchors from that of learned spatial representations.
\item \textbf{A controlled test of measurement geometry.} With hidden targets, estimator and anchor count held fixed in 22 previously evaluated patients, excluding anchors from a central region reduced median paired Pearson correlation by 0.1541 and increased mean squared error by 45.0\% (median paired relative change) for inverse-distance interpolation at 20\% coverage. All 22 patients deteriorated; uniform interpolation showed the same direction.
\item \textbf{Multi-cohort evidence for sparse molecular reconstruction.} Using 20\% distributed observations, reconstruction achieved median patient correlations of 0.359 and 0.416 in the original three-patient Wu and 22-patient Bassiouni evaluations, respectively. Comparisons across three measurement budgets and twenty molecular estimators characterize the dependence on coverage and reconstruction rule without selecting an external-data winner.
\item \textbf{Matched evidence for context-dependent graph gains.} Within both frozen backbones, GCN-assisted reconstruction improved correlation and errors over matched local prediction in all 22 Bassiouni patients, while molecular interpolation remained stronger overall. Neighborhood-mean and shuffled controls delimit the interpretation of these gains.
\end{itemize}

\section{Related Work}
\label{sec:related_work}

\Needspace{5\baselineskip}
\noindent\textbf{Graph representation learning and geometric inductive biases.}
Graph convolutional networks combine node attributes with local graph structure, while GraphSAGE learns inductive representations by aggregating neighborhood features~\cite{kipf2017,hamilton2017}. The message-passing formulation provides a broader framework for relating these architectures through learned messages, aggregation and node updates~\cite{gilmer2017}. Subsequent developments address complementary representational requirements: GraphGPS combines local message passing with positional or structural encodings and global attention, whereas EGNNs use invariant messages and equivariant coordinate updates to respect Euclidean transformations~\cite{rampasek2022,satorras2021}. These principles motivate ways to represent tissue geometry, but architectural expressivity and symmetry do not by themselves establish useful molecular prediction. GraphPathology uses these established principles as controlled representation alternatives within a reconstruction framework whose distinguishing focus is the separation of learned context, available molecular measurements and anchor placement.

\Needspace{5\baselineskip}
\noindent\textbf{Relational and graph-based learning for general images.}
Visual representation learning also benefits from treating interactions among image regions explicitly. Non-local neural networks aggregate features across spatially separated positions, providing a mechanism for information exchange beyond a local convolutional neighborhood~\cite{wang2018nonlocal}. Vision GNN represents image patches as graph nodes and alternates graph-based feature aggregation with feature transformations for recognition and detection~\cite{han2022vig}. These approaches demonstrate that image structure need not be restricted to a regular convolutional grid and motivate examination of both local and nonlocal tissue context. They nevertheless optimize visual tasks rather than reconstruction from experimentally observed molecular anchors. GraphPathology differs by evaluating visual relations at registered tissue locations and testing whether their representations improve gene-expression reconstruction beyond the same measured molecular support.

\Needspace{5\baselineskip}
\noindent\textbf{Graph learning in computational histopathology.}
Graph-based pathology methods represent tissue through biologically interpretable entities or spatially arranged image patches. HACT-Net combines cell- and tissue-level graphs to model hierarchical tissue composition for histopathological classification~\cite{pati2022hact}. Patch-GCN represents whole-slide images as spatial patch graphs and aggregates their contextual features for patient-survival prediction~\cite{chen2021patchgcn}. Both illustrate how relationships among local structures can support tissue-level inference, although their node definitions, supervision and outputs differ. These studies motivate contextual histological representations without implying that improvements in classification or prognosis transfer directly to molecular reconstruction. In contrast, GraphPathology predicts a fixed expression panel at withheld spatial locations and evaluates histological context against a molecular-only predictor supplied with the identical same-section anchors.

\Needspace{5\baselineskip}
\noindent\textbf{Graph learning for spatial domains and tissue organization.}
Spatial-transcriptomics methods have used graphs to combine expression similarity, physical proximity and histological information. SpaGCN integrates gene expression, spatial coordinates and histology for spatial-domain identification and analysis of spatially variable genes~\cite{hu2021spagcn}. STAGATE learns low-dimensional representations with an adaptive graph-attention autoencoder, while GraphST uses graph-based self-supervised contrastive learning for spatial clustering, integration and deconvolution~\cite{dong2022stagate,long2023graphst}. SPACEL extends spatial analysis to cell-type deconvolution, domain identification and alignment of tissue sections~\cite{spacel2023}. These tasks establish the value of spatial relationships for interpreting measured molecular profiles. GraphPathology instead evaluates expression prediction at sites whose complete target vectors are unavailable, restricting molecular messages to observed anchors and explicitly testing the effect of their placement.

\Needspace{5\baselineskip}
\noindent\textbf{Graph-guided reconstruction of spatial expression.}
Graph structure has also been used directly for transcriptome reconstruction. GNTD models spatial expression as a tensor indexed by spatial coordinates and genes, combining nonlinear tensor decomposition with graph regularization informed by spatial neighborhoods and gene functional relationships~\cite{song2023gntd}. Its formulation demonstrates that both spatial and molecular relational priors can guide recovery from sparse expression data. Graph-guided completion therefore provides an important precedent for molecular reconstruction, while the observation model determines whether missingness concerns individual expression entries or complete spatial profiles. GraphPathology distinguishes itself through complete-vector withholding at selected locations, separate histological and anchor-support pathways, and fixed-target placement comparisons, without using a gene-interaction network as a reconstruction input.

\Needspace{5\baselineskip}
\noindent\textbf{Predicting spatial expression from histological images.}
ST-Net and DeepSpaCE learn relationships between histological appearance and spatial gene expression, establishing image-based prediction as a complementary approach to direct molecular profiling~\cite{he2020,monjo2022}. Contextual models extend this principle: Hist2ST combines convolutional feature extraction, Transformer context and graph-based neighborhood modeling; THItoGene combines richer local representations with global and graph-attention components; and FmH2ST integrates foundation-model representations with spatial and semantic graph information~\cite{hist2st2022,thitogene2024,fmh2st2025}. These methods motivate the study of morphology at multiple contextual scales, but their published results are not an equal-input benchmark for the present sparse-anchor task. CMRCNet also investigates cross-modal reconstruction and contrastive learning for expression prediction~\cite{liu2026}. GraphPathology differs by making observed same-section expression an explicit inference resource and measuring the incremental contribution of image context against an anchor-matched molecular baseline.

\Needspace{5\baselineskip}
\noindent\textbf{Multimodal reconstruction and spatial super-resolution.}
Methods integrating molecular measurements with histology address a range of reconstruction resolutions and information settings. XFuse uses a generative fusion model to infer higher-resolution expression maps, TESLA integrates histology and expression to enhance molecular maps and also evaluates masked-spot recovery, and iStar uses hierarchical image features to predict expression at finer spatial resolution~\cite{bergenstrahle2022,tesla2023,istar2024}. PanoSpace further combines low-resolution ST, high-resolution histology and matched single-cell RNA sequencing for continuous, single-cell-level tissue mapping~\cite{panospace2026}. These methods connect spatial reconstruction to resolution enhancement and multimodal integration, with different assumptions about measured tissue and available molecular references. GraphPathology's distinguishing emphasis is reconstruction at the retained assay locations and controlled separation of anchor geometry, estimator choice and added histological information, rather than increased output resolution or inferred cell identities.

\Needspace{5\baselineskip}
\noindent\textbf{Sparse-measurement learning and acquisition strategies.}
Sparser2Sparse addresses ST imputation through within-sample sparser-to-sparse learning, natural-image co-learning and data-consistent reconstruction~\cite{sparser2026}. SCR$^2$-ST combines single-cell prior information, reinforcement-learning-based active sampling and a regression--retrieval predictor to improve acquisition and prediction under limited budgets~\cite{scr2st2026}. These approaches directly share the motivation of recovering molecular information from restricted measurements. Conventional inverse-distance interpolation also provides an established spatial estimator against which the added value of learned components can be assessed~\cite{shepard1968}. The relevant distinction is therefore not sparse acquisition alone, but the question asked of the acquisition design. GraphPathology complements acquisition-optimized systems by isolating a prespecified anchor-placement effect at fixed measurement count, estimator and hidden targets, rather than proposing an optimized sampling policy.

\Needspace{5\baselineskip}
\noindent\textbf{Benchmarking, generalization and uncertainty.}
HEST-1k provides paired histology--expression resources for evaluating computational representations, while cross-study benchmarking by Wang and colleagues examines prediction quality, generalizability and translational considerations across histology-based expression models~\cite{jaume2024,wang2025}. More broadly, structured validation is important when observations exhibit spatial or hierarchical dependence~\cite{roberts2017}. TISSUE addresses a different aspect of reliability by estimating uncertainty in spatial-expression predictions and incorporating it into downstream analyses~\cite{sun2024}. Patient-level uncertainty in an aggregate performance contrast is not interchangeable with calibrated uncertainty for an individual reconstructed expression value. GraphPathology complements these efforts through patient-resolved, anchor-matched comparisons and identical-target placement controls, reporting reconstruction errors and descriptive patient-bootstrap intervals without claiming calibrated prediction intervals or superiority to the cited systems.

\section{Results}
\subsection{Study design separated information access and dataset roles}
\begin{figure}[!htbp]
\centering
\includegraphics[width=\textwidth,height=0.72\textheight,keepaspectratio]{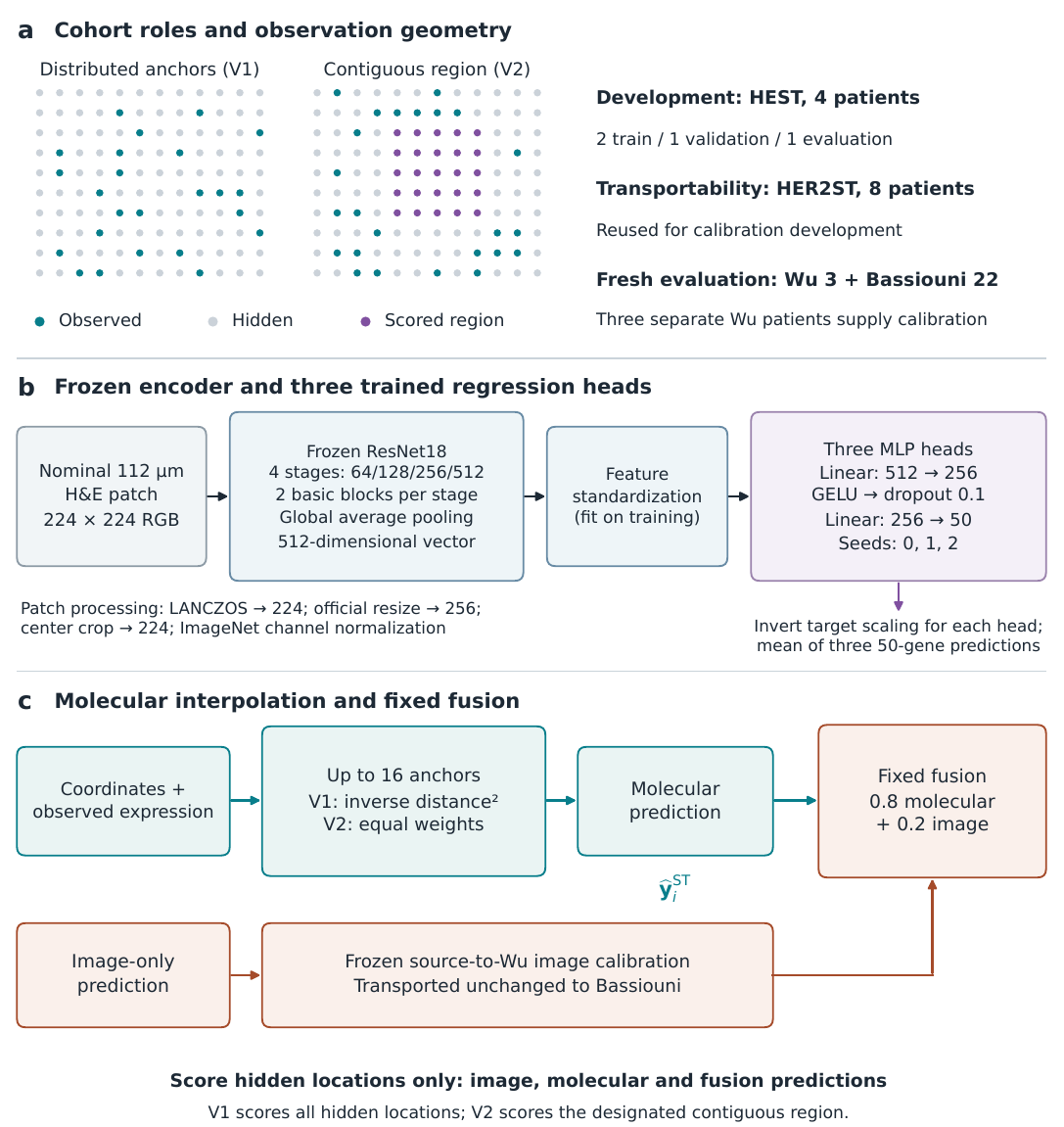}
\caption{\textbf{Original study design and neural-image/molecular reconstruction architecture.} \textbf{a,} Cohort roles and illustrative observation geometries. Teal locations are observed molecular anchors; gray locations are hidden; purple locations form the contiguous scored region. The synthetic spatial examples illustrate 20\% coverage and are not patient measurements. The Wu calibration and evaluation groups contain different patients. \textbf{b,} The frozen ImageNet ResNet18 maps histology to a 512-dimensional vector after global average pooling; the ImageNet classifier is removed. External nominal 112-\um{} patches are first resized to $224\times224$ with LANCZOS, then processed by the official resize-to-256, central-crop-to-224 and channel-normalization transform. Feature scaling fitted on the training patients precedes each $512\rightarrow256\rightarrow50$ MLP, with GELU activation and dropout 0.1 after the hidden linear layer. The three heads use seeds 0, 1 and 2. Their predictions are inverse-scaled to the common target units and averaged. \textbf{c,} Molecular interpolation uses up to 16 observed neighbors, with inverse-distance-squared weighting for distributed evaluation (V1) and equal weighting for contiguous regional completion (V2). Fixed fusion combines this output with the calibrated image prediction using weights 0.8 and 0.2. The uncalibrated image ensemble is also evaluated separately. Calibration uses the development-training and Wu calibration patients and is transported unchanged to Bassiouni. Evaluation uses hidden locations only. V1 and V2 differ in weighting and scoring support as well as observation geometry.}
\label{fig:overview}
\end{figure}

The task used a fixed 50-gene panel and paired spatial expression and histology (Fig.~\ref{fig:overview}; Table~\ref{tab:cohorts}). Four HEST patients supplied 35,536 locations: two patients for fitting, one for validation and one for development evaluation. Eight HER2ST patients (3,455 locations) supported transportability and calibration development; their reuse precludes treating them as untouched external validation. The original external evaluation comprised three held-out Wu patients and the first numbered section from each of 22 Bassiouni patients, totaling 25 patients and 38,484 locations. All 50 genes were retained externally; HER2ST used a fixed 40-gene intersection.

The additional placement controls and neural comparisons reused these already evaluated cohorts and one development patient. They are exploratory follow-up analyses, not new blind replication. Image-only predictors receive histology and permitted geometry, molecular estimators receive coordinates and same-section anchor vectors, and fusion or residual predictors receive both. Thus image-only versus molecular comparisons quantify the value of measured anchors under unequal information access; incremental image value requires comparison with the same anchors supplied to both methods. The public external images were downsampled, containing only 15/25 native pixels in Wu and 24--29 in Bassiouni across the nominal 112-\um{} patch.

\begin{figure}[!htbp]
\centering
\includegraphics[width=\textwidth,height=0.78\textheight,keepaspectratio]{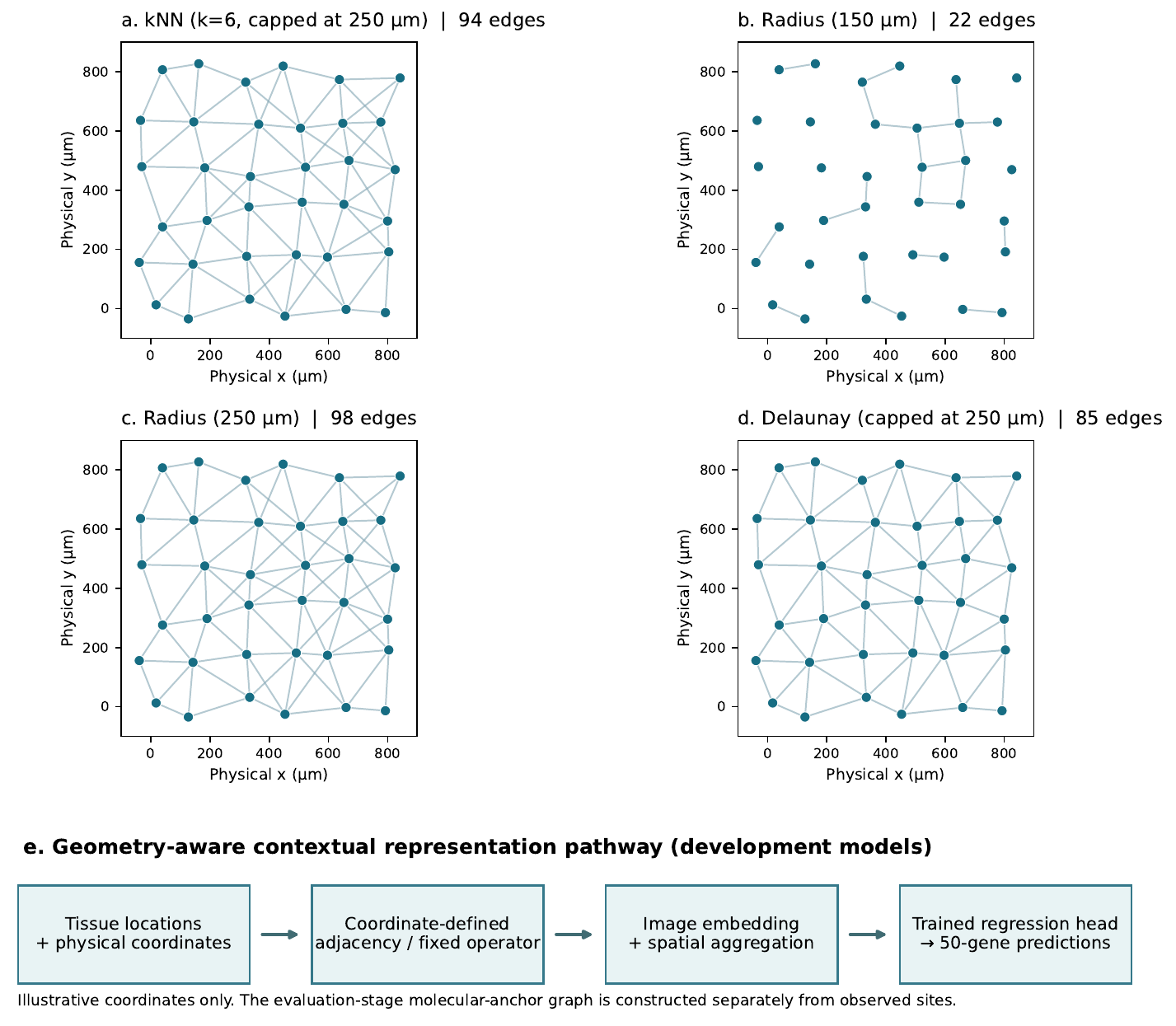}
\caption{\textbf{Geometry-aware graph construction and contextual representations.} \textbf{a--d,} The same 36 synthetic tissue locations are connected by (a) a symmetrized six-nearest-neighbor graph with a 250-\um{} distance cap, (b,c) 150- and 250-\um{} radius neighborhoods, or (d) a 250-\um{}-capped Delaunay graph. Coordinates and axes are identical across panels and all retained edges are displayed. Symmetrization means that a node can have more than six neighbors. \textbf{e,} Geometry determines a fixed adjacency/aggregation operator acting on the frozen image embeddings before an expression-regression head. The molecular-reconstruction stage uses a separate observation-dependent graph joining hidden locations to up to 16 measured molecular anchors. Synthetic points illustrate the architecture only, not a fitted patient graph or a performance result.}
\label{fig:graph_architecture}
\end{figure}
\begin{table}[!htbp]
\caption{\textbf{Cohorts and their analytical roles.} Counts refer to retained locations and patient-level study roles, not independent spot replicates. The same eight HER2ST patients were first inspected for transportability and subsequently used for calibration development. The Wu study contains 15,601 locations in all six patients; only 8,481 locations from its three evaluation patients enter the original external evaluation. All 22 Bassiouni patients were evaluated without new training or calibration. ``Pixels per patch'' refers to the native image information in a nominal 112-\um{} field of view, before resizing. The primary panel was fixed in development.}
\label{tab:cohorts}
\centering\small
\begin{tabularx}{\textwidth}{@{}>{\RaggedRight\arraybackslash}p{21mm}>{\RaggedRight\arraybackslash}p{26mm}>{\RaggedRight\arraybackslash}p{16mm}>{\RaggedRight\arraybackslash}p{21mm}>{\RaggedRight\arraybackslash}p{14mm}Y@{}}
\toprule
Cohort & Platform & Patients, total / evaluation & Locations, total / evaluation & Genes & Analytical role and image quality \\
\midrule
Development (HEST) & Targeted Xenium & 4 / 1 & 35,536 / 20,761 & 50/50 & Two training, one validation, one primary evaluation patient; conditional four-patient sensitivity analyses. Estimated 0.21--0.36~\um{}/pixel. \\
HER2ST & Sequencing ST & 8 / 8 & 3,455 / 3,455 & 40/50 & Initial transportability, then calibration development on the same patients. Approximately 0.69--0.92~\um{}/pixel. \\
Wu & Visium & 6 / 3 & 15,601 / 8,481 & 50/50 & Three calibration and three original evaluation patients. Only 15 or 25 native pixels per patch. \\
Bassiouni & Visium & 22 / 22 & 30,003 / 30,003 & 50/50 & First numbered section per patient; original external evaluation with no cohort-specific fitting. Only 24--29 native pixels per patch. \\
\bottomrule
\end{tabularx}
\end{table}

\FloatBarrier

\subsection{Distributed molecular observations supported reconstruction across cohorts}
\begin{figure}[!htbp]
\centering
\includegraphics[width=\textwidth,height=0.60\textheight,keepaspectratio]{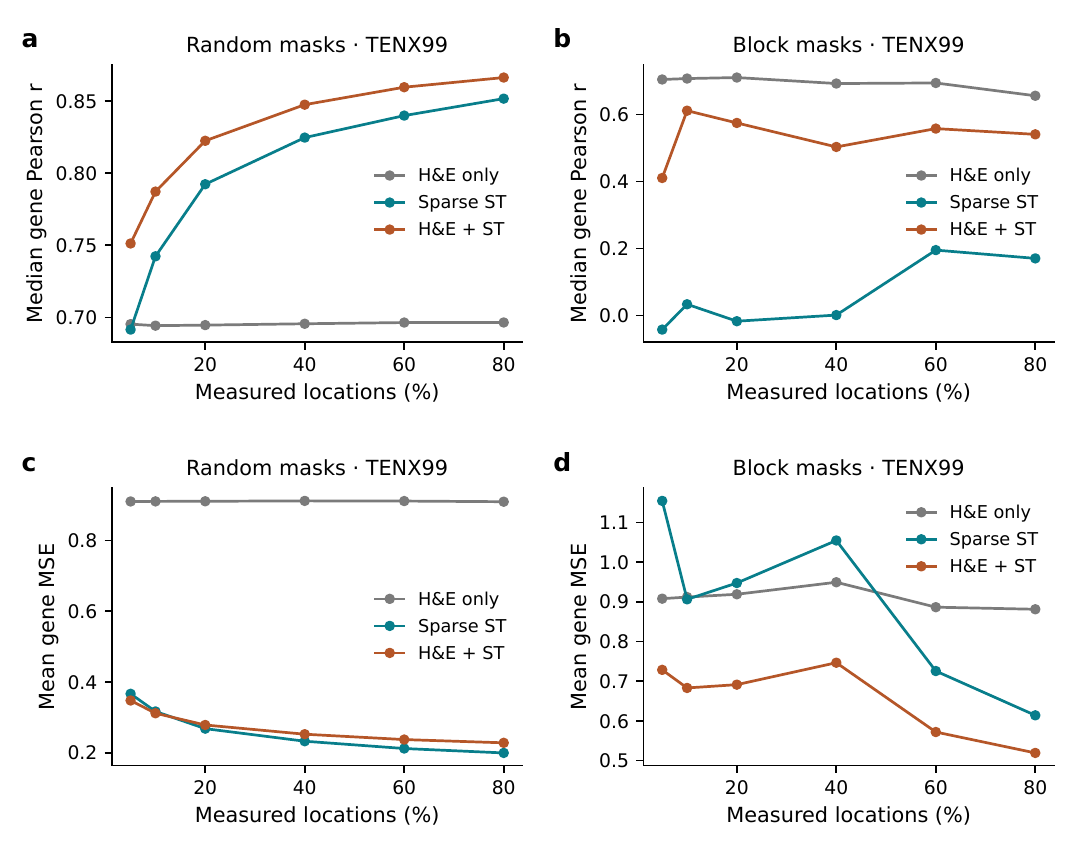}
\caption{\textbf{Development reconstruction depends on molecular coverage and masking regime.} Results are from the single primary development patient, TENX99, using the frozen 50-gene panel and ResNet18 representation. \textbf{a,c,} Distributed random observation. \textbf{b,d,} The earlier development block-mask regime. Panels a and b show median gene Pearson correlation; panels c and d show mean gene MSE across measurement fractions. The image-only, selected molecular interpolation and fixed fusion curves summarize three frozen mask replicates per condition; curves are means over masks, not patient confidence intervals. Targets and error units are unchanged across methods. The development block masks differ from the external V2 forbidden-region protocol and should not be treated as directly equivalent geometry tests. At 20\% distributed coverage, fusion has higher correlation than interpolation but slightly higher MSE, illustrating that spatial association and calibration are distinct properties. The image-only comparator receives no observed molecular anchors.}
\label{fig:development}
\end{figure}

At 20\% distributed coverage in TENX99, median gene Pearson correlations were 0.695 for image-only prediction, 0.792 for interpolation and 0.822 for fixed fusion; corresponding mean squared errors (MSEs) were 0.911, 0.268 and 0.279 (Fig.~\ref{fig:development}; Table~\ref{tab:performance}). The correlation increase from fusion therefore did not imply an error reduction. These are one-patient development results.

\begin{figure}[!htbp]
\centering
\includegraphics[width=\textwidth,height=0.70\textheight,keepaspectratio]{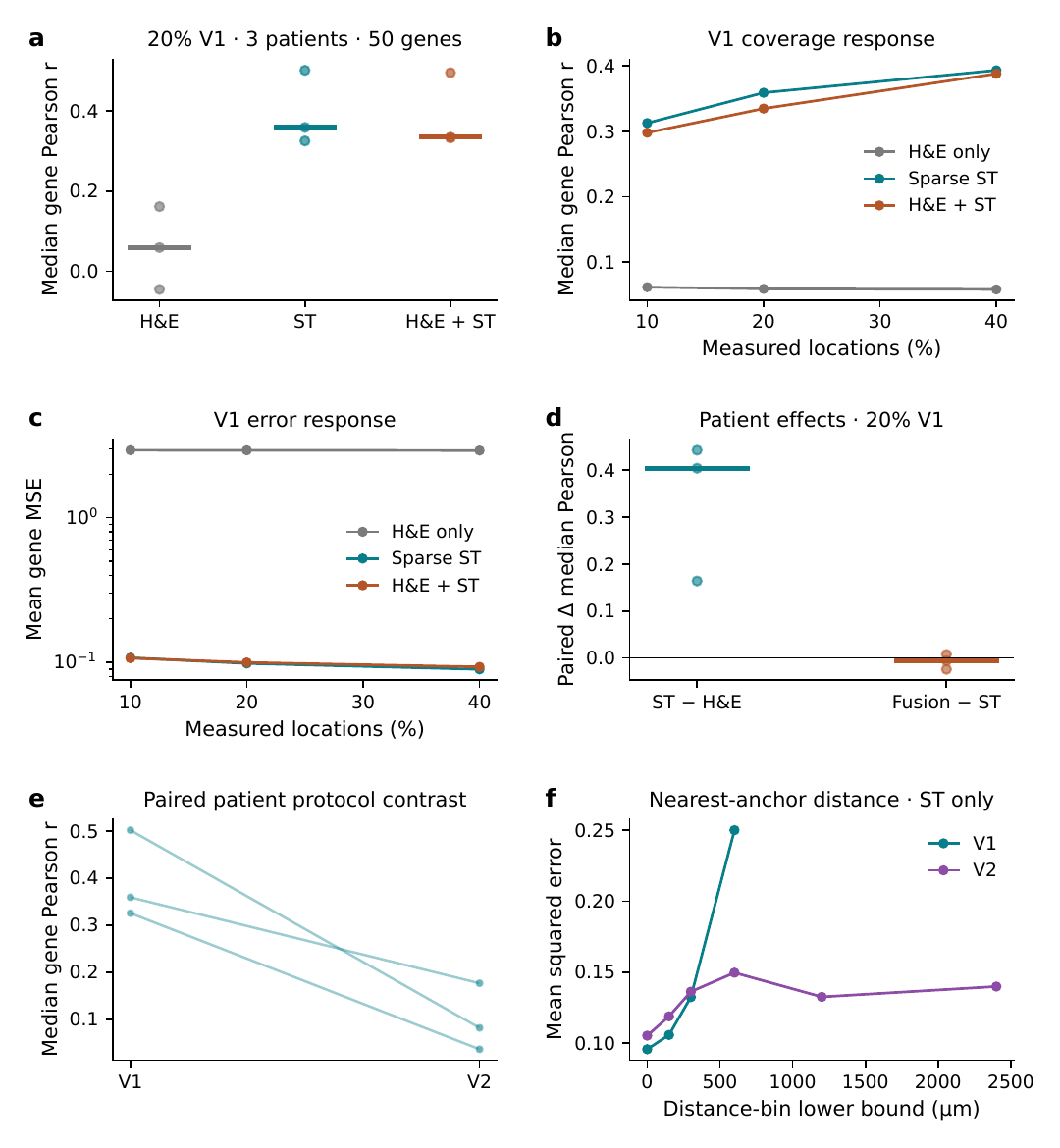}
\caption{\textbf{Fresh evaluation in three held-out Wu patients supports interpolation but not robust image fusion.} The frozen 50-gene panel was evaluated in patients 1160920F, CID4290 and CID44971, none of whom contributed to Wu calibration. \textbf{a,} Patient-level correlation summaries at 20\% distributed coverage. \textbf{b,c,} Median patient correlation and MSE across 10\%, 20\% and 40\% distributed coverage. \textbf{d,} Paired patient correlation effects for interpolation versus image-only prediction and fusion versus interpolation. \textbf{e,} Paired distributed (V1) and contiguous-region (V2) molecular correlations at 20\%. \textbf{f,} Distance-resolved molecular MSE at 20\%. Patient summaries average ten random mask seeds within geometry and then geometry means; cohort summaries are medians, not pooled spot estimates. Three fixed contiguous geometries are used for V2. Each patient is a biological replicate; seeds are not. V1 uses inverse-distance-squared interpolation and V2 uses a nearest-anchor mean, so panel e is a protocol contrast. The public image inputs contained only 15 or 25 native pixels per 112-\um{} patch, limiting conclusions about morphology. Error units are common development-standardized log1p counts.}
\label{fig:wu}
\end{figure}

\begin{figure}[!htbp]
\centering
\includegraphics[width=\textwidth,height=0.70\textheight,keepaspectratio]{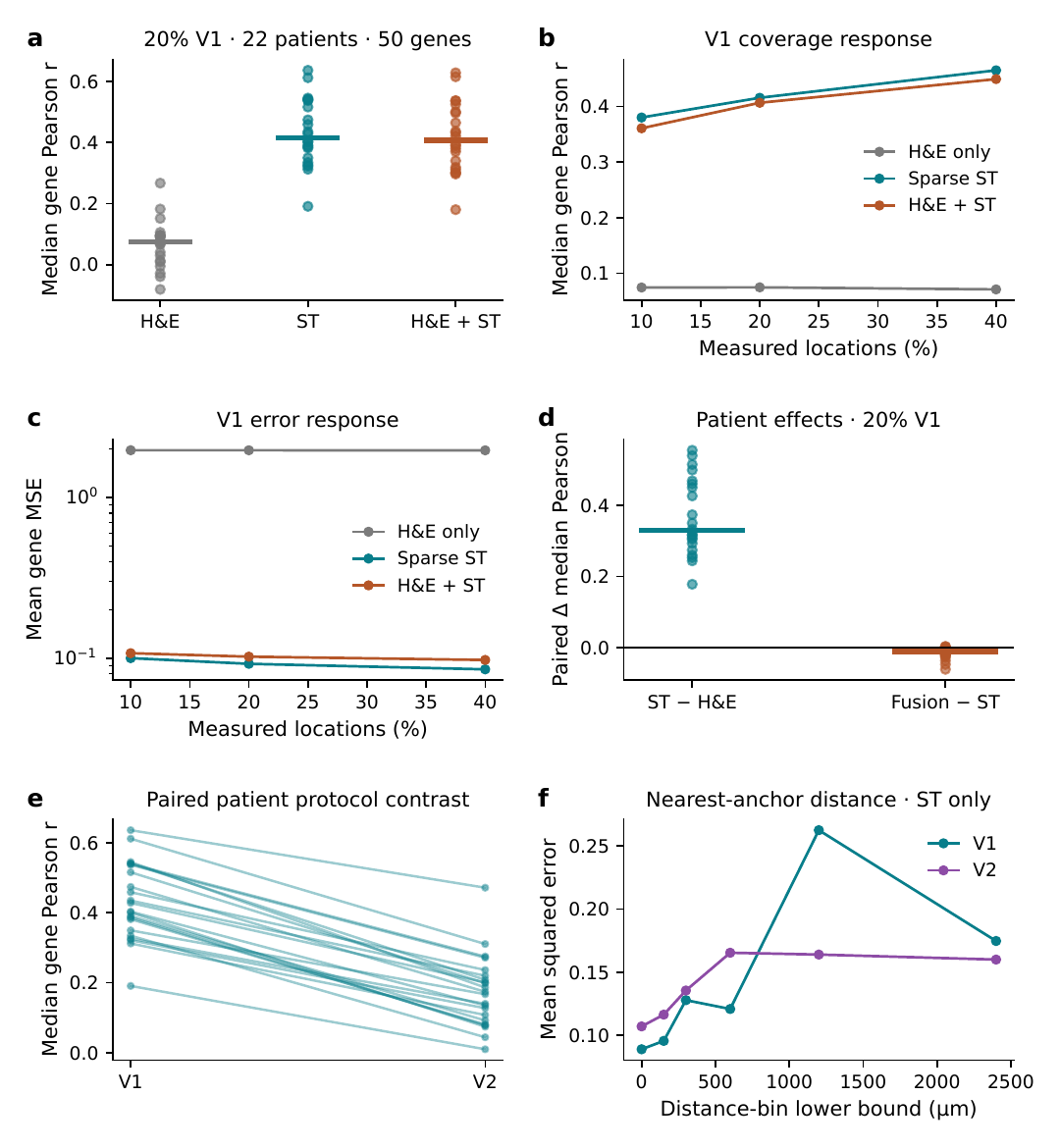}
\caption{\textbf{Independent 22-patient replication confirms the coverage response and weaker contiguous completion.} One first-numbered section per Bassiouni patient was selected before evaluation, yielding 30,003 locations and the same 50 primary genes. No patient from this study was used for new training or calibration. \textbf{a,} Patient correlation summaries at 20\% distributed coverage. \textbf{b,c,} Correlation and MSE across measurement fractions. \textbf{d,} Paired interpolation--image and fusion--interpolation correlation effects. \textbf{e,} Paired V1/V2 molecular correlations at 20\%. \textbf{f,} Distance-resolved molecular MSE at 20\%. Each patient summary averages ten mask seeds within geometry and then the three V2 geometry means where applicable; cohort curves use patient medians. All 22 patients improved in correlation and MSE between 10\% and 40\% V1 coverage and deteriorated in correlation, MSE and MAE under V2 relative to V1 at 20\%. V1/V2 differ in scoring support and interpolation weighting as well as missing-region design. Image fusion uses the frozen Wu calibration and 0.2 image weight. Public histology contained only 24--29 native pixels per 112-\um{} patch. Distance curves have varying occupied-bin support and are not calibrated uncertainty intervals. No p-values or patient confidence intervals are implied.}
\label{fig:bassiouni}
\end{figure}

In the original Wu evaluation, the median of patient-level median gene correlations was 0.359 for interpolation and 0.059 for image-only prediction. Corresponding MSEs were 0.0987 and 2.9155, and mean absolute errors (MAEs) were 0.2179 and 1.5578 (Fig.~\ref{fig:wu}). Bassiouni interpolation achieved correlation 0.416, MSE 0.0923 and MAE 0.2033, compared with image-only values 0.075, 1.9673 and 1.2120 (Fig.~\ref{fig:bassiouni}). Every evaluation patient favored interpolation in correlation, MSE and MAE at each tested budget. All 22 Bassiouni patients met the predefined joint support criterion, which was descriptive rather than a significance test.

\begin{table}[!htbp]
\caption{\textbf{Reconstruction at 20\% distributed measurement coverage.} Each entry is a cohort median of patient-level summaries after averaging mask seeds within geometry and geometries within patient, except that the development row represents the single primary patient. Pearson $r$ is the median correlation across panel genes; MSE and MAE are patient means across genes and hidden locations before the cohort median. All methods within a cohort use identical target units and observation masks. Different gene panels and assay count distributions preclude treating these values as a pooled cross-cohort ranking. Fusion rules and their calibration are frozen but differ between the development and external protocols; see Methods and Supplementary Information. The image-only branch has no same-section molecular anchors.}
\label{tab:performance}
\centering\small
\begin{tabularx}{\textwidth}{@{}Yrrlrrr@{}}
\toprule
Cohort & $n$ & Genes & Method & Pearson $r$ & MSE & MAE \\
\midrule
Development & 1 & 50 & Image only & 0.695 & 0.9110 & 0.7606 \\
 &  &  & Spatial interpolation & 0.792 & 0.2684 & 0.3904 \\
 &  &  & Fixed fusion & 0.822 & 0.2787 & 0.4173 \\
\addlinespace[5pt]
HER2ST, zero-shot & 8 & 40 & Image only & 0.143 & 4.3665 & 1.9376 \\
 &  &  & Spatial interpolation & 0.194 & 0.0680 & 0.1730 \\
 &  &  & Fixed fusion & 0.255 & 0.3481 & 0.5153 \\
\addlinespace[5pt]
Wu, original & 3 & 50 & Image only & 0.059 & 2.9155 & 1.5578 \\
 &  &  & Spatial interpolation & 0.359 & 0.0987 & 0.2179 \\
 &  &  & Fixed fusion & 0.335 & 0.1002 & 0.2321 \\
\addlinespace[5pt]
Bassiouni, original & 22 & 50 & Image only & 0.075 & 1.9673 & 1.2120 \\
 &  &  & Spatial interpolation & 0.416 & 0.0923 & 0.2033 \\
 &  &  & Fixed fusion & 0.407 & 0.1022 & 0.2288 \\
\bottomrule
\end{tabularx}
\end{table}

Increasing distributed coverage from 10\% to 40\% improved correlation and MSE in every patient. Bassiouni correlations at 10/20/40\% were 0.380/0.416/0.465, with MSEs 0.1001/0.0923/0.0852; Wu values were 0.313/0.359/0.393 and 0.1080/0.0987/0.0899 (Supplementary Table~S9). These budget-specific protocols use nested anchor sets but different hidden sets. They neither identify a universally sufficient budget nor compare an identical target set across coverages. Earlier sampling comparisons, including label-free spatial coverage sampling, are retained in Supplementary Fig.~S8.

\subsection{Controlled anchor placement altered reconstruction in all 22 patients}
The controlled experiment fixed a central region $R$ and a hidden subset $H$, then changed only the eligible placement of an equal number of anchors. Distributed anchors could occupy any non-$H$ location; region-excluded anchors were restricted outside $R$. Within each comparison, the estimator, gene panel, hidden targets and total anchor count were identical. This design isolates placement for the specified central-region sampling rule.

\begin{figure}[!htbp]
\centering\includegraphics[width=\textwidth]{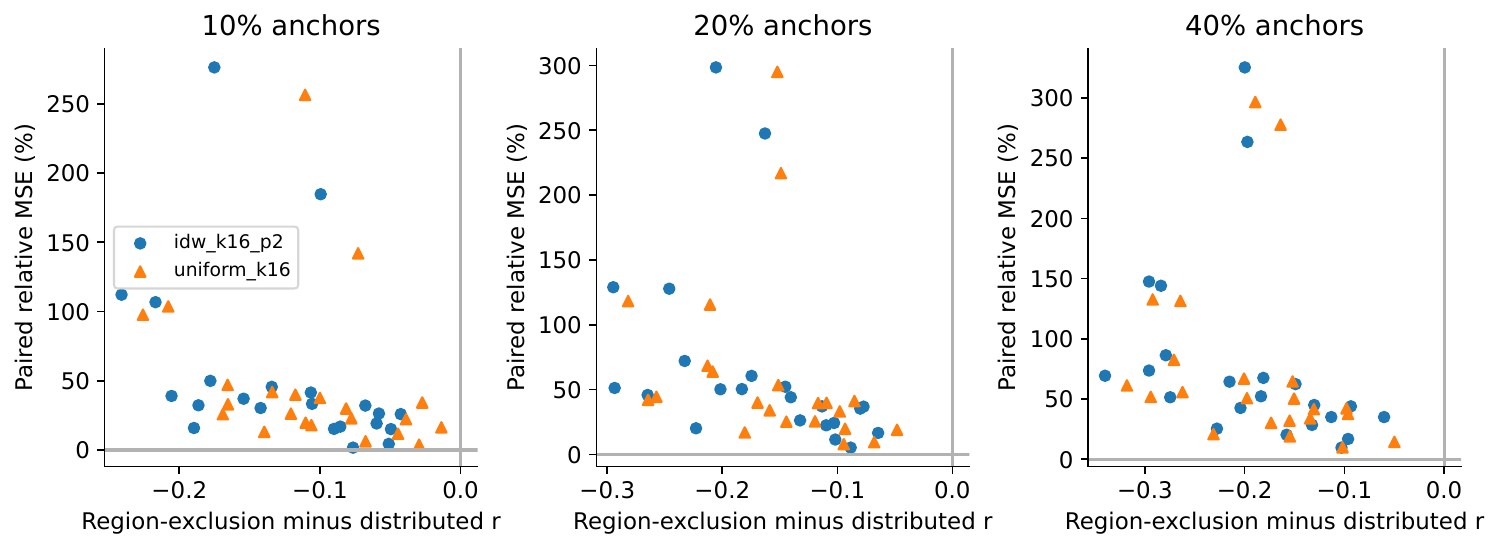}
\caption{\textbf{Identical-target placement effects in all 22 patients.} Each point is one patient, averaged over ten mask seeds; shapes distinguish fixed IDW and uniform k16. Both anchor placements score the same H with equal anchor counts. Panels retain all three budgets. These are not historical V2-minus-V1 effects.}
\label{fig:extension_placement}
\end{figure}

At 20\% coverage, region exclusion worsened Pearson, MSE and MAE in all 22 Bassiouni patients for both fixed estimators (Fig.~\ref{fig:extension_placement}; Table~\ref{tab:extension_placement}). For inverse-distance weighting (IDW), the median paired correlation change was $-0.1541$ (descriptive 95\% patient-bootstrap interval $-0.2056$ to $-0.1096$), relative MSE increased 45.0\% (29.8--52.2\%) and MAE increased 0.0335 (0.0266--0.0508). Uniform interpolation gave a correlation change of $-0.1502$ ($-0.1805$ to $-0.1094$), relative MSE increase 39.9\% (29.2--53.5\%) and MAE increase 0.0326 (0.0234--0.0462). All three budgets and all patient effects are retained. These intervals describe uncertainty across the 22 patients; they are not p-values or evidence of a universal geometry effect.
\FloatBarrier
\begin{table}[!htbp]
\caption{\textbf{Controlled Bassiouni placement effects.} Region-exclusion minus distributed anchors at 20\% coverage, identical H and anchor counts. Median paired patient effects with descriptive 95\% percentile intervals from 10,000 patient resamples. Positive error changes deteriorate. All 22 patients are retained; masks are averaged within patient. The full 20-estimator and three-budget exports accompany the paper.}
\label{tab:extension_placement}
\centering\footnotesize
\begin{tabular}{@{}lrr@{}}
\toprule
Operator & Effect & Median [95\% interval] \\
\midrule
IDW & $\Delta r$ & -0.1541 [-0.2056, -0.1096] \\
IDW & MSE\% & +44.9864 [+29.8116, +52.2347] \\
IDW & $\Delta$ MAE & +0.0335 [+0.0266, +0.0508] \\
Uniform & $\Delta r$ & -0.1502 [-0.1805, -0.1094] \\
Uniform & MSE\% & +39.9128 [+29.2387, +53.4974] \\
Uniform & $\Delta$ MAE & +0.0326 [+0.0234, +0.0462] \\
\bottomrule
\end{tabular}
\end{table}
\FloatBarrier

The three previously evaluated Wu patients corroborated the direction: at 20\% coverage, region exclusion reduced IDW correlation by 0.106--0.245 and increased relative MSE by 20.1--45.4\%; uniform reconstruction lost 0.093--0.188 in correlation and increased MSE by 21.7--38.9\% (Fig.~\ref{fig:journal_controls}b; Supplementary Table~S12). Each patient is reported individually. The original distributed-versus-regional protocol comparison is distinct: it changes geometry, estimator and scored support, rather than placement alone.

\subsection{Matched spatial baselines separated estimator choice from sampling design}
\begin{figure}[!htbp]
\centering\includegraphics[width=\textwidth]{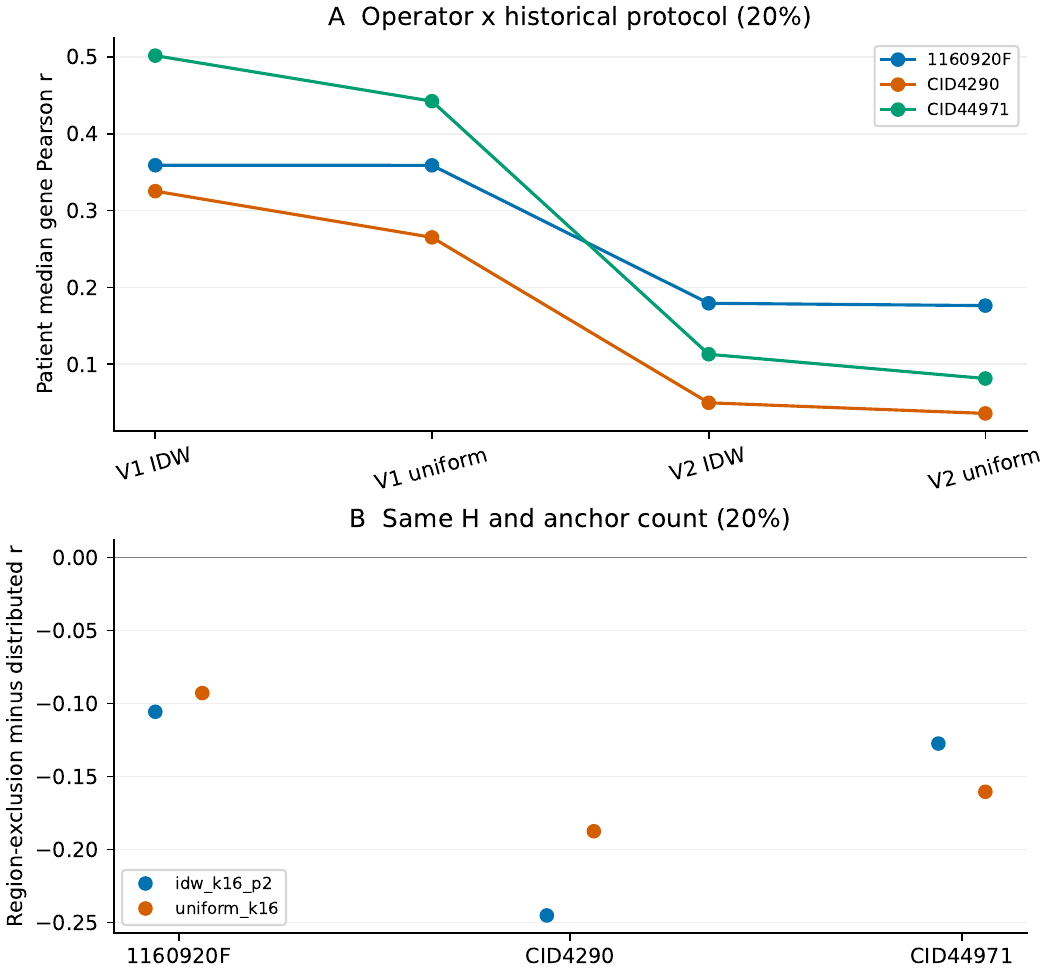}
\caption{\textbf{Matched estimator and placement controls in previously inspected Wu patients.} (a) Both operators share each historical mask and scored set; V2 values average the three geometries. (b) Region-exclusion minus distributed correlation on the identical H, with equal total anchor count and ten random seeds. The three individual patients are shown; masks and genes are not biological replicates. All points use 20\% coverage and the frozen 50-gene panel.}
\label{fig:journal_controls}
\end{figure}

\begin{figure}[!htbp]
\centering\includegraphics[width=\textwidth]{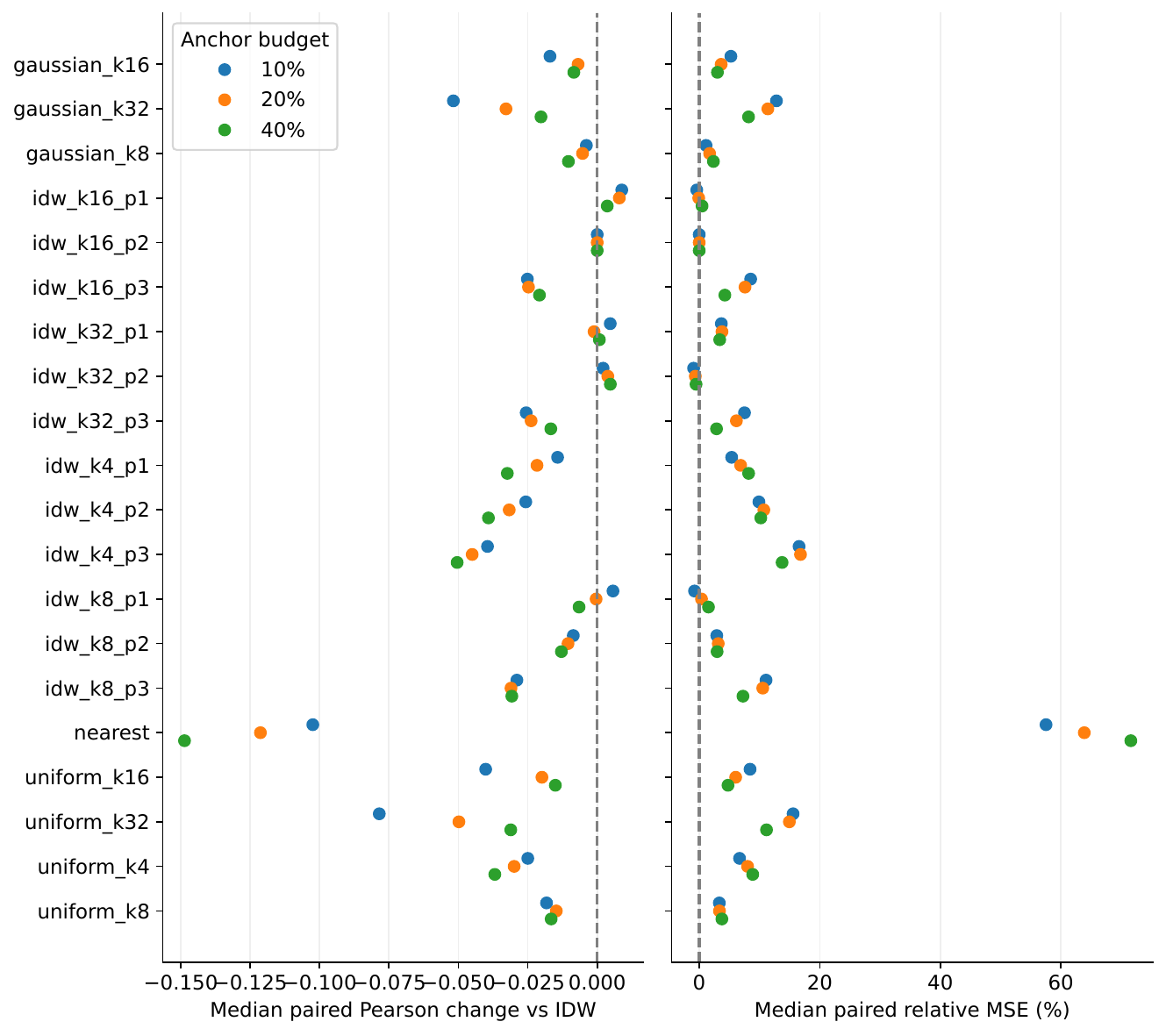}
\caption{\textbf{Complete matched-information baseline grid in 22 previously inspected Bassiouni patients.} Every declared estimator and budget is shown. Points are medians of paired patient effects relative to fixed IDW k16/p2 after averaging ten original mask trajectories. Correlation increases and error decreases improve. No estimator was selected from these outcomes.}
\label{fig:extension_baselines}
\end{figure}

We compared twenty fixed molecular estimators on identical distributed masks in Wu and Bassiouni at each of three coverage levels: 10\%, 20\% and 40\%. The grid included nearest-anchor, uniform, IDW and Gaussian reconstruction; all declared estimators are shown, without choosing an outcome-selected winner (Fig.~\ref{fig:extension_baselines}; Supplementary Table~S10 and source tables). Separately, IDW and uniform interpolation were crossed with the original distributed and regional masks (Fig.~\ref{fig:journal_controls}a; Supplementary Table~S11). This separates the effect of changing the estimator from the controlled placement comparison, while retaining the original protocol-level results. Predetermined gene maps include poorly reconstructed and nearly constant fields (Supplementary Figs.~S13,S14,S26 and S27).
\FloatBarrier
\FloatBarrier

\subsection{Incremental value of graph context across matched visual backbones}
\begin{figure}[p]
\centering\includegraphics[width=\textwidth]{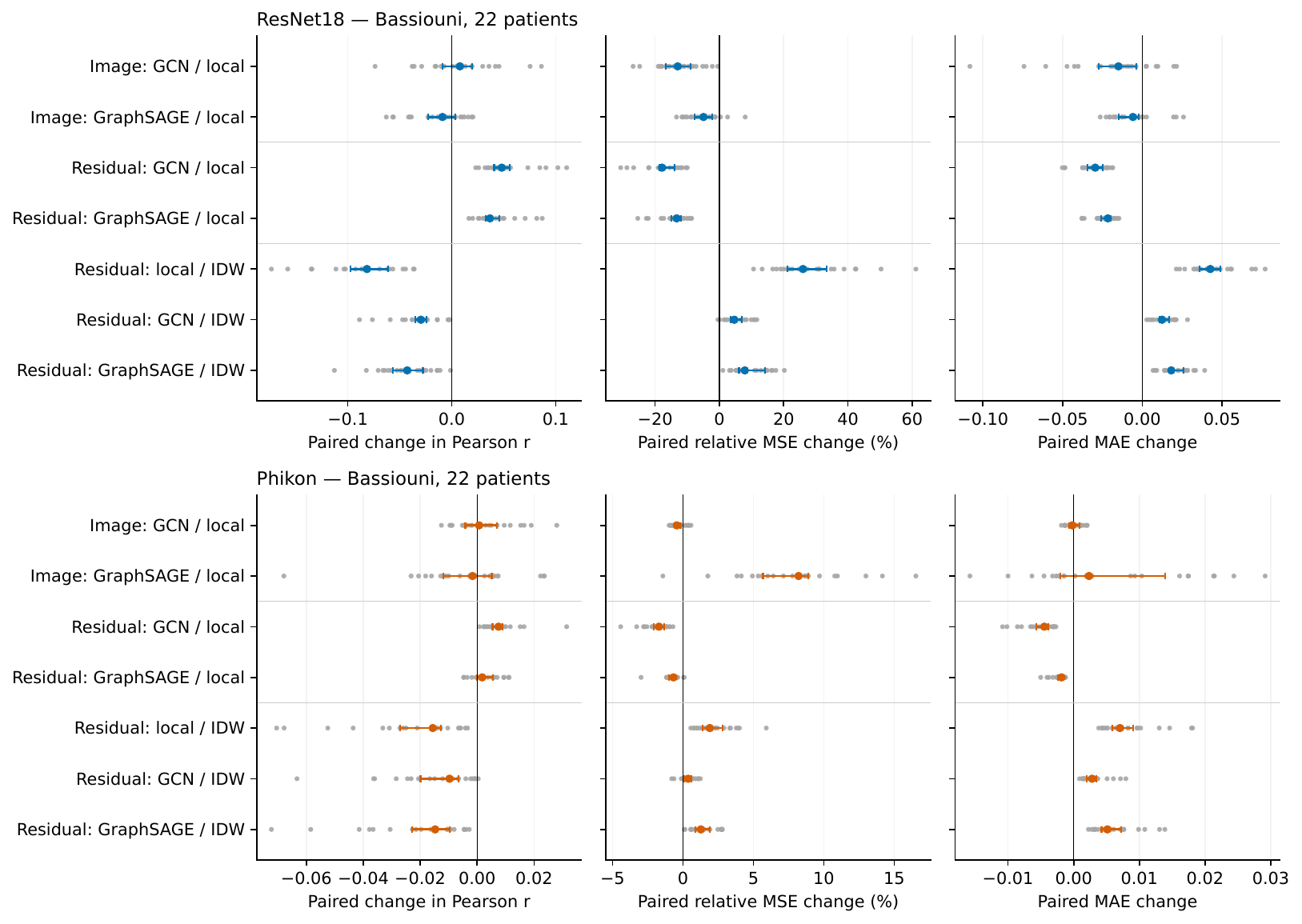}
\caption{\textbf{Graph context and molecular support answer different questions.} Primary 20\% distributed comparisons in 22 Bassiouni patients, separately for ResNet18 and Phikon. Gray points are paired patient effects after mask/training-seed averages; colored points and bars are medians and descriptive 95\% intervals from 10,000 patient resamples. Image rows compare graph versus matched local prediction without same-section expression. Residual rows use identical observed molecular anchors and weights; comparison with local-assisted reconstruction and with anchor-only IDW are separate. Positive Pearson changes and negative error changes favor the candidate. Axis scales differ across panels. All prespecified seeds are retained; model selection used TENX95 only. The simple neighborhood-mean and shuffled controls are reported separately and qualify topology-specific interpretation.}
\label{fig:matched_backbone_effects}
\end{figure}

The matched comparisons separated expression prediction from the incremental value of histology when molecular anchors were available. Within each frozen backbone, local and graph heads used the same registered images, targets and evaluation supports, with equal validation-only selection opportunities (Table~\ref{tab:matched_backbone}; Fig.~\ref{fig:matched_backbone_effects}). Patient effects average mask and training seeds before pairing; the three Wu patients and the single development patient remain separate from Bassiouni.

\paragraph{Image-based prediction.}
Graph context did not produce a uniform image-only advantage. In Bassiouni, ResNet18 GCN versus its matched local predictor changed median paired correlation by $+0.0079$, relative MSE by $-12.97$\% and MAE by $-0.0150$; 14/22, 22/22 and 16/22 patients improved, respectively. GraphSAGE reduced MSE and MAE but decreased correlation ($-0.0089$), a tradeoff. With Phikon, GCN produced small median changes ($+0.0007$ correlation, $-0.43$\% MSE and $-0.0002$ MAE), with correlation improved in 11/22 patients and its descriptive interval spanning zero. Phikon GraphSAGE decreased median correlation by 0.0016 and increased relative MSE by 8.21\% and MAE by 0.0023. In Wu, both Phikon graph heads had small positive median correlation changes but worse median errors; the three individual effects are retained in the source tables.

\paragraph{Reconstruction with identical molecular anchors.}
Adding the same observed-residual reconstruction rule changed the comparison. GCN improved correlation, MSE and MAE over its matched local-assisted predictor in all 22 Bassiouni patients for both encoders. ResNet18 median paired changes were $+0.0481$, $-17.84$\% and $-0.0295$, respectively; Phikon changes were smaller, at $+0.0075$, $-1.70$\% and $-0.0045$. The corresponding three endpoint gains also occurred in all three Wu patients. GraphSAGE improved all three endpoints in all 22 Bassiouni patients with ResNet18; with Phikon, MSE improved in 20/22 and MAE in 22/22, while correlation improved in 15/22 and its interval included zero. The common-valid-gene sensitivity gave the same median GCN correlation gains for both backbones. Thus context could improve a local visual reconstruction without yielding a uniform architecture- or endpoint-independent advantage.

\paragraph{Comparison with molecular interpolation.}
Those gains did not establish an advantage over molecular support alone. Against the identical-anchor IDW reference in Bassiouni, residual GCN gave median paired correlation changes of $-0.0297$ with ResNet18 and $-0.0097$ with Phikon, relative MSE increases of 4.64\% and 0.37\%, and MAE increases of 0.0124 and 0.0028. GraphSAGE also worsened all three cohort-median endpoints versus IDW in both backbones. Phikon GCN in Wu slightly improved median correlation and MSE versus IDW, but MAE worsened in all three patients. Such results are error tradeoffs, not joint superiority. Development-patient improvements against interpolation therefore remain development-specific evidence rather than consistent external gains.

\paragraph{Aggregation and spatial controls.}
Neither graph head improved the three cohort-median endpoints over the neighborhood-mean control in either backbone. Spatial GraphSAGE outperformed its shuffled control on all three median endpoints in both encoders, whereas the ResNet18 GCN shuffled-control comparison traded higher correlation for worse errors. These comparisons use independently validation-selected heads and do not isolate an inference-only edge intervention. Taken together, the controls support testing neighboring image information, but do not establish an advantage of the tested graph normalization or concatenation over simple neighborhood averaging. Complete controls, alternative coverage and placement conditions, undefined-gene counts and common-valid-gene sensitivities remain available in the machine-readable results.

\begin{table}[p]
\centering
\caption{\textbf{Matched graph-versus-local and anchor-reference comparisons.} Primary 20\% distributed coverage uses the same three mask seeds and hidden sites, with three training seeds averaged within patient. Both backbones received the same two-rate validation-only selection opportunity. Entries separate image-only prediction (A) from observed-residual reconstruction (B). Each metric pair gives candidate/reference cohort medians; effects are medians of paired patient changes, so they need not equal differences of these marginal medians. Brackets give improved/valid patients for that endpoint. Relative MSE is computed within patient after seed averaging. External image outputs use the same frozen C9 T1 map once; no same-section molecular values enter A, and all B comparators use identical anchors. Bassiouni intervals appear in Fig.~\ref{fig:matched_backbone_effects}; Wu effects are reported individually in the source tables. Local denotes the non-graph MLP on the same backbone.}
\label{tab:matched_backbone}
\scriptsize\setlength{\tabcolsep}{2pt}\renewcommand{\arraystretch}{1.08}
\begin{tabular}{lrrrrrr}\toprule
Candidate/reference & $r_C/r_R$ & MSE$_C$/MSE$_R$ & MAE$_C$/MAE$_R$ & $\Delta r$ [$n$] & $\Delta$MSE\% [$n$] & $\Delta$MAE [$n$]\\\midrule
\multicolumn{7}{l}{\textbf{A. Image-only prediction}}\\
\multicolumn{7}{l}{ResNet18; Wu, $n=3$}\\
GCN/local & 0.113/0.051 & 0.2486/0.2727 & 0.3786/0.3873 & 0.0621 [3/3] & -10.81 [2/3] & -0.0087 [2/3]\\
GraphSAGE/local & 0.073/0.051 & 0.2532/0.2727 & 0.3750/0.3873 & 0.0225 [2/3] & -4.88 [3/3] & -0.0098 [3/3]\\
\multicolumn{7}{l}{ResNet18; Bassiouni, $n=22$}\\
GCN/local & 0.070/0.062 & 0.3133/0.3376 & 0.4119/0.3997 & 0.0079 [14/22] & -12.97 [22/22] & -0.0150 [16/22]\\
GraphSAGE/local & 0.050/0.062 & 0.3201/0.3376 & 0.3962/0.3997 & -0.0089 [8/22] & -4.92 [19/22] & -0.0059 [17/22]\\
\multicolumn{7}{l}{Phikon; Wu, $n=3$}\\
GCN/local & 0.068/0.053 & 0.2333/0.2323 & 0.3441/0.3436 & 0.0020 [2/3] & 0.40 [1/3] & 0.0016 [0/3]\\
GraphSAGE/local & 0.056/0.053 & 0.2542/0.2323 & 0.3456/0.3436 & 0.0036 [2/3] & 9.39 [0/3] & 0.0092 [0/3]\\
\multicolumn{7}{l}{Phikon; Bassiouni, $n=22$}\\
GCN/local & 0.025/0.027 & 0.3009/0.3010 & 0.3790/0.3788 & 0.0007 [11/22] & -0.43 [16/22] & -0.0002 [12/22]\\
GraphSAGE/local & 0.023/0.027 & 0.3152/0.3010 & 0.3774/0.3788 & -0.0016 [9/22] & 8.21 [1/22] & 0.0023 [9/22]\\
\midrule
\multicolumn{7}{l}{\textbf{B. Observed-residual reconstruction}}\\
\multicolumn{7}{l}{ResNet18; Wu, $n=3$}\\
GCN/local & 0.336/0.290 & 0.0982/0.1161 & 0.2251/0.2513 & 0.0455 [3/3] & -14.29 [3/3] & -0.0234 [3/3]\\
GraphSAGE/local & 0.326/0.290 & 0.1018/0.1161 & 0.2316/0.2513 & 0.0357 [3/3] & -11.05 [3/3] & -0.0180 [3/3]\\
local/IDW & 0.290/0.357 & 0.1161/0.0985 & 0.2513/0.2176 & -0.0732 [0/3] & 17.85 [0/3] & 0.0316 [0/3]\\
GCN/IDW & 0.336/0.357 & 0.0982/0.0985 & 0.2251/0.2176 & -0.0317 [1/3] & 4.15 [1/3] & 0.0082 [0/3]\\
GraphSAGE/IDW & 0.326/0.357 & 0.1018/0.0985 & 0.2316/0.2176 & -0.0503 [1/3] & 6.19 [0/3] & 0.0137 [0/3]\\
\multicolumn{7}{l}{ResNet18; Bassiouni, $n=22$}\\
GCN/local & 0.392/0.345 & 0.0971/0.1194 & 0.2211/0.2506 & 0.0481 [22/22] & -17.84 [22/22] & -0.0295 [22/22]\\
GraphSAGE/local & 0.380/0.345 & 0.1015/0.1194 & 0.2284/0.2506 & 0.0368 [22/22] & -13.27 [22/22] & -0.0216 [22/22]\\
local/IDW & 0.345/0.418 & 0.1194/0.0922 & 0.2506/0.2027 & -0.0816 [0/22] & 26.01 [0/22] & 0.0427 [0/22]\\
GCN/IDW & 0.392/0.418 & 0.0971/0.0922 & 0.2211/0.2027 & -0.0297 [0/22] & 4.64 [1/22] & 0.0124 [0/22]\\
GraphSAGE/IDW & 0.380/0.418 & 0.1015/0.0922 & 0.2284/0.2027 & -0.0428 [0/22] & 7.92 [0/22] & 0.0182 [0/22]\\
\multicolumn{7}{l}{Phikon; Wu, $n=3$}\\
GCN/local & 0.353/0.348 & 0.0980/0.0983 & 0.2189/0.2214 & 0.0045 [3/3] & -0.46 [3/3] & -0.0015 [3/3]\\
GraphSAGE/local & 0.352/0.348 & 0.0982/0.0983 & 0.2209/0.2214 & 0.0006 [3/3] & -0.09 [2/3] & -0.0005 [3/3]\\
local/IDW & 0.348/0.357 & 0.0983/0.0985 & 0.2214/0.2176 & -0.0042 [1/3] & 0.49 [1/3] & 0.0026 [0/3]\\
GCN/IDW & 0.353/0.357 & 0.0980/0.0985 & 0.2189/0.2176 & 0.0004 [2/3] & -0.04 [2/3] & 0.0010 [0/3]\\
GraphSAGE/IDW & 0.352/0.357 & 0.0982/0.0985 & 0.2209/0.2176 & -0.0040 [1/3] & 0.49 [1/3] & 0.0019 [0/3]\\
\multicolumn{7}{l}{Phikon; Bassiouni, $n=22$}\\
GCN/local & 0.403/0.390 & 0.0924/0.0938 & 0.2059/0.2103 & 0.0075 [22/22] & -1.70 [22/22] & -0.0045 [22/22]\\
GraphSAGE/local & 0.400/0.390 & 0.0933/0.0938 & 0.2084/0.2103 & 0.0018 [15/22] & -0.67 [20/22] & -0.0019 [22/22]\\
local/IDW & 0.390/0.418 & 0.0938/0.0922 & 0.2103/0.2027 & -0.0156 [0/22] & 1.91 [0/22] & 0.0070 [0/22]\\
GCN/IDW & 0.403/0.418 & 0.0924/0.0922 & 0.2059/0.2027 & -0.0097 [1/22] & 0.37 [4/22] & 0.0028 [0/22]\\
GraphSAGE/IDW & 0.400/0.418 & 0.0933/0.0922 & 0.2084/0.2027 & -0.0147 [0/22] & 1.27 [0/22] & 0.0051 [0/22]\\
\bottomrule
\end{tabular}
\end{table}

\FloatBarrier

\begin{figure}[!htbp]
\centering\includegraphics[width=\textwidth]{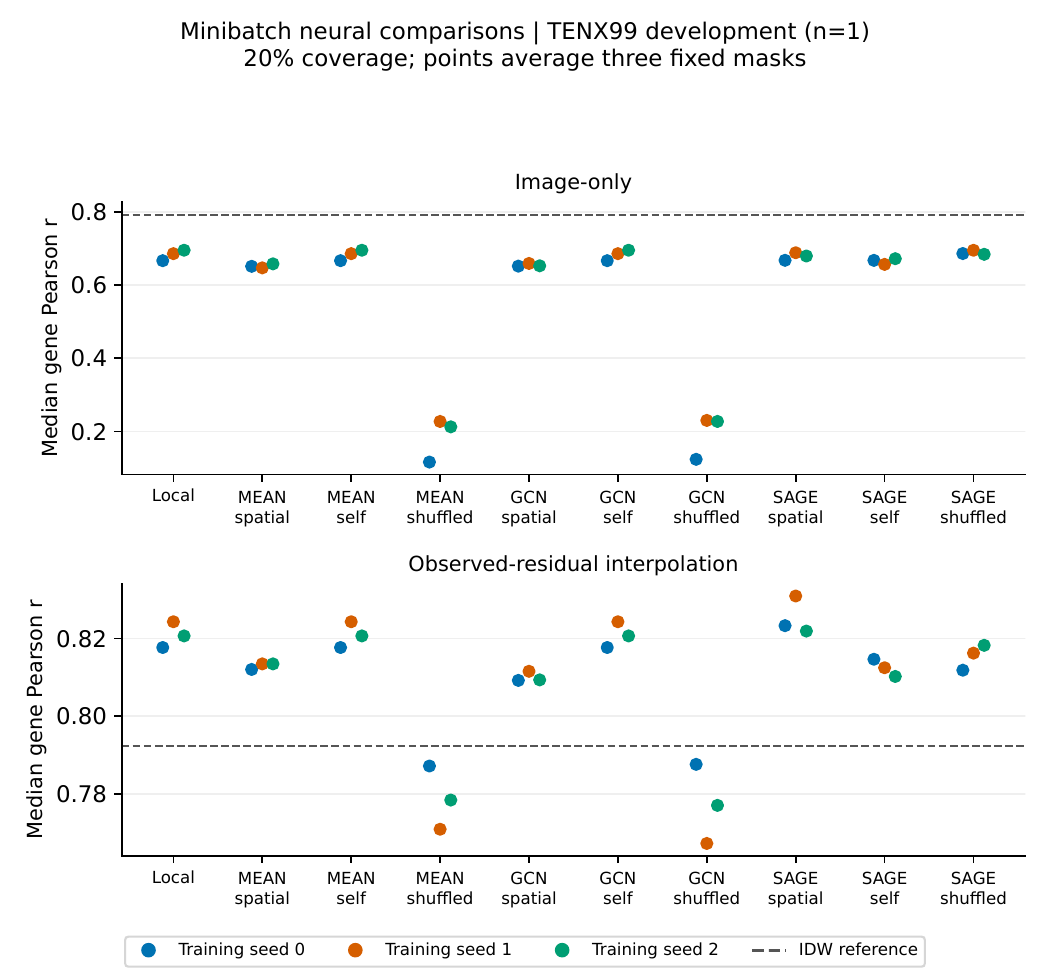}
\caption{\textbf{Minibatch neural comparisons in the development patient.} Three training seeds for each local, mean, GCN and GraphSAGE configuration, including self-only and node-shuffled controls; all evaluated on TENX99, a single previously used development patient. Points average three fixed masks at 20\% coverage. Dashed lines give identical-mask IDW reference values. Top: image-only prediction; bottom: the protocol-specified exploratory residual-anchor baseline. Color identifies training seed, not an independent patient. Panels use different y-ranges. IDW receives measured expression unavailable to the image-only branch; residual versus IDW tests incremental morphology. These fits were not tuned on Wu or Bassiouni outcomes.}
\label{fig:journal_learned}
\end{figure}

\paragraph{Development-patient context.}
The GNN comparisons included local multilayer perceptrons, fixed mean/GCN/GraphSAGE representations, bounded nonlocal models and fixed-distance or latent-coordinate message passing. Their architectural and optimization regimes are distinguished in Methods. In the 30-head minibatch development analysis, image-only seed-mean correlations for local, GCN and GraphSAGE predictors were 0.682, 0.654 and 0.678, with MSEs 0.9114, 0.8999 and 0.8859. Residual-anchor correlations were 0.821, 0.810 and 0.825, with MSEs 0.2567, 0.2510 and 0.2338; the identical-mask IDW reference was 0.792 and 0.2684. These one-patient results did not establish a topology-specific advantage (Fig.~\ref{fig:journal_learned}; Supplementary Table~S13). The full seed comparisons, including shuffled controls, remain available.
\FloatBarrier

\subsection{Nonlocal and latent-coordinate updates did not improve molecular interpolation}

\begin{figure}[!htbp]
\centering\includegraphics[width=\textwidth]{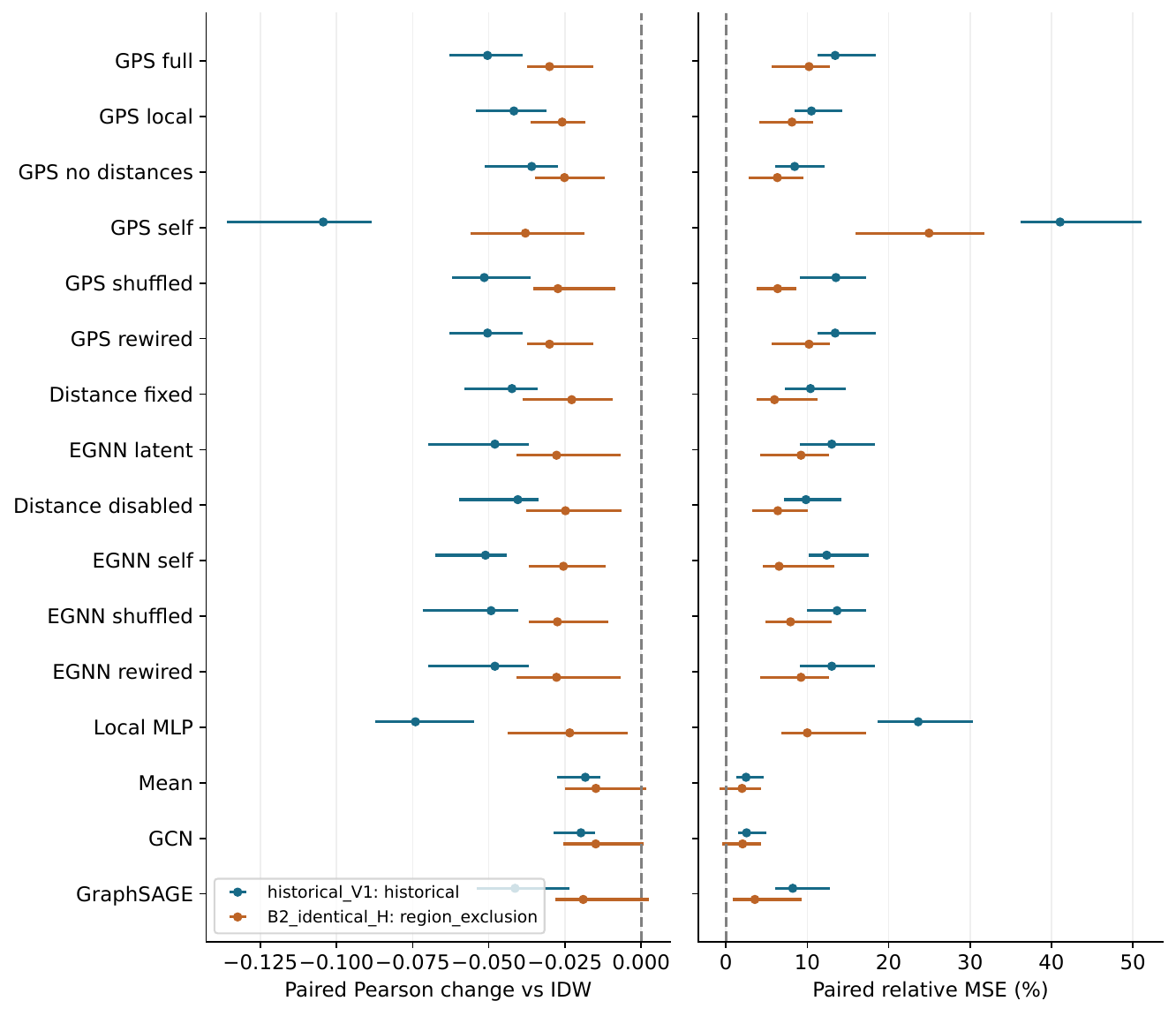}
\caption{\textbf{All 16 full-section configurations versus molecular IDW in 22 Bassiouni patients.} Exploratory residual route at 20\% coverage; colors distinguish historical distributed V1 and identical-H region exclusion. Bars are descriptive 95\% patient percentile intervals for median paired effects after mask/training-seed averages. Positive correlation and negative relative MSE improve. The two designs have different scored support and are not subtracted here. All models are shown regardless of direction; nominally equal parameter counts do not imply equal effective function classes.}
\label{fig:extension_neural}
\end{figure}

The full-section architectural comparison retained all 48 fits alongside the 30 frozen minibatch fits in TENX99, Wu 3 and Bassiouni 22, using identical masks and information routes. Architectural definitions and permitted inputs are specified in Methods. In the primary 20\% distributed Bassiouni comparison, the residual-anchor GraphGPS-inspired model changed paired patient correlation versus IDW by $-0.0504$ (95\% interval $-0.0629$ to $-0.0389$), relative MSE by $+13.4$\% (11.3--18.4\%) and MAE by $+0.0244$ (0.0204--0.0295). The latent-coordinate EGNN gave $-0.0480$ ($-0.0698$ to $-0.0368$), $+13.0$\% (9.1--18.3\%) and $+0.0252$ (0.0188--0.0310), respectively. Neither primary model improved any of these endpoints in any of the 22 patients. Under identical-$H$ region exclusion, their median relative MSE increases were 10.2\% and 9.2\% (Fig.~\ref{fig:extension_neural}).

Component comparisons also failed to establish the proposed gains: the full nonlocal model worsened correlation and errors relative to its local counterpart, and latent-coordinate updates did not improve the fixed-distance model (Table~\ref{tab:extension_controls}). Same-information conventional neural comparisons, all routes, individual Wu effects, training seeds and distance-bin support are retained in Supplementary Figs.~S15--S25,S28 and Tables~S14--S17. Conditional geometric and reach checks confirm the implemented properties; they do not establish biological long-range association or full-image equivariance. These negative findings concern the tested training budget, frozen features and low-resolution cross-cohort inputs.

\FloatBarrier
\FloatBarrier
\begin{table}[!htbp]
\caption{\textbf{Predefined architecture contrasts in recovered Bassiouni patients.} At 20\% historical V1 coverage, the exploratory residual route uses identical measured anchors for both models. Effects average all three training seeds and three mask seeds within each patient before pairing; relative MSE is normalized by the reference patient mean MSE after these averages. Intervals resample 22 whole patients, not genes, masks or training seeds. They are descriptive, unadjusted intervals, not simultaneous significance claims. Other routes, controls, budgets and individual Wu effects are fully exported.}
\label{tab:extension_controls}
\centering\footnotesize
\begin{tabular}{@{}lrr@{}}
\toprule
Candidate vs control & Effect & Median [95\% interval] \\
\midrule
GPS full vs GPS local & $\Delta r$ & -0.0050 [-0.0074, -0.0044] \\
GPS full vs GPS local & MSE\% & +2.7228 [+2.0494, +3.6169] \\
GPS full vs GPS local & $\Delta$ MAE & +0.0039 [+0.0030, +0.0054] \\
GPS full vs GPS no distances & $\Delta r$ & -0.0121 [-0.0135, -0.0088] \\
GPS full vs GPS no distances & MSE\% & +4.9138 [+3.6647, +5.7686] \\
GPS full vs GPS no distances & $\Delta$ MAE & +0.0064 [+0.0056, +0.0078] \\
EGNN latent vs Distance fixed & $\Delta r$ & -0.0048 [-0.0074, -0.0008] \\
EGNN latent vs Distance fixed & MSE\% & +1.6673 [+0.7107, +2.8664] \\
EGNN latent vs Distance fixed & $\Delta$ MAE & +0.0026 [+0.0014, +0.0046] \\
EGNN latent vs Distance disabled & $\Delta r$ & -0.0065 [-0.0096, -0.0024] \\
EGNN latent vs Distance disabled & MSE\% & +1.9736 [+1.0100, +3.3448] \\
EGNN latent vs Distance disabled & $\Delta$ MAE & +0.0028 [+0.0020, +0.0053] \\
\bottomrule
\end{tabular}
\end{table}
\FloatBarrier

\subsection{Regional completion, distance and calibration defined failure regimes}
The original contiguous-completion protocol was substantially harder than distributed reconstruction. At 20\% coverage, Bassiouni correlation fell from 0.416 to 0.171, MSE increased from 0.0923 to 0.1496 and MAE from 0.2033 to 0.2624. Every patient deteriorated; median paired correlation change was $-0.267$ and relative MSE increase 52.8\%. Wu correlation fell from 0.359 to 0.0815, with lower correlation and higher errors in every patient (Supplementary Table~S9). Because the interpolation rule and scored tissue also change, these effects must not be attributed solely to geometry. A withheld region may still lie inside the anchors' convex hull.

\begin{figure}[!htbp]
\centering
\includegraphics[width=\textwidth,height=0.60\textheight,keepaspectratio]{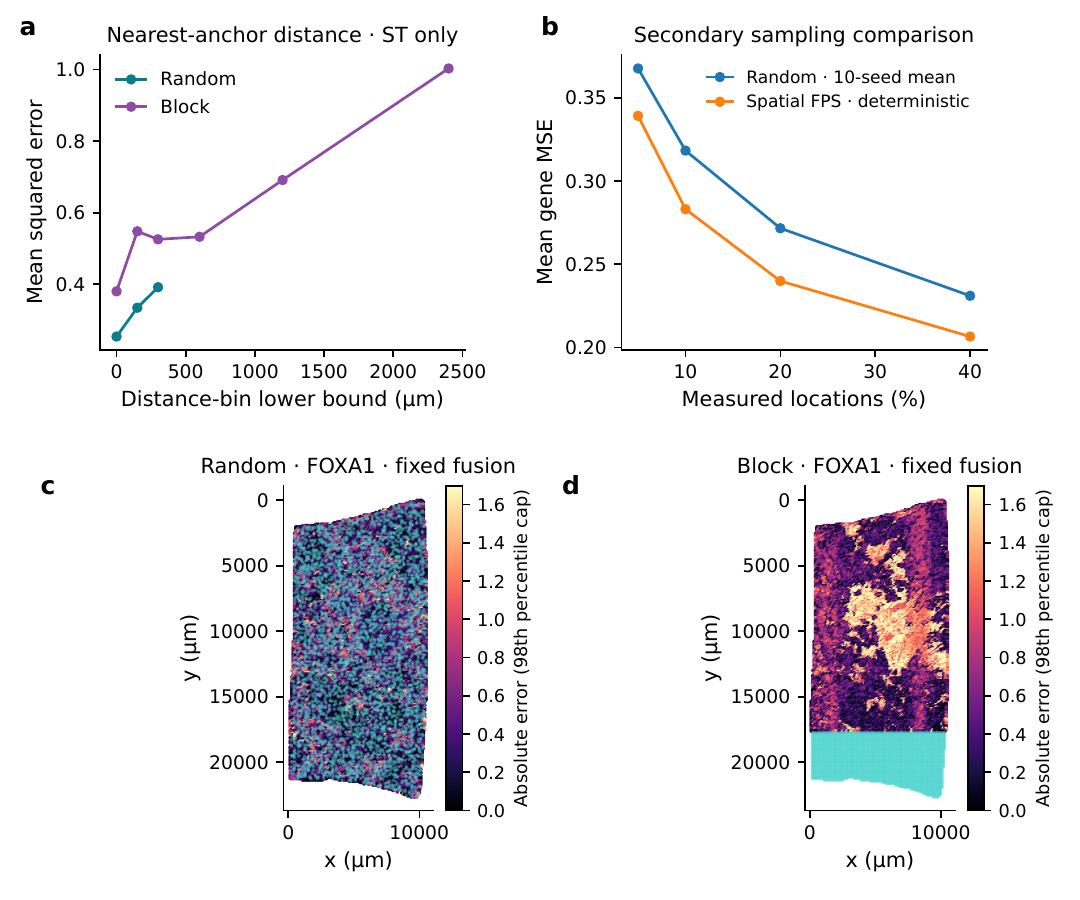}
\caption{\textbf{Spatial diagnostics reveal heterogeneous error beyond an overall prediction score.} Development results use TENX99 and the frozen 50-gene panel. \textbf{a,} Mean gene MSE versus nearest-anchor distance for molecular interpolation at 20\% coverage under distributed and development block masks; only occupied distance bins are shown. \textbf{b,} Development comparison of random and spatial farthest-point sampling, shown as MSE at matched measured-location budgets. These are descriptive single-patient sampling diagnostics, not a demonstration of universally optimal acquisition. \textbf{c,d,} Absolute-error maps for the preselected gene \textit{FOXA1}, fixed fusion, 20\% coverage and replicate 0 under random and block masks. Both maps use the same error scale, capped at the 98th percentile of the pooled preselected map-source values, which also include the archived capture-window condition; no locations are removed. Missing-region shape, available anchors and scored tissue differ between the maps. The figure illustrates why distance-bin support and spatial errors should accompany global correlation and why development geometry findings require independent evaluation.}
\label{fig:geometry}
\end{figure}

Distance-resolved errors were heterogeneous (Fig.~\ref{fig:geometry}). At 20\% Bassiouni coverage, farthest-bin MSE exceeded nearest-bin MSE in 15/22 patients under distributed interpolation and 20/22 under regional completion; strictly increasing curves occurred in only 6 and 13 patients. Wu distributed curves increased across occupied bins, while regional curves were nonmonotonic. Bin support and tissue composition differ, precluding a universal reliability cutoff (Supplementary Table~S7).

\FloatBarrier

Fixed image fusion also failed to improve external reconstruction consistently. Bassiouni correlation, MSE and MAE at 20\% distributed coverage were 0.407, 0.1022 and 0.2288, versus 0.416, 0.0923 and 0.2033 for interpolation. Only 3/22 patients improved in correlation and none in either error. The median paired correlation effect was $-0.0108$ (95\% interval $-0.0187$ to $-0.0081$), with relative MSE increase 10.7\% (6.6--14.5\%). Wu showed correlation improvement in 1/3 patients and no error improvement. The original anchor-versus-image paired correlation gain was 0.329 (0.304--0.438), a contrast with different information access (Table~\ref{tab:journal_uncertainty}). HER2ST calibration reduced scale-related error but did not establish robust joint correlation/error gains (Supplementary Figs.~S1,S9 and S10). An affine transform followed by normalized interpolation and its inverse left molecular predictions unchanged, as expected algebraically (Supplementary Methods).
\begin{table}[!htbp]
\caption{\textbf{Patient-level uncertainty reanalysis of original Bassiouni exports.} Median paired effects at 20\% V1 coverage, with 10,000 patient-resampling percentile 95\% intervals (seed 20261101). Negative error changes improve; positive correlation changes improve. These are descriptive exploratory intervals, not simultaneous bands or p-values. Anchors versus image is an unequal-information contrast; fusion versus anchors tests incremental image contribution. All original point estimates and source bytes are unchanged.}
\label{tab:journal_uncertainty}
\centering\footnotesize
\begin{tabular}{@{}lrrrr@{}}
\toprule
Contrast & Effect & Median & 95\% interval & Improved \\
\midrule
Anchors minus image & $\Delta r$ & 0.3293 & [0.3044, 0.4382] & 22/22 \\
Anchors minus image & Relative MSE (\%) & -95.4534 & [-96.2444, -94.0201] & 22/22 \\
Anchors minus image & $\Delta$ MAE & -1.0074 & [-1.1341, -0.9108] & 22/22 \\
Fusion minus anchors & $\Delta r$ & -0.0108 & [-0.0187, -0.0081] & 3/22 \\
Fusion minus anchors & Relative MSE (\%) & 10.7078 & [6.6170, 14.5048] & 0/22 \\
Fusion minus anchors & $\Delta$ MAE & 0.0238 & [0.0171, 0.0278] & 0/22 \\
\bottomrule
\end{tabular}
\end{table}
\FloatBarrier

Anchor-vector and anchor-position permutations degraded molecular reconstruction in predefined controls; image permutation supplied no comparable evidence for a useful image contribution (Supplementary Figs.~S11 and S12; Table~S4). These related controls and exact hidden-label invariance checks support computational traceability without supplying additional biological replicates.

\section{Discussion}
Measurement placement was a determinant of reconstruction quality even when the amount of observed expression was held fixed. The 22-patient comparison shows that equal budgets can produce different performance when anchors are excluded from the region containing the scored targets. Its strength is the fixed hidden set and estimator, which separate placement from changes in tissue support or reconstruction rule. The result applies to the specified central-region design; it does not establish a universally optimal sampling policy or biological correlation length.

Distributed molecular anchors nevertheless supported useful reconstruction of the predefined panel across two breast-cancer studies. The coverage response suggests that nearby same-section measurements carry information that a transported image predictor may not recover. This is a task-specific advantage of measured expression, not an equal-information demonstration that interpolation outperforms all image-based systems. Reliability should be assessed on the intended acquisition design: favorable distributed masking does not ensure accurate completion of a large unmeasured region~\cite{roberts2017}.

The matched-backbone results refine the role of graph context. With identical molecular anchors, a GCN improved on its matched local visual predictor across both encoders in all 22 Bassiouni patients, although the magnitude depended strongly on the representation. This is a benefit relative to a non-graph visual reconstruction, not evidence of superiority to molecular interpolation. Image-only gains were less consistent, and some correlation improvements accompanied worse errors. Neighborhood averaging remained a strong control, preventing attribution of the observed benefit specifically to the tested graph operators.

The broader architectural comparisons retain an important negative finding: nonlocal access and latent-coordinate updates did not establish the proposed external advantage over the same molecular reference. Functional attention reach and conditional equivariance are software properties, rather than evidence of learned molecular relationships. Together, the two comparisons distinguish useful context relative to a local predictor from incremental value beyond measured expression, and support retaining strong molecular and aggregation baselines.

Cross-cohort histology remains a substantial limitation. Public images were heavily downsampled, encoders were frozen, and fitting used only two source patients. Cross-backbone differences include pretraining, feature dimension and effective physical field of view; they cannot be attributed solely to adding graphs. The frozen transported calibration was shared within comparisons rather than re-estimated for each model on evaluation labels. Differences in capture depth, support and composition remain after mathematical normalization. Calibration can change image predictions, but cannot create a new interpolation benefit when a per-gene affine transform is inverted around the same normalized weighted average. Native-resolution representations, broader source training and genuinely independent evaluation would be needed to establish whether morphology adds reproducible incremental information.

The scope is also limited by retrospective masking, a development-selected 50-gene panel (40 genes in HER2ST), small Wu patient numbers and restriction to breast tissue. The exploratory controls reused evaluated cohorts. Genes, sites and repeated masks do not increase biological sample size, and descriptive patient intervals do not establish clinical generalizability. Distance alone did not define a reliable error threshold; occupied-bin support and patient-specific failures should remain visible before reconstructed expression is used downstream. Approaches such as TISSUE motivate explicit uncertainty assessment~\cite{sun2024}, but calibrated predictive uncertainty was not evaluated here.

Coverage is a count of measured locations, not a measured financial saving. If a full assay has fixed cost $C_0$ and per-location cost $c$ for $N$ locations, the conditional ratio is
\begin{equation}
\frac{C_{\mathrm{full}}}{C_{20\%}}=\frac{C_0+cN}{C_0+0.2cN}\leq5.
\label{eq:cost_ratio}
\end{equation}
Only a purely location-dependent component could fall fivefold at 20\% coverage. Selective capture, laboratory expenditure, sequencing requirements and downstream accuracy were not prospectively tested. Future studies could selectively profile a subset of tissue locations and reconstruct the predefined expression panel at the remaining locations, while evaluating placement, reconstruction error and realized costs prospectively.

GraphPathology therefore supports an information-explicit account of sparse reconstruction: measured molecular support can be useful, its placement matters under controlled conditions, and contextual gains over local predictors must be distinguished from improvement over the same molecular support. Prospective sampling and independent native-resolution cohorts are the next tests of practical value.
\section{Methods}
\subsection{Data, target panel and preprocessing}
All experiments used previously released paired spatial expression and histology data. The four development samples were obtained through HEST~\cite{jaume2024}: NCBI783 and NCBI785 were the training patients, TENX95 was the validation patient and TENX99 was the primary development evaluation patient. The cohort comprised 7,015 training locations, 7,760 validation locations and 20,761 primary evaluation locations. Earlier conditional leave-one-patient-out analyses reused frozen gene panels and development-selected configurations and were interpreted as sensitivity analyses, not unbiased external cross-validation.

HER2ST~\cite{andersson2021} contributed one patient-specific section for each of A1--H1. Initial evaluation was followed by patient-disjoint calibration development using the same inspected cohort; these two analytical roles are distinguished throughout. In the Wu local Visium cohort~\cite{wu2021}, patients 1142243F, CID4465 and CID4535 supplied calibration statistics, and patients 1160920F, CID4290 and CID44971 remained held out. The Bassiouni evaluation cohort was obtained from GSE210616~\cite{bassiouni2023}. The source study includes 43 sections from 22 patients. We selected the first numbered section for every published patient from metadata before target evaluation, yielding BASS\_P01--P09, P11--P15 and P17--P24. Gaps in identifiers reflect source numbering rather than outcome-dependent exclusion. The source study's own reference and validation groups were not used as calibration groups in this reconstruction study.

Metadata, repository identifiers, prior-use records and available asset hashes were checked to exclude reused or reprocessed development samples. Such checks support cohort independence but cannot conclusively establish patient identity across unrelated anonymized repositories. Cohort roles and sample counts are summarized in Table~\ref{tab:cohorts}; identifiers and exact panels are supplied in Supplementary Tables~S1 and S2.

The primary panel contained the 50 genes ranked using development-training expression variance; ranking did not use evaluation expression. The development assay provided 280 harmonized biological genes, and larger 250- and 280-gene settings were used in earlier development analyses. The original external studies retained all 50 genes. HER2ST retained the frozen 40-gene intersection without replacing missing genes. The optional 278/280 intersection in the Bassiouni metadata was not used to claim a secondary whole-transcriptome experiment.

For the second replication, the source filtered feature-barcode H5 files, tissue positions, scale-factor JSON files and tissue images were acquired for each selected section. Barcode joins and published in-tissue flags defined included locations. No new expression-derived quality threshold, tissue-class annotation or deconvolution output was used to select evaluation locations. Counts were checked for finite, nonnegative integer values. For count $c_{ig}$ at location $i$ and gene $g$, the evaluation target was
\begin{equation}
 y_{ig}=\frac{\log(1+c_{ig})-\mu_g^{\mathrm{dev}}}{\sigma_g^{\mathrm{dev}}},
 \label{eq:target}
\end{equation}
where the centering and scaling constants were the pooled location mean and population standard deviation frozen from development-training patients; a standard deviation below $10^{-8}$ was replaced by 1. No destination library-size or capture-area correction was fitted. Identical target units were used for all methods within each comparison. These common mathematical units do not imply cross-platform calibration of absolute molecular abundance. Supplementary Fig.~S2 summarizes cross-cohort patterns without pooling panels; the gene-level results are provided in Supplementary Table~S3.

\subsection{Registration and frozen image features}\label{sec:registration}
Visium array coordinates were mapped to image coordinates by a least-squares fit to the source hexagonal lattice with 100-\um{} spacing. Coordinate residuals, image scale and anisotropy were checked against the frozen qualification settings, and overlays were inspected before outcome evaluation. These tests concern location consistency and do not establish cell-level registration. Images were not super-resolved.

The Wu and Bassiouni images contained only 15/25 and 24--29 native pixels, respectively, across the nominal 112-\um{} patch width. Resizing does not recover morphology absent from these public inputs. Image quality was therefore recorded as a limitation of the image comparator, not as a restriction on spatial interpolation that uses only coordinates and observed expression.

At each external location, an RGB patch covered a nominal $112$-\um{} square, rounded to the available native pixels. The patch was resized with LANCZOS to $224\times224$ pixels. The torchvision \texttt{IMAGENET1K\_V1} preprocessing then resized the shorter side to 256 pixels using bilinear interpolation, took a central $224\times224$ crop, rescaled intensities to $[0,1]$, and normalized RGB channels using means $(0.485, 0.456, 0.406)$ and standard deviations $(0.229, 0.224, 0.225)$. The second resize and central crop were retained explicitly; $112$~\um{} denotes the nominal extracted patch, rather than the final uncropped field seen by the network. Development embeddings used the supplied $224\times224$ HEST patches with the same torchvision transform. No stain normalization, random image augmentation or encoder fine-tuning was applied.

The frozen ResNet18~\cite{he2016} comprised a $7\times7$ convolutional stem with stride 2, batch normalization, a rectified-linear activation and $3\times3$ max pooling with stride 2, followed by four residual stages. The stages contained two basic blocks each, with channel widths 64, 128, 256 and 512. Each block used two $3\times3$ convolutions with batch normalization and a rectified-linear activation after the first convolution and after shortcut addition. Identity or projected shortcuts matched the feature dimensions. Global average pooling produced a 512-dimensional vector; the original ImageNet classifier was replaced by the identity operation. All encoder parameters and batch-normalization statistics remained fixed during feature extraction.

The final central crop retains approximately 0.875 of the extracted patch linear field.

For the matched-backbone comparison, we used the frozen Phikon ViT-B/16 histopathology encoder and its published preprocessing pipeline~\cite{dosovitskiy2021vit,filiot2023}. Its official ViTImageProcessor consumes the same registered RGB patches, resizes to $224\times224$, rescales by $1/255$ and uses ImageNet channel means and standard deviations, without augmentation or stain normalization. The 12-layer encoder preserves the full patch field and returns the 768-dimensional final-layer-normalized classification token without a pooler. ViT attention relates tokens within a patch; the tissue graph adds relationships between registered locations. The ResNet18 transform retains $224/256=0.875$ of each patch width after resizing and center cropping; its effective field therefore differs from Phikon's. Within each backbone, all matched heads use identical input and embedding bytes. Cross-backbone differences include field of view, pretraining and representation dimension and are not attributed solely to architecture class. Phikon extraction used float32 scaled dot-product attention with four concurrent eight-patch forwards, one CPU intra-operation thread per forward, and no encoder gradient computation. Source-only resource benchmarks and exact serial/concurrent checks preceded extraction; interrupted partial caches and resource revisions were retained.

\subsection{Coordinate-defined graph representations}
\label{sec:graph_methods}
Each retained location is a node with physical position $q_i\in\mathbb R^2$ in micrometers and a frozen image embedding. Following message-passing principles~\cite{gilmer2017}, graphs were constructed independently within each section from physical coordinates, without expression labels. The candidates were symmetrized six-nearest-neighbor adjacency capped at 250~\um{}, radius graphs at 150 or 250~\um{}, and Delaunay adjacency capped at 250~\um{}. Symmetrization takes the union of directed nearest-neighbor edges, so final degree need not equal six. The binary adjacency $A$ had no self edges. With $\widetilde A=A+I$, degree matrix $\widetilde D$ of $\widetilde A$, and degree matrix $D$ of $A$, the three fixed operators were
\[
 P=\widetilde D^{-1}\widetilde A,\qquad
 Q=\widetilde D^{-1/2}\widetilde A\widetilde D^{-1/2},\qquad
 R=D^{-1}A.
\]
For $R$, an isolated node has a zero neighbor aggregate. The input to the MLP was $PX$ for mean pooling, $QX$ for GCN and $[X,RX]$ for mean-GraphSAGE, where $X$ contains standardized frozen image embeddings and brackets denote concatenation. Each predictor therefore applies one fixed graph aggregation followed by the one-hidden-layer MLP with the regime-specific dimensions and scaling specified below. These single-step predictors used neither learned adjacency nor attention; the additional message-passing models are defined below.

Graph choice was selected at seed 0 by the highest validation median gene Pearson correlation, then lower MSE and deterministic graph-name order. All three 50-gene predictors selected capped Delaunay adjacency. Mean pooling and GCN used width 128; GraphSAGE used width 67 to approximately match the initial local head's parameter count after doubling the input dimension. With ResNet18, these models contain 72,114 and 72,075 trainable head parameters, respectively; all use dropout 0.1. Supplementary Table~S8 summarizes the selected dimensions.

\subsection{Neural predictors and distinct training regimes}
\label{sec:experiments}
All neural fitting used NCBI783/NCBI785 and validation-only stopping on TENX95. The following regimes are distinct; parameter similarity does not imply identical update schedules or effective capacity.

\paragraph{Original primary image ensemble.}
Each expression head was a one-hidden-layer MLP with the exact sequence
\begin{equation}
512 \xrightarrow{\mathrm{linear}}256
\xrightarrow{\mathrm{GELU}}256
\xrightarrow{\mathrm{dropout}\;(0.1)}256
\xrightarrow{\mathrm{linear}}50,
\label{eq:head}
\end{equation}
where GELU denotes the Gaussian error linear unit. Both linear layers included biases and used the default PyTorch initialization, uniform on $[-1/\sqrt{d_{\mathrm{in}}},1/\sqrt{d_{\mathrm{in}}}]$. The output was linear, with no nonnegativity constraint. Each head contained 144,178 trainable parameters. Feature-wise centering and scaling, $S_x$, were fitted to the pooled 7,015 training locations from NCBI783 and NCBI785. A separate per-gene scaler, $S_y$, was fitted to their already normalized expression targets. Both used population standard deviations, replacing scales below $10^{-8}$ with 1. If $f_{\theta_s}$ denotes the head trained with seed $s\in\{0,1,2\}$ and $\phi_{\mathrm{R18}}$ the frozen encoder, the image prediction in the common target units was
\begin{equation}
\widehat{\boldsymbol y}^{\mathrm{image}}_i
=\frac{1}{3}\sum_{s=0}^{2}
S_y^{-1}\!\left[f_{\theta_s}\!\left(S_x\!\left(\phi_{\mathrm{R18}}(x_i)\right)\right)\right].
\label{eq:image_ensemble}
\end{equation}
Dropout was disabled at inference. The three inverse-scaled predictions were averaged without fitting ensemble weights.

Heads minimized mean squared error in the additionally standardized target space using AdamW, learning rate $10^{-3}$, weight decay $10^{-4}$, $(\beta_1,\beta_2)=(0.9, 0.999)$ and $\epsilon=10^{-8}$. Weight decay applied to all head parameters, including biases. Each epoch shuffled the training locations into minibatches of 512, retaining the final partial batch, for at most 80 epochs. Training used CPU float32 computation and no learning-rate scheduler. Early stopping used the TENX95 validation patient: an improvement required validation MSE to decrease by more than $10^{-4}$, and training stopped after 10 consecutive epochs without such an improvement. The best qualifying parameters were restored. Hidden widths $\{128,256\}$ and dropout probabilities $\{0, 0.1\}$ were compared at seed 0; the configuration with the highest validation median gene Pearson correlation was selected, with lower validation MSE breaking ties. This configuration-selection criterion was distinct from the within-fit stopping criterion. The selected width-256, dropout-0.1 heads were fitted with seeds 0, 1 and 2; their retained epochs were 12, 11 and 12, respectively. No external patient was used to select or refit these heads.

\paragraph{Additional minibatch heads.}
Thirty additional local, mean, GCN and GraphSAGE heads (three seeds per specified configuration/control) used the following settings. They were fitted separately from the original primary ensemble and were later frozen as references.
A single learned linear/GELU/dropout/readout layer followed mean or symmetric GCN aggregation; GraphSAGE concatenated self and neighbor mean. Widths were 128 for local/mean/GCN and 67 for SAGE (72,114 versus 72,075 parameters). Spatial, self-only and degree-preserving node-shuffled controls shared each family's nominal parameters, labels and training settings. SAGE self-only concatenated X with X and has a smaller effective function class. AdamW used lr $10^{-3}$, decay $10^{-4}$, batch size 512, dropout 0.1, maximum 80 epochs, patience 10 and minimum validation improvement $10^{-4}$; TENX95 alone selected stopping. Seeds 0/1/2 were all retained. Deterministic float32 CPU execution used four torch/BLAS threads; Local, mean-self and GCN-self identity controls used the same inputs. Their float32 weight/prediction comparisons allowed rtol=atol=$32\epsilon_{\mathrm{float32}}=3.8147\times10^{-6}$; this tolerance did not relax float64 molecular replay or exact hidden-label invariance. Further identity diagnostics are in Supplementary Methods.

The matched-backbone follow-up reused the local, mean, GCN, GraphSAGE and shuffled GCN/GraphSAGE configurations from this minibatch regime (three seeds each), retaining the remaining historical controls separately. Phikon used local/mean/GCN width 128 and GraphSAGE width 66, giving 104,882 and 104,792 parameters; the within-backbone capacity mismatch was below 0.1\%. Approximate parameter matching does not equalize hidden width, functional class or inference cost. Separate population feature scalers were fitted on NCBI783/NCBI785 for each encoder; both used common C0 targets without another head-target scaler. The encoders stayed frozen.

A documented validation-guided refinement followed the initial Phikon fits, 17 of which selected the first epoch while later fitting loss declined and validation error rose. Before any Phikon evaluation outcomes, every backbone/configuration was assigned the same two learning-rate candidates, $10^{-3}$ and $10^{-4}$, with the minibatch and stopping rules above. Eighteen existing ResNet18 fits and 18 initial Phikon fits supplied the first candidate; 36 additional fits supplied the smaller rate. One rate per backbone/configuration was selected by mean best qualifying TENX95 MSE across seeds 0/1/2; exact ties favored the smaller rate. All three seeds at that rate entered the primary evaluation. All 72 candidate states, histories and validation losses were preserved, including unselected trials. No test outcome selected rates, architectures, seeds or fusion. This adaptive source/validation refinement is exploratory follow-up, not formal preregistration. Selected/run epochs, actual update counts, input/state hashes and observed costs appear in Supplementary Table~S18 and the manifests. No new state replaces a missing historical binary.

\paragraph{Full-section comparison.}
The 48 additional fits crossed 16 configurations with seeds 0/1/2. AdamW used learning rate $10^{-3}$, weight decay $10^{-4}$, maximum 80 epochs, patience 10 and minimum validation-MSE improvement $10^{-4}$. Each epoch made one full-section update per source training patient in seeded random order. Local/mean/GCN widths were 128 and concatenated GraphSAGE width 67. These baselines match the new models' update budget, whereas the 30 frozen minibatch heads do not. Both additional regimes trained directly in the common development target units, without the primary ensemble's extra head-target scaler. Model selection retained every seed; TENX99 was used only for evaluation. Parameter counts, selected epochs, histories and computational costs are in Supplementary Tables S15 and S16. Initial development searches, Phikon representation comparisons~\cite{filiot2023} and conditional leave-one-patient-out fits have separate specifications in Supplementary Methods S6.

\subsection{Bounded nonlocal and distance-aware message passing}
This implementation draws on GraphGPS principles rather than reproducing the reference implementation. Its blocks use a $512$-to-$64$ projection, two 64-wide blocks, four attention heads, dropout 0.1, pre-LayerNorm, local linear messages gated by eight relative-distance radial features, parallel bounded attention, residual additions and a two-layer GELU feed-forward path. The local graph is Delaunay with a $250$\,\um{} cap. Global keys/values are at most 32 deterministic coordinate-only farthest-point landmarks, giving $O(NL)$ attention storage with $L\leq32$; there is no dense N-by-N allocation. The nonlocal-disabled model attends self plus up to seven nearest graph neighbors, padding unused slots with self. Thus full/local changes attention range and key count as well as connectivity, despite identical 121,018 nominal parameters. Local radial features use log(1+d/250) with eight centers 0--1 and width 0.15; attention relative bias uses eight centers 0--4 and width 0.5. The distance-disabled version removes explicit gates/biases but retains coordinate-defined graph/landmark selection and is not coordinate-free.

The invariant fixed-coordinate and EGNN alternatives use two 64-wide message blocks. With centered normalized $p_i=(q_i-\overline q)/250$, where $q_i$ is the physical position in micrometers, $m_{ij}=\phi_e(h_i,h_j,\mathrm{RBF}(\|p_i-p_j\|^2))$, node updates aggregate the mean message through $\phi_h$, using SiLU, LayerNorm and dropout 0.1. Eight radial centers span 0--4 with width 0.5. EGNN additionally updates private coordinates as
\[
p_i^{l+1}=p_i^l+\frac{1}{\max(1,|\mathcal N_i|)}\sum_{j\in\mathcal N_i}0.1\tanh\!\big(\phi_x(m_{ij})\big)(p_i^l-p_j^l).
\]
Each $\phi_x$ is a bias-free 64-to-1 map initialized at standard deviation 0.001. Only the first coordinate update affects the second layer's scalar readout; the last update has no downstream scalar loss. EGNN has 87,282 nominal parameters versus 87,154 for fixed distance (128 additional coordinate coefficients, of which the last 64 do not affect prediction). Both are E(2)-compatible conditional on frozen features, so their accuracy difference cannot identify equivariance itself. Latent positions never change original masks, anchors or IDW distances.

Self-only and coordinate/graph-shuffled variants retain nominal parameter tensors but have different effective function classes; inactive attention or edge parameters do not make them capacity-equivalent. Conjugating the graph and assigned coordinates together preserves edge-length geometry while breaking image-to-location assignment. Degree/length rewiring preserves each node's degree and global 50\,\um{} edge-length bins below the same 250\,\um{} cap, targeting $E/4$ accepted double swaps with at most $20E$ attempts. Actual successes are reported, including infeasible or weak rewiring. No hidden expression enters any graph, landmark or distance encoding.

\subsection{Observation masks and controlled placement}
For a section with $N$ retained locations, the observed set $\mathcal O$ contained $\lfloor Nb+0.5\rfloor$ anchors at budget $b\in\{0.10, 0.20, 0.40\}$. Every hidden location lacked its entire expression vector; coordinates and image features remained available. Primary random trajectories used seeds 0--9 with nested prefixes, so higher budgets added anchors without discarding earlier observations. All trajectories were frozen without using expression values.

In distributed interpolation (V1), anchors were eligible throughout the tissue and all unobserved locations were scored. In contiguous regional completion (V2), a forbidden region consisted of the nearest 40\% of locations to a center determined from the 25th, 50th or 75th percentile of the first coordinate principal-component rank. The three centers defined three fixed geometries. Anchors were sampled outside the forbidden region, and only the forbidden region was scored. All other unobserved labels were also removed from model inputs. Thus the same nominal budget refers to the full retained section even when evaluation is restricted to a region.

A deterministic spatial farthest-point policy was retained as a secondary sampling analysis. It started from the eligible location farthest from the eligible-coordinate centroid and selected locations without expression labels. It was not optimized on external outcomes. Earlier development block and capture-window masks differ from the final regional-completion definition and are labeled separately in Fig.~\ref{fig:development} and the source tables.

The controlled identical-target placement experiment (B2) used only the predeclared central PCA-coordinate region $R$ (40\% of locations). $H$ was a fixed random 20\% subset of $R$ (about 8\% of all locations), selected using SHA 256(``H|patient'') first eight bytes, little-endian. Distributed anchors were drawn from all non-H sites; region-exclusion anchors from outside R. Seeds 0--9 gave nested prefixes with equal total counts at every budget. R, H, barcodes and anchor trajectories were persisted before scoring. Both estimators scored exactly H; no expression-based geometry or exclusion was used. Seed means preceded geometry means and patient pairing. Both primary error and correlation effects and nearest-anchor distances were exported, with occupied-bin errors and support.

The additional neural matrix used three original random masks per budget for TENX99, and original V1/central-V2 geometry 1 masks with seeds 0/1/2 externally. All 78 fits also used B2 seeds 0/1/2 at 10/20/40\% budgets. The matched-backbone study used these exact same mask and scoring supports, retaining their hashes. This three-mask subset is distinct from the ten-seed, three-geometry molecular baseline analysis.

\subsection{Molecular interpolation, calibration and residual prediction}
For every unmeasured location $i\in\mathcal U$, we define directed support edges $i\rightarrow j$ to at most 16 measured locations $j\in\mathcal O$, identified by the smallest coordinate distance. The resulting observation-dependent edge-weight matrix $W_{\mathcal U\mathcal O}$ has row-normalized positive weights and zero entries for unavailable anchors. The molecular reconstruction operator is
\begin{equation}
\widehat{Y}^{\mathrm{ST}}_{\mathcal U}=W_{\mathcal U\mathcal O}Y_{\mathcal O}.
\label{eq:molecular_graph}
\end{equation}
The operator has no explicit maximum-distance cutoff, permitting cross-gap transport of observed expression. It is not a learned attention matrix, and absence of a cutoff is not evidence of reliable long-range inference. The distinct V1 and V2 weights and scoring supports are defined next.

The molecular predictor used the up to 16 nearest observed anchors, $\mathcal N_{16}(i;\mathcal O)$. For distinct observed and hidden coordinates, distributed interpolation used normalized inverse-distance-squared weights, a local form of inverse-distance interpolation~\cite{shepard1968}:
\begin{equation}
 \widehat y_{ig}^{\mathrm{ST}}=\sum_{j\in\mathcal N_{16}(i;\mathcal O)}w_{ij}y_{jg},\qquad
 w_{ij}=\frac{\|q_i-q_j\|^{-2}}{\sum_{k\in\mathcal N_{16}(i;\mathcal O)}\|q_i-q_k\|^{-2}}.
 \label{eq:interp}
\end{equation}
For the frozen V2 protocol, the same neighborhood size was used with equal weights $w_{ij}=1/|\mathcal N_{16}|$. These rules were fixed during the Wu evaluation and were not selected on Bassiouni outcomes. Fewer than 16 anchors, if available, were handled by using all available anchors. Euclidean distances were rounded to six decimal places in micrometers and floored at $10^{-6}$~\um{} before inverse-square weighting; the normalized weights therefore remained finite at coincident coordinates.

The image-only prediction was denoted $\widehat y^{\mathrm{image}}$. The final fusion was
\begin{equation}
 \widehat y_{ig}^{\mathrm{fusion}}=0.8\widehat y_{ig}^{\mathrm{ST}}+
 0.2\widetilde y_{ig}^{\mathrm{image}},
 \label{eq:fusion}
\end{equation}
where the image calibration was
\begin{equation}
\widetilde y^{\mathrm{image}}_{ig}
=a^{\mathrm W}_g+b^{\mathrm W}_g
\frac{\widehat y^{\mathrm{image}}_{ig}-a^{\mathrm S}_g}{b^{\mathrm S}_g}.
\label{eq:calibration}
\end{equation}
Here $\mathrm S$ denotes the two development-training patients and $\mathrm W$ the three Wu calibration patients. Moments were computed from all retained targets of the designated source and calibration patients, rather than their sparse anchor subsets or image predictions. Each patient contributed equal total weight: for a domain with $P$ patients and $n_p$ locations in patient $p$,
\begin{equation}
 a_g=\frac{1}{P}\sum_{p=1}^{P}\frac{1}{n_p}\sum_{i=1}^{n_p}y_{pig},
 \qquad
 b_g=\sqrt{\frac{1}{P}\sum_{p=1}^{P}\frac{1}{n_p}\sum_{i=1}^{n_p}(y_{pig}-a_g)^2}.
\label{eq:moments}
\end{equation}
A scale below $10^{-8}$ was replaced by 1. This scale includes variation about the shared domain center; it is not the mean of patient-specific standard deviations. All moments used the common target units in Eq.~\ref{eq:target}. Bassiouni evaluation reused these source and Wu constants without fitting new moments or fusion weights. The calibrated image term used in fusion therefore differs from the raw image-only comparator reported in Table~\ref{tab:performance}.

Per-gene affine transformation, normalized interpolation and inverse transformation are algebraically equivalent to interpolation in the original units (Supplementary Note S1). Calibration therefore changes the image term, not the underlying normalized molecular operator.

For additional graph comparisons, $g_{ig}$ denotes the gene-$g$ image prediction at site $i$ in the common target space, calibrated by Eq.~\ref{eq:calibration} for both external cohorts. Alongside image-only $g$ and fixed fusion, we evaluated observed-residual interpolation,
\begin{equation}
\widehat y_{ig}=g_{ig}+\sum_{j\in\mathcal O}w_{ij}\,[y_{jg}-g_{jg}].
\label{eq:observed_residual}
\end{equation}
Only observed residual vectors enter this predictor; no regression coefficients are learned from the evaluation section. It is distinct from the development ridge residual correction, which fits gene-specific coefficients to image and molecular graph summaries using training patients (Supplementary Methods S6). Neither should be interchanged with fixed image fusion. HER2ST patient-disjoint calibration and fusion-family selection are also specified in Supplementary Methods S6; the final external image weight and Wu-derived moments remained fixed.

The matched-estimator grid used all ten original random V1 trajectories at each of 10/20/40\% coverage for nearest anchor; uniform k=4,8,16,32; IDW at those k with p=1,2,3; and Gaussian k=8,16,32. We retained cKDTree ordering, six-decimal distances and a $10^{-6}$~\um{} inverse-weight floor. Gaussian bandwidth was the median observed-anchor distance to its kth other observed neighbor, with k limited by available anchors; it never used hidden expression. Every declared estimator is reported. The operator-by-protocol comparison (B1) held each original mask and scored set fixed while changing IDW k16/p2 versus uniform k16, for both V1 and all three V2 geometries.

\subsection{Metrics, patient effects and uncertainty}
All expression metrics were computed on hidden locations only; the full observed expression vector was available at each anchor and the full target vector was withheld at every other location. Observed rows were selected before molecular prediction, so hidden expression could not influence anchor values or interpolation weights. For each gene, Pearson and Spearman correlations summarized spatial association, while MSE and MAE summarized prediction error. Panel Pearson values were medians across genes with defined correlations. Constant-target or constant-prediction correlations were left undefined rather than assigned zero; such genes did not pass positive-correlation threshold counts. Panel MSE and MAE were means of per-gene errors. In formulae, for the scored hidden set $\mathcal H$,
\begin{equation}
 \mathrm{MSE}=\frac{1}{G|\mathcal H|}\sum_{g=1}^{G}\sum_{i\in\mathcal H}(\widehat y_{ig}-y_{ig})^2,\qquad
 \mathrm{MAE}=\frac{1}{G|\mathcal H|}\sum_{g=1}^{G}\sum_{i\in\mathcal H}|\widehat y_{ig}-y_{ig}|.
 \label{eq:errors}
\end{equation}
Within each patient and geometry, metrics were averaged over mask seeds; geometry means were then averaged equally. Headline cohort values were medians of these patient summaries. Tables label an MSE value as a cohort median of patient mean MSEs, not a pooled mean over all spots. Paired effects were calculated within patients before taking their median. This distinction is necessary because a median paired effect need not equal the difference or ratio of marginal medians.

The patient was the biological unit. Seeds described computational or mask variability, not additional patients.  The original external-evaluation support criterion required a median gene Pearson gain of at least 0.02 over image-only prediction, relative mean MSE reduction of at least 2\%, and non-worsening MAE. The cohort rule required at least $\lceil2N/3\rceil$ patients meeting the joint criterion at at least two budgets, positive median benefits at every budget, and the same support threshold for correlation/MSE improvement from 10\% to 40\%. For 22 patients this threshold was 15. The limited native image resolution restricts interpretation of the image comparisons even when the molecular support criterion is satisfied.

For the additional neural comparisons, mask means were formed within geometry, geometry means within patient and training seeds averaged within patient before paired effects. Relative MSE is the candidate-minus-reference patient mean MSE divided by the reference mean MSE after these averages. Bassiouni intervals resampled 22 paired patients 10,000 times (seed 20261101) and took percentile 95\% limits for the median effect. They are descriptive, unadjusted intervals, not p-values or simultaneous bands. Cohorts were never pooled; Wu effects are individual, and no population interval was inferred from TENX99. Raw Spearman was secondary because near-tied ranks are sensitive to floating-point differences. Full defined-correlation counts accompany the panel summaries.

For each hidden location, the nearest-anchor distance was $d_i=\min_{j\in\mathcal O}\|q_i-q_j\|$. Distances were grouped into the frozen bins $[0,150)$, $[150,300)$, $[300,600)$, $[600,1200)$, $[1200,2400)$ and $[2400,\infty)$~\um{}. Error summaries were reported for occupied bins; the Bassiouni binwise correlations required at least 20 scored locations, while the Wu distance summaries contained errors only. Empty bins were omitted, not filled with zero. Counts, geometry and run identities were retained to distinguish sparse support from low error. Nearest-versus-farthest contrasts and monotonicity summaries were descriptive properties of the occupied-bin curves, not estimates of a universal biological length scale.

For graph-versus-local and anchor-reference contrasts, the primary Pearson summary retained each run's defined-gene median. A separately labelled sensitivity analysis intersected the finite Pearson genes of the paired methods within each mask/training-seed run, computed both medians on that common set, and applied the same geometry, mask and training-seed averaging before patient pairing. Undefined and common-valid gene counts were exported; undefined correlations were never replaced by zero. Correlation gains with worse MSE or MAE were reported as tradeoffs.

\subsection{Controls and reproducibility}
In the second replication, control patient BASS\_P01, 20\% coverage, random-mask seed 0, V1 geometry 0 and V2 geometry 1 were designated before evaluation. Three control seeds permuted observed expression-vector assignments, anchor-position assignments or image predictions. The first two disrupt related molecular--location associations; they were not interpreted as independent biological evidence. Every saved prediction was also repeated after all unobserved expression was replaced with NaNs and with independent random values. Observed/hidden barcodes, mask identities, configurations and hashes accompanied the predictions.

Analyses used Python, NumPy~\cite{harris2020}, SciPy~\cite{virtanen2020}, pandas, PyTorch and related scientific libraries. Reproduction instructions and environment specifications are maintained with the code. Model training and original external evaluation used recorded configurations and checksums; source-data tables provide the numerical outputs underlying the published figures.

Graphs, anchor trajectories, feature scalers and landmark selection were label-free or fitted on the designated source patients only. Replacing every unobserved label, including unscored labels, with NaNs and independent random values had to leave fixed/residual predictions exactly unchanged. Image-only B2 metrics were also identical across placements and budgets because $H$ was fixed. Input identities and barcode/coordinate joins were checked against original records; independently derived coordinates agreed with preserved label-free coordinates within $10^{-8}$\,\um{}, which governed exact neighbor ties. Numerical source tables, environment records and fitted-state hashes are retained in the repository; the detailed audit counts are summarized in the supplementary source guide.

The original external results were frozen before the additional controls. The latter followed timestamped local protocols, retained all specified configurations/seeds and reused previously inspected patients; they were exploratory, not formally preregistered. No evaluation outcome selected a new architecture, seed, calibration map or fusion coefficient. Original primary heads and newly fitted states have separate identities and availability records.

Conditional geometric tests applied rotations, reflections, translations and node permutations at float64/float32, recording scalar and physical latent errors separately. Float 64 graph-model tolerances were $10^{-8}$; float32 and preaggregated baseline tolerances were $10^{-4}$. Physical latent tolerances included the 250-\um{} normalization. Recomputed-graph tests used nondegenerate coordinates without distance ties. These tests condition on frozen image embeddings and documented graph/landmark handling, not invariance of the complete image pipeline. Attention diagnostics recorded physical/assigned connection fractions and actual mass beyond 250/600/1200\,\um{} per head/layer/seed/patient. A coordinate-selected remote landmark feature was zeroed to test functional dependency, not biological correlation. The degree/length control accepted no swaps in 64/66 recovered-cohort patient--seed graphs, with at most one swap in the remaining graphs; those cases supply little independent comparison.

\paragraph{AI-assisted computational tools.}
The study was conceived and directed by the author, who established the research objectives, methodological framework, and experimental protocols. OpenAI Codex assisted with software implementation, computational experiments, and model training, while ChatGPT supported manuscript drafting and editing. The author reviewed the analyses and retains full responsibility for the scientific findings, interpretations, and conclusions.

\section{Ethics}
This work reanalyzed publicly released, anonymized datasets without new recruitment, tissue collection or experimental intervention. The original studies describe their ethics and consent procedures~\cite{andersson2021,wu2021,bassiouni2023}.
\section{Data availability}
Development data were drawn from HEST~\cite{jaume2024}. HER2ST data are available at \url{https://zenodo.org/records/4751624}, Wu data at \url{https://zenodo.org/records/4739739}, and Bassiouni data under GEO accession GSE210616, \url{https://www.ncbi.nlm.nih.gov/geo/query/acc.cgi?acc=GSE210616}. The selected Bassiouni inputs have been recovered and verified against original checksums. Source tables for all manuscript figures and tables accompany the journal package. Full per-gene exports and trained head-state bundles, including the matched-backbone follow-up, are currently author-local with file/checksum manifests in the repository; no public deposit or DOI is claimed. Raw tissue images and complete prediction caches are not redistributed with the manuscript.
\section{Code availability}
Code, protocols and reproduction instructions are available at \url{\CodeRepository}. Publication figures and documents can be regenerated from the recorded exports. Full replay of the original image predictions remains limited by three missing historical primary image-head binaries. Newly fitted journal heads are distinct from those missing binaries. Candidate and validation-selected state manifests specify author-local availability; neither a manifest nor an author-review bundle constitutes public deposition.
\section{Acknowledgements}
We acknowledge the investigators who generated and released the source datasets. Figures were drawn from numerical exports using Matplotlib 3.10.8 and 3.11.2 (\url{https://matplotlib.org/}). The original conceptual spatial schematic is synthetic; gene maps show measured data, anchors, predictions and errors without inferred tissue annotations.
\section{Funding}
No funding to disclose.
\section{Author contributions}
Truong-Son Hy was responsible for study conception and design, methodology, intellectual oversight, evaluation of analyses and validation, interpretation, visualization, and manuscript writing and revision. Computational and editorial tool assistance is disclosed in Methods.
\section{Competing interests}
Truong-Son Hy is an Editorial Board Member of \textit{Scientific Reports}. The author declares no other competing interests.
\bibliographystyle{naturemag}
\bibliography{main}

\clearpage
\appendix
\setcounter{section}{0}\setcounter{figure}{0}\setcounter{table}{0}
\renewcommand{\thesection}{S\arabic{section}}
\renewcommand{\thesubsection}{S\arabic{section}.\arabic{subsection}}
\renewcommand{\thefigure}{S\arabic{figure}}
\renewcommand{\thetable}{S\arabic{table}}

\begin{center}
{\Large\bfseries Supplementary Information\par}
\vspace{10pt}
{\large\bfseries \PaperTitle\par}
\vspace{12pt}
\AuthorNames\par
\AuthorAffiliations\par
\Correspondence\par
\end{center}
\section*{Contents and source-data guide}
This document provides supporting figures and tables, cohort-resolved graph comparisons, computational diagnostics and detailed development methods. Source tables retain all tested estimators, masks, training seeds and defined-correlation counts. Patients are biological replicates; masks and training seeds describe computational variation. The main Methods distinguish the original evaluation, retrospective controls and the two additional neural training regimes.
\clearpage
\section{Supplementary notes}
\subsection{Affine interpolation identity}
For a gene with transform $T(y)=(y-\mu)/\sigma$, $\sigma>0$, and interpolation weights satisfying $\sum_j w_{ij}=1$, we have
\begin{align*}
T^{-1}\!\left(\sum_j w_{ij}T(y_j)\right)
&=\sigma\sum_jw_{ij}\frac{y_j-\mu}{\sigma}+\mu\\
&=\sum_j w_{ij}y_j.
\end{align*}
Consequently, applying and inverting a fixed per-gene affine transform does not create a distinct molecular interpolation method when weights are unchanged. In the Wu evaluation, the raw and transformed predictions differed by at most $2.66\times10^{-15}$, consistent with roundoff. Tiny numerical changes can break tied ranks and alter raw Spearman calculations; such changes were logged as numerical rank sensitivity, not counted as harmonization benefits. Calibration can still alter an image branch or a learned fusion model; that is a different comparison. The manuscript therefore omits duplicate molecular interpolation curves and makes no claim of an affine-harmonization advantage for molecular interpolation itself.

\subsection{Dataset roles, masking and interpretation}
The four-patient HEST cohort was used for development. HER2ST was first evaluated under a frozen protocol and subsequently used for calibration development; those two views reuse the same eight patients. The Wu cohort supplied three calibration and three evaluation patients. The 22 Bassiouni patients were evaluated without new cohort-specific fitting. Counts of independent evaluation patients must not add repeated HER2ST analyses or mask replicates as new patients.

Distributed evaluation reveals molecular vectors at sampled anchors and scores all remaining locations. The external regional-completion protocol instead withholds a contiguous 40\% region, samples anchors outside it, and scores that region only. V1 uses inverse-distance-squared interpolation and V2 uses an equal-weight nearest-anchor mean. Their contrast is therefore a protocol-level difference, not an isolated geometry intervention. The earlier development block/capture masks are also not identical to the external forbidden-region design.

Public external H\&E images are downsampled, and sparse molecular interpolation receives same-section expression unavailable to the image-only baseline. Image-resolution limitations and unequal available information must accompany comparisons with histology. An H\&E-fusion failure under these frozen inputs does not establish that native-resolution histology is generally uninformative.

\subsection{Aggregation and metric interpretation}
The primary external summaries compute gene-level metrics on hidden locations, summarize genes within each run, average random seeds within geometry, average geometries within patient, and finally summarize patients. Unless a figure legend states otherwise, external cohort markers are medians across patients. Several development and transportability plots instead show arithmetic means; their legends state this explicitly. A difference between two cohort medians is not a median paired patient effect.

Mean squared error and mean absolute error use the unchanged development-standardized log1p-count target. Shared numerical units do not remove capture-depth or assay differences. Correlations are undefined when the target or prediction range is at most $10^{-12}$; they are excluded from correlation averages but not counted as successes in threshold fractions. Spearman correlations use average ranks for ties. Mask and model seeds describe computational variation, not independent biological replication. Descriptive patient-bootstrap intervals are reported for the 22-patient comparisons (main Tables 3--5); they are not p-values. Wu patients are reported individually, with no primary three-patient bootstrap claim.
\clearpage
\section{Supplementary figures}
\begin{figure}[!htbp]
\centering
\includegraphics[width=\textwidth,height=0.65\textheight,keepaspectratio]{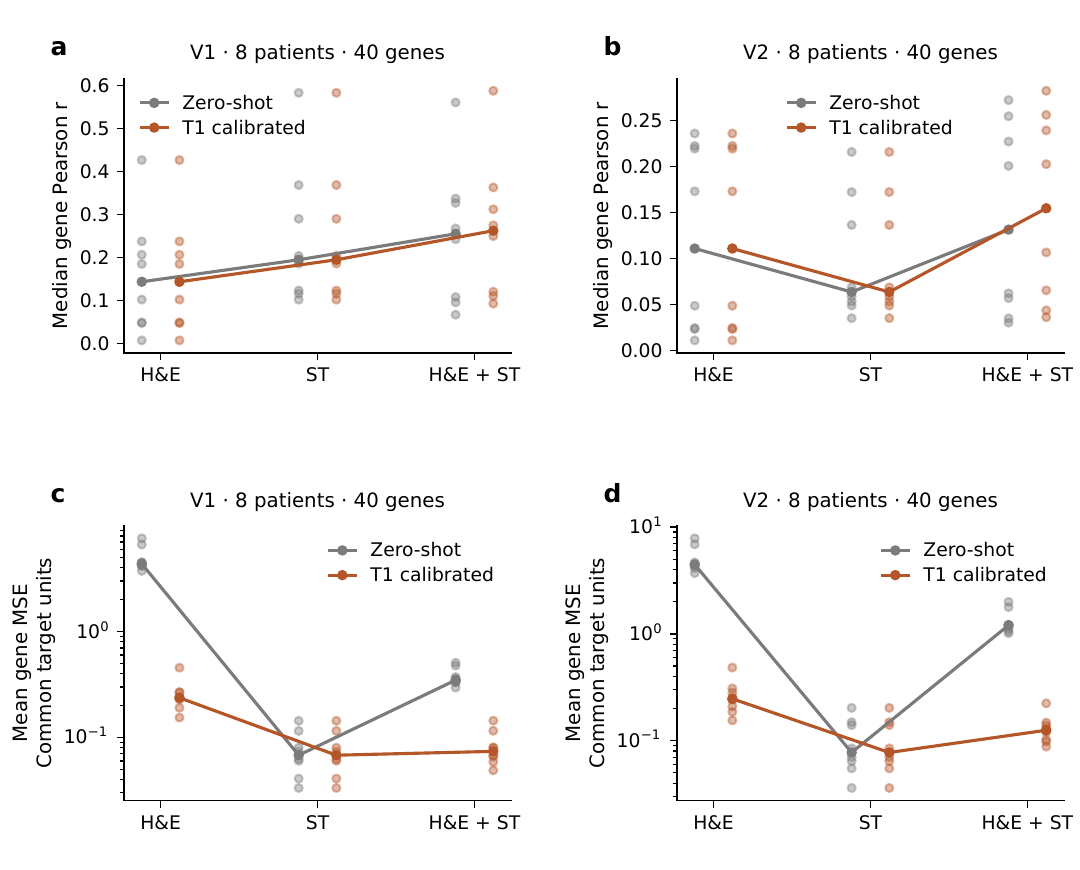}
\caption{\textbf{HER2ST transportability and target calibration.} Eight HER2ST patients and the frozen 40-gene intersection are shown at 20\% random measured-location coverage. \textbf{a,c,} Distributed interpolation (V1). \textbf{b,d,} Contiguous regional completion (V2). Panels a/b show median gene Pearson correlation and c/d show mean gene MSE in common target units on a logarithmic axis. Dots are patient summaries after averaging seeds and geometries; connected markers are medians across patients. T0 denotes the uncalibrated zero-shot analysis, whereas T1 denotes the calibrated target mapping with fixed fusion. Image weights were 0.25 for V1 and 0.5 for V2; these development weights differ from the final external weight of 0.2. These are the same patients, not independent replication groups. The T1 fixed-fusion analysis differs from the later fold-selected fusion results in Supplementary Fig.~S10. Affine transformation and inverse transformation leave molecular interpolation unchanged; apparent image/fusion calibration effects are not a new interpolation advantage.}
\label{fig:s1}
\end{figure}\clearpage
\begin{figure}[p]
\centering
\includegraphics[width=\textwidth,height=0.80\textheight,keepaspectratio]{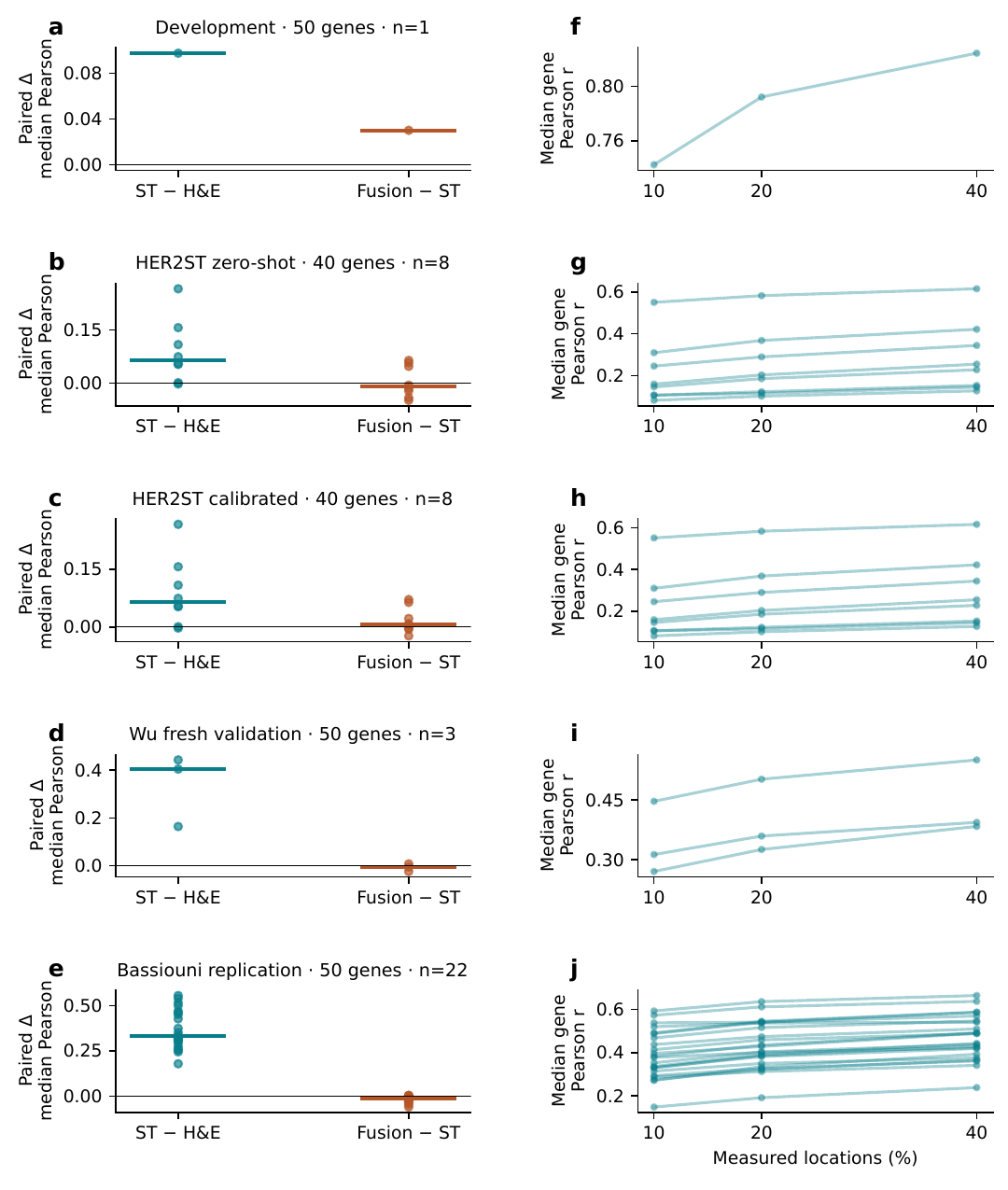}
\caption{\textbf{Cross-cohort synthesis without pooling different panels.} Rows correspond to the development primary patient ($n=1$), HER2ST zero-shot ($n=8$), HER2ST calibrated ($n=8$, the same patients), original Wu evaluation ($n=3$) and original Bassiouni evaluation ($n=22$). The development and original evaluation cohorts use 50 genes; HER2ST uses 40. \textbf{a--e,} Left-column paired patient correlation effects at 20\% distributed coverage for molecular interpolation minus image-only prediction and fusion minus molecular interpolation; bars are median paired effects. \textbf{f--j,} Right-column individual patient molecular correlation curves at 10\%, 20\% and 40\% distributed coverage. The panels preserve differences in analytical role and gene panel and do not define a pooled cross-cohort score. HER2ST calibration is development, not an additional independent validation cohort.}
\label{fig:s2}
\end{figure}\clearpage
\begin{figure}[p]
\centering
\includegraphics[width=\textwidth,height=0.65\textheight,keepaspectratio]{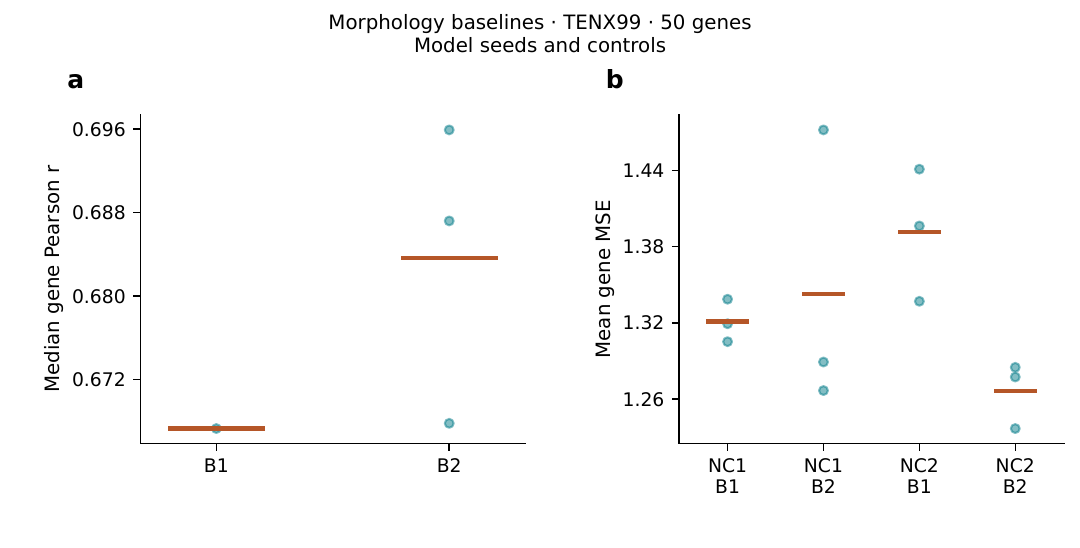}
\caption{\textbf{Morphology-only development baselines and controls.} The morphology-only development analyses use the single primary development patient TENX99 and 50 genes. \textbf{a,} Median gene Pearson correlation for ridge regression (B1) and the small multilayer perceptron (B2). The training-mean baseline has constant predictions and undefined correlations and is not drawn as zero. \textbf{b,} MSE under training-target-row permutation (NC1) and training-embedding-row permutation (NC2). Dots represent saved model-seed runs and bars are arithmetic means of displayed runs, not independent patient estimates. The controls disrupt image--expression pairing while preserving validation and evaluation data. This initial image-only head differs from the frozen local MLP selected in the representation comparison and used for external replication.}
\label{fig:s3}
\end{figure}\clearpage
\begin{figure}[p]
\centering
\includegraphics[width=\textwidth,height=0.65\textheight,keepaspectratio]{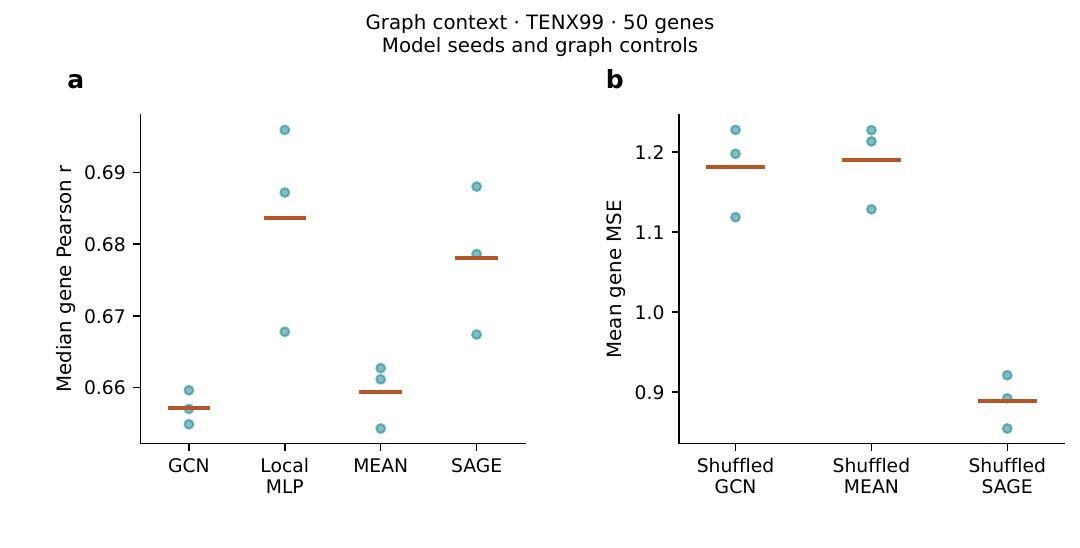}
\caption{\textbf{Single-scale graph context did not by itself establish spatial specificity.} The single-scale graph comparison uses TENX99 and the 50-gene panel. \textbf{a,} Median gene correlation for the frozen local MLP, neighborhood mean pooling plus MLP, GCN and GraphSAGE. \textbf{b,} Mean gene MSE for the archived shuffled-neighborhood runs. Dots are training-seed runs and bars are their arithmetic means. A graph model outperforming a local predictor does not establish that the correct spatial adjacency caused the gain; generic aggregation or model parameterization can also contribute. The results use the original fitted models without additional graph selection or local-predictor refitting.}
\label{fig:s4}
\end{figure}\clearpage
\begin{figure}[p]
\centering
\includegraphics[width=\textwidth,height=0.65\textheight,keepaspectratio]{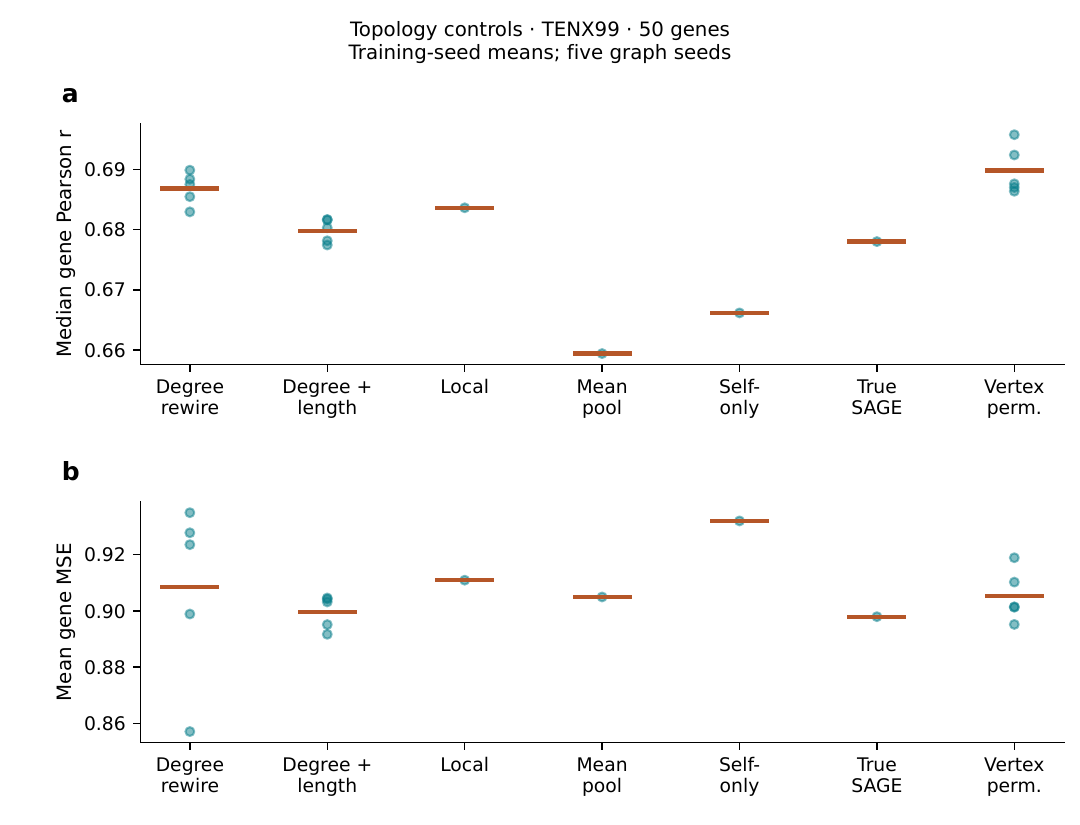}
\caption{\textbf{Matched graph controls limited claims about exact topology.} The matched graph controls use one development patient, TENX99, and 50 genes. \textbf{a,} Median gene Pearson correlation. \textbf{b,} Mean gene MSE. Local, self-only, mean-pooling and true-spatial GraphSAGE results accompany vertex-permuted, degree-rewired and degree-plus-edge-length-rewired controls. Training seeds are averaged within each graph randomization seed; the randomized controls have five graph seeds. Dots are these averages and bars are arithmetic means. Random graph seeds are not biological replicates. The experiment did not establish a reproducible advantage of the true spatial topology. This figure records a negative/mixed development result, not evidence that topology is irrelevant in every representation or tissue.}
\label{fig:s5}
\end{figure}\clearpage
\begin{figure}[p]
\centering
\includegraphics[width=\textwidth,height=0.65\textheight,keepaspectratio]{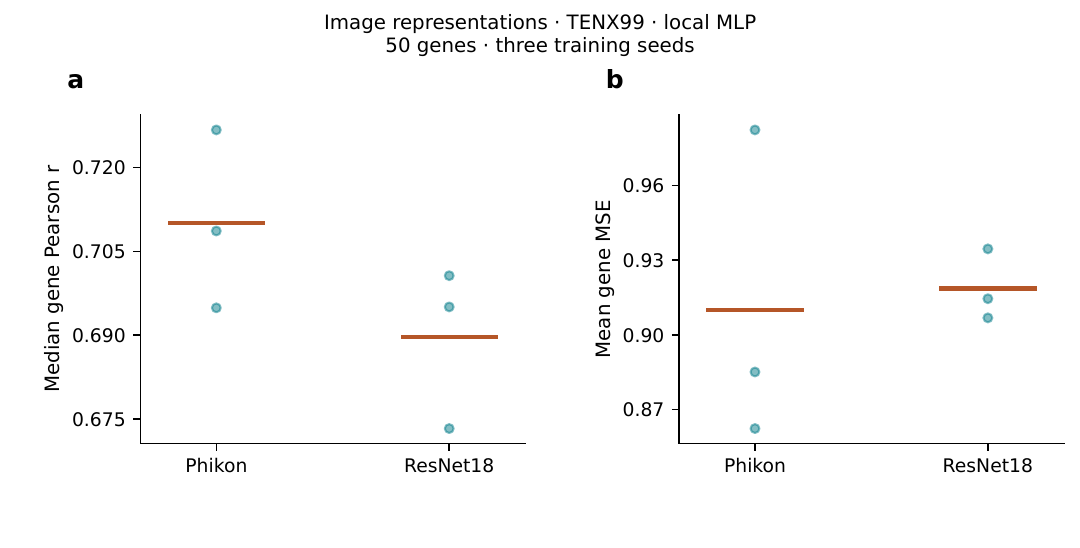}
\caption{\textbf{Pathology representation effects in the development patient.} The local MLP representation comparison evaluates frozen ImageNet ResNet18 and Phikon representations in TENX99 for 50 genes. \textbf{a,} Median gene Pearson correlation. \textbf{b,} Mean gene MSE. Each dot is one of three training seeds; bars are arithmetic means. The panel-specific improvement did not establish general superiority on larger target panels or reproducible topology-specific gains. The original frozen Bassiouni evaluation retained the ResNet18 image branch. Phikon was subsequently evaluated in Wu and Bassiouni patients in the separate exploratory matched-backbone study. This plot remains a one-patient development comparison, not external encoder validation.}
\label{fig:s6}
\end{figure}\clearpage
\begin{figure}[p]
\centering
\includegraphics[width=\textwidth,height=0.65\textheight,keepaspectratio]{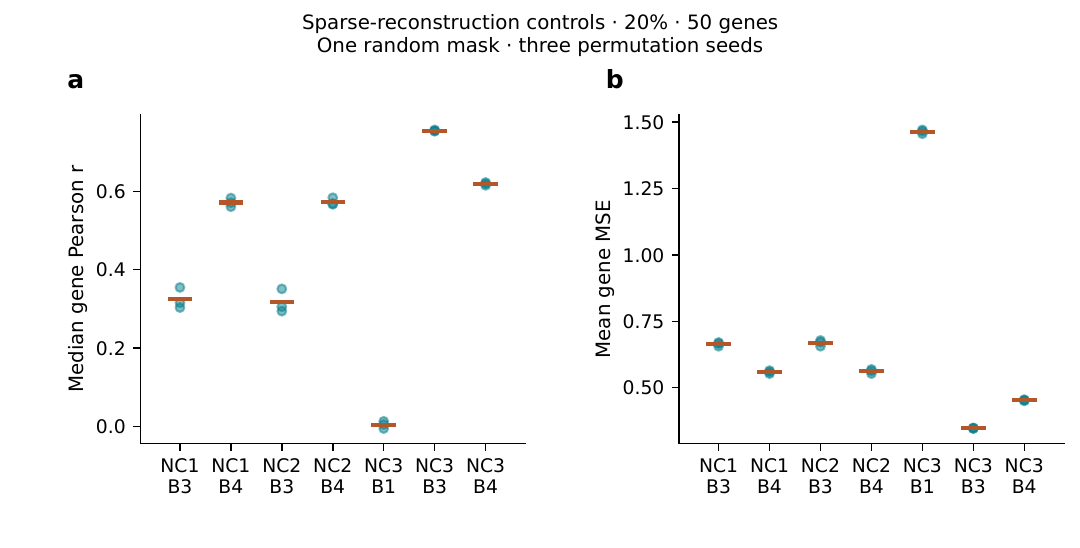}
\caption{\textbf{Sparse-reconstruction development controls.} The sparse-reconstruction controls use the ResNet18 branch, TENX99, 50 genes and 20\% measured-location coverage. In this figure, B1 denotes image-only prediction, B3 image--molecular fusion and B4 learned propagation. Controls include anchor-expression permutation (NC1), anchor-position permutation (NC2) and image-location permutation (NC3). Model labels are specific to this experiment. \textbf{a,} Median gene Pearson correlation. \textbf{b,} Mean gene MSE. All controls use the same random observation mask (replicate 0); dots represent three permutation seeds and bars are arithmetic means. These distributions are control diagnostics rather than estimates of a geometry effect. Separate integrity checks establish invariance to corruption of hidden labels. Related anchor-assignment disruptions must not be counted as independent biological mechanisms.}
\label{fig:s7}
\end{figure}\clearpage
\begin{figure}[p]
\centering
\includegraphics[width=\textwidth,height=0.65\textheight,keepaspectratio]{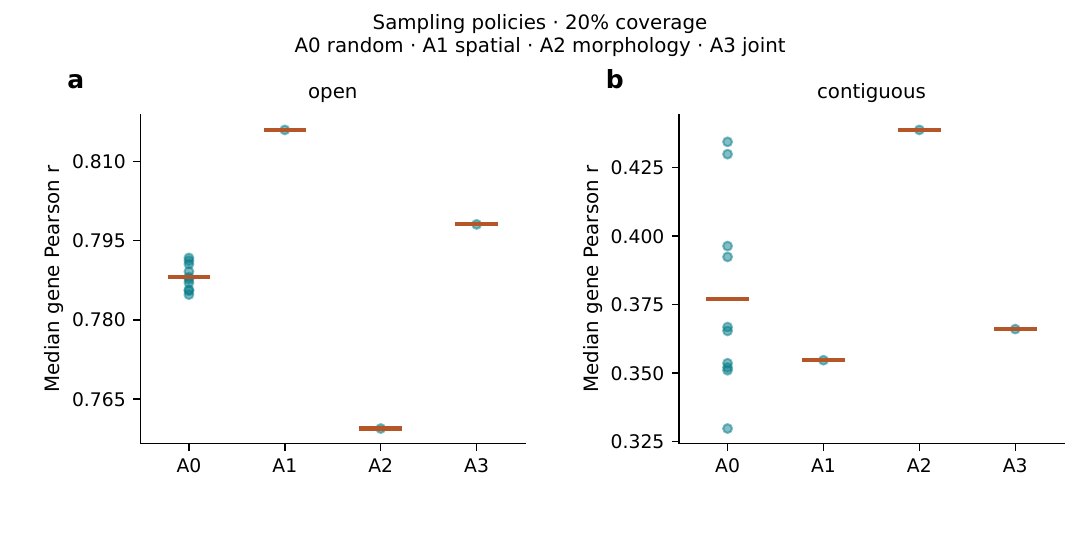}
\caption{\textbf{Sampling effects depended on the reconstruction regime.} The sampling comparison uses interpolation-only results for TENX99, the ResNet18 reference branch, 50 genes and 20\% coverage. \textbf{a,} Open-tissue interpolation. \textbf{b,} Contiguous missing-region evaluation. A0 is random sampling with ten seeds; A1 is spatial farthest-point sampling, A2 morphology-space farthest-point sampling, and A3 a locked joint policy. A1--A3 are deterministic given their stored initialization. Dots are policy runs and bars are arithmetic means. This limited baseline diagnostic is not a complete summary of all learned/fusion families and does not demonstrate universally optimal sampling.}
\label{fig:s8}
\end{figure}\clearpage
\begin{figure}[p]
\centering
\includegraphics[width=\textwidth,height=0.65\textheight,keepaspectratio]{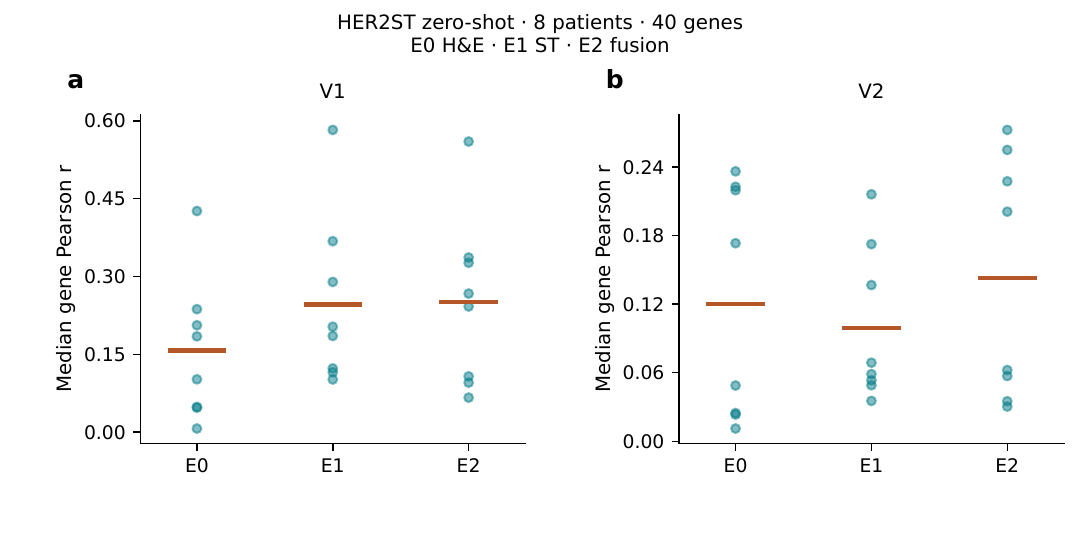}
\caption{\textbf{HER2ST patient distributions under the frozen zero-shot protocol.} The eight HER2ST patients, 40 shared primary genes and 20\% random-anchor coverage are summarized separately for \textbf{a,} distributed interpolation and \textbf{b,} contiguous regional completion. E0 denotes H\&E-only prediction, E1 molecular interpolation and E2 frozen fusion, with image weights 0.25 for V1 and 0.5 for V2. Dots are patient means across seeds/geometries. Bars are arithmetic patient means, unlike the median summaries in Supplementary Fig.~S1 and the primary tables. These zero-shot results and the subsequent calibration analyses use the same patients; HER2ST therefore does not supply an untouched final validation cohort.}
\label{fig:s9}
\end{figure}\clearpage
\begin{figure}[p]
\centering
\includegraphics[width=\textwidth,height=0.65\textheight,keepaspectratio]{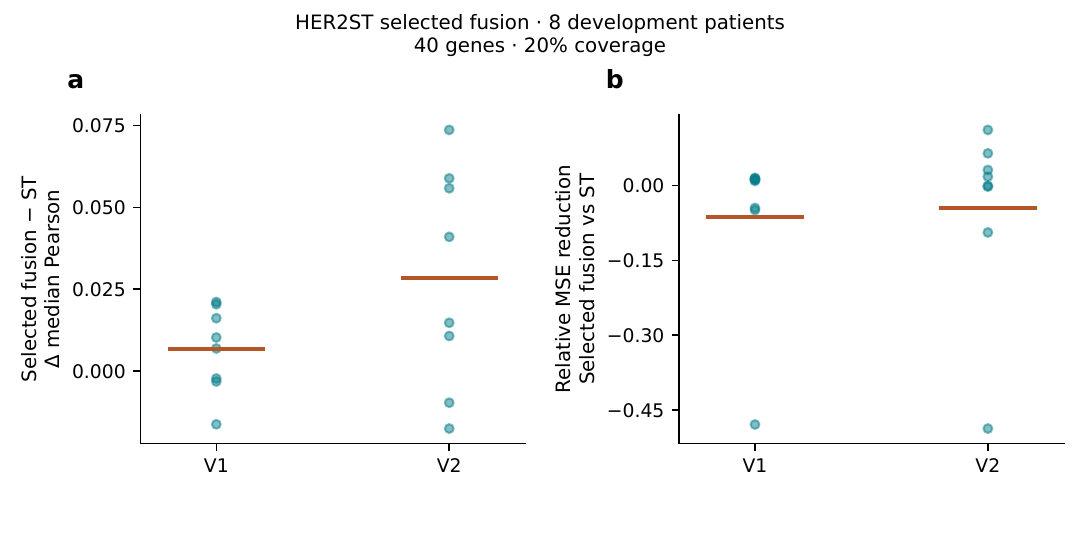}
\caption{\textbf{Patient-disjoint calibration did not establish robust fusion superiority.} The calibration-development analysis includes eight HER2ST patients, the frozen 40-gene intersection and 20\% coverage. \textbf{a,} Selected calibrated fusion minus molecular interpolation in Pearson correlation. \textbf{b,} Relative MSE reduction compared with molecular interpolation; positive values indicate lower fusion error. Dots are patient/fold results grouped by V1/V2 and bars are arithmetic means. The selected fusion rule was determined within patient-disjoint training/calibration folds, not from the displayed outer-patient outcomes. Joint correlation and error improvements were not consistent across patients. These selected-fusion results should not be interchanged with the T1 fixed-fusion target study shown in Supplementary Fig.~S1.}
\label{fig:s10}
\end{figure}\clearpage
\begin{figure}[p]
\centering
\includegraphics[width=\textwidth,height=0.65\textheight,keepaspectratio]{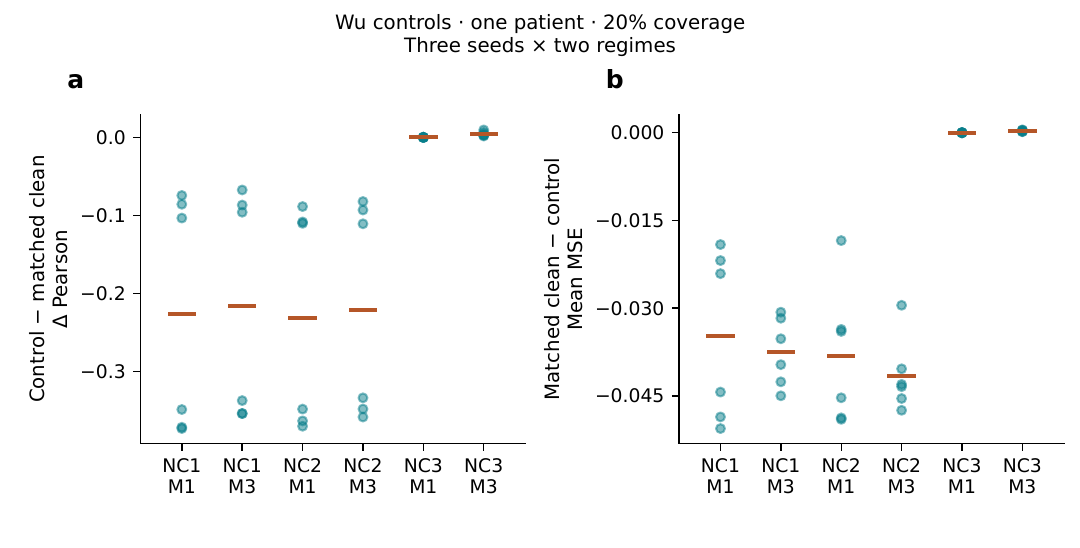}
\caption{\textbf{Matched negative controls in original Wu evaluation.} The predeclared held-out patient 1160920F, 50 genes, 20\% coverage and random mask seed 0 were used, with V1 geometry 0 and V2 geometry 1. Three permutation seeds were evaluated for NC1 anchor-vector reassignment, NC2 anchor-position reassignment and NC3 image-prediction permutation. \textbf{a,} Control minus matched-clean median gene Pearson correlation. \textbf{b,} Matched-clean minus control MSE, so negative values indicate increased control error. Molecular interpolation (M1) and fixed image fusion (M3) are shown. Dots are matched control settings and bars are means; they are not independent patients. Molecular controls disrupt the correct spatial anchor assignments, whereas image permutation provides no robust supporting evidence for the frozen image contribution.}
\label{fig:s11}
\end{figure}\clearpage
\begin{figure}[p]
\centering
\includegraphics[width=\textwidth,height=0.65\textheight,keepaspectratio]{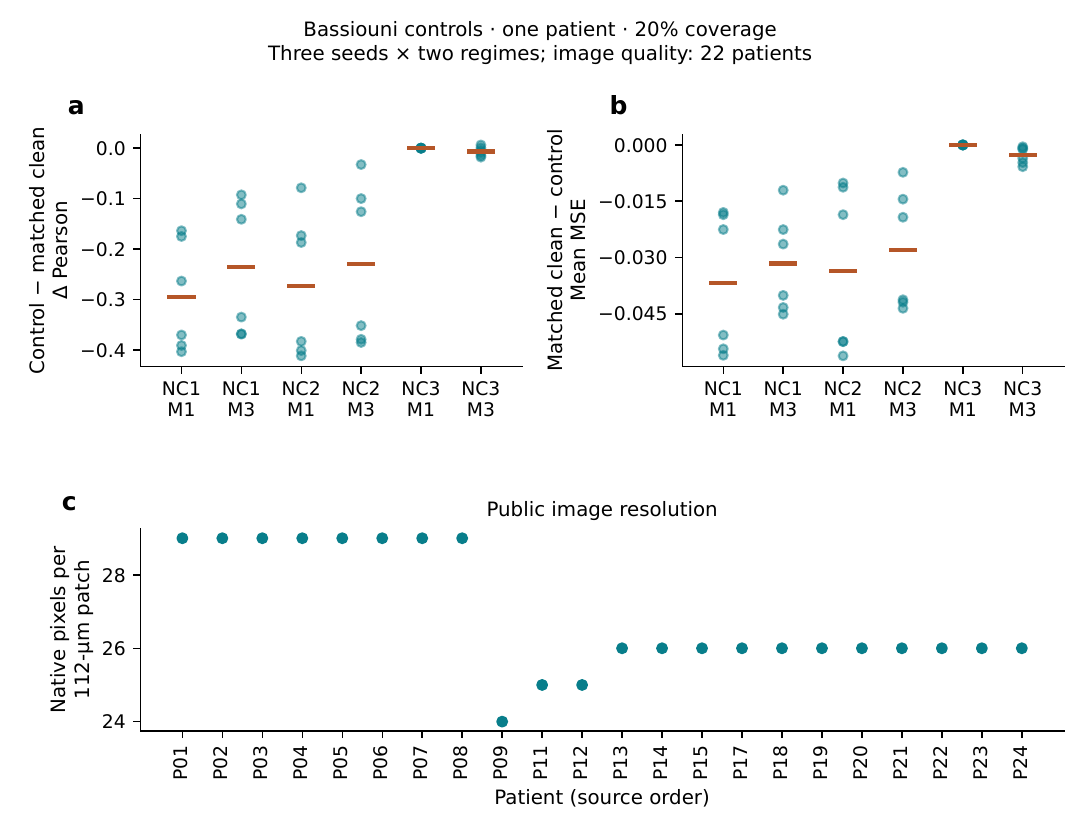}
\caption{\textbf{Replication controls and native image resolution.} The independent-replication controls use the predeclared Bassiouni patient BASS\_P01, 50 genes, 20\% coverage, random mask seed 0, V1 geometry 0 and V2 geometry 1. \textbf{a,b,} Matched Pearson and MSE effects for M1/M3 under three control seeds, with definitions and signs as in Supplementary Fig.~S11. Dots are matched control settings and bars are arithmetic means. \textbf{c,} Native pixels per nominal 112-\um{} patch for all 22 patients in source order. The supplied images provide only 24--29 native pixels before resizing for the encoder. This is image-input quality control, not expression-prediction performance. It explains why the image comparisons cannot establish the limits of native-resolution histopathology. Coordinate overlays for all patients are recorded in the archived qualification audit.}
\label{fig:s12}
\end{figure}\clearpage
\section{Supplementary tables}
The tables below provide compact views of the accompanying machine-readable data and definitions needed to interpret the supplementary figures. Table~S2 is rounded for presentation; the digital files retain unrounded exported values. Large gene-level and distance-resolved tables are supplied as separate files rather than reproduced in full in this PDF. Mask, model and permutation seeds are computational replicates and do not increase the number of independent patients.
\begin{table}[htbp]
\caption{\textbf{Primary gene panel.} The 50 genes are listed in the order fixed by variance ranking in the development-training patients. The original Wu and Bassiouni evaluation retain all 50 genes. HER2ST analyses use the 40-gene intersection with this panel; unavailable genes are not replaced. The full data file also includes the larger development panels and cohort-specific availability records: \protect\path{source_data/SuppTable1_gene_panels.csv}.}
\label{tab:primary_gene_panel}
\centering\small
\begin{tabular}{rlrl}\toprule
Order & Gene & Order & Gene\\\midrule
1 & FOXA1 & 26 & LUM \\
2 & SCD & 27 & MYO5B \\
3 & KRT7 & 28 & SFRP4 \\
4 & ERBB2 & 29 & KRT14 \\
5 & POSTN & 30 & TOP2A \\
6 & CEACAM6 & 31 & IL7R \\
7 & SERHL2 & 32 & AR \\
8 & ANKRD30A & 33 & KRT5 \\
9 & EPCAM & 34 & PCLAF \\
10 & ABCC11 & 35 & KLF5 \\
11 & TACSTD2 & 36 & SFRP1 \\
12 & GATA3 & 37 & TRAF4 \\
13 & FASN & 38 & CXCR4 \\
14 & MLPH & 39 & MYH11 \\
15 & S100A14 & 40 & TENT5C \\
16 & CDH1 & 41 & TCIM \\
17 & CD9 & 42 & NARS \\
18 & KRT8 & 43 & ACTG2 \\
19 & CCND1 & 44 & PDK4 \\
20 & DSP & 45 & CD3E \\
21 & SERPINA3 & 46 & TRAC \\
22 & LYPD3 & 47 & JUP \\
23 & TFAP2A & 48 & LYZ \\
24 & TPD52 & 49 & CENPF \\
25 & PTGDS & 50 & CCDC80 \\
\bottomrule\end{tabular}\end{table}\clearpage
\begin{table}[p]
\caption{\textbf{Patient-resolved reconstruction metrics.} Compact view of the 25 original external-evaluation patients at 20\% distributed coverage, using 50 genes. Values average mask seeds within patient. $r$ is the median gene Pearson correlation; MSE is mean gene squared error in the common target units. I denotes image-only, ST molecular interpolation, and F fixed image fusion. Complete exported patient summaries for the available cohorts, regimes, budgets and methods are supplied in \protect\path{source_data/SuppTable2_per_patient.csv}.}
\label{tab:patient_resolved_metrics}
\centering\small
\begin{tabular}{lrrrrrr}\toprule
Patient & $r_{\mathrm I}$ & $r_{\mathrm{ST}}$ & $r_{\mathrm F}$ & MSE$_{\mathrm I}$ & MSE$_{\mathrm{ST}}$ & MSE$_{\mathrm F}$\\\midrule
\multicolumn{7}{l}{\textit{Wu}}\\
1160920F & -0.045 & 0.359 & 0.335 & 0.846 & 0.082 & 0.095\\
CID4290 & 0.161 & 0.325 & 0.333 & 2.915 & 0.099 & 0.100\\
CID44971 & 0.059 & 0.502 & 0.496 & 2.949 & 0.115 & 0.118\\
\multicolumn{7}{l}{\textit{Bassiouni}}\\
BASS\_P01 & 0.077 & 0.382 & 0.370 & 1.304 & 0.087 & 0.106\\
BASS\_P02 & 0.009 & 0.459 & 0.431 & 1.067 & 0.093 & 0.127\\
BASS\_P03 & 0.030 & 0.545 & 0.537 & 2.585 & 0.081 & 0.089\\
BASS\_P04 & -0.039 & 0.334 & 0.312 & 1.882 & 0.113 & 0.125\\
BASS\_P05 & 0.096 & 0.636 & 0.628 & 1.581 & 0.133 & 0.155\\
BASS\_P06 & -0.005 & 0.312 & 0.297 & 1.567 & 0.069 & 0.072\\
BASS\_P07 & 0.013 & 0.191 & 0.180 & 2.684 & 0.092 & 0.097\\
BASS\_P08 & 0.095 & 0.403 & 0.393 & 2.120 & 0.088 & 0.095\\
BASS\_P09 & 0.075 & 0.391 & 0.380 & 1.763 & 0.120 & 0.139\\
BASS\_P11 & -0.081 & 0.474 & 0.464 & 2.052 & 0.104 & 0.109\\
BASS\_P12 & 0.182 & 0.435 & 0.436 & 1.748 & 0.113 & 0.120\\
BASS\_P13 & 0.151 & 0.612 & 0.615 & 2.228 & 0.104 & 0.119\\
BASS\_P14 & 0.041 & 0.540 & 0.536 & 1.458 & 0.126 & 0.154\\
BASS\_P15 & 0.093 & 0.386 & 0.389 & 1.777 & 0.101 & 0.115\\
BASS\_P17 & 0.095 & 0.428 & 0.421 & 1.773 & 0.088 & 0.096\\
BASS\_P18 & 0.090 & 0.516 & 0.496 & 2.391 & 0.083 & 0.086\\
BASS\_P19 & 0.068 & 0.537 & 0.501 & 2.431 & 0.068 & 0.078\\
BASS\_P20 & 0.067 & 0.326 & 0.319 & 2.407 & 0.096 & 0.098\\
BASS\_P21 & 0.075 & 0.401 & 0.340 & 2.607 & 0.092 & 0.108\\
BASS\_P22 & -0.029 & 0.322 & 0.300 & 2.149 & 0.087 & 0.090\\
BASS\_P23 & 0.267 & 0.541 & 0.525 & 2.166 & 0.054 & 0.061\\
BASS\_P24 & 0.106 & 0.350 & 0.302 & 1.626 & 0.048 & 0.052\\
\bottomrule\end{tabular}
\end{table}\clearpage
\begin{table}[htbp]
\caption{\textbf{Per-gene results and metric definitions.} The accompanying digital table, \protect\path{source_data/SuppTable3_per_gene.csv.gz}, contains the exported Wu and Bassiouni gene-level results, with run, patient, panel, mask and method identifiers. Compression changes storage only. Every metric is calculated using the hidden locations scored by the corresponding protocol. The definitions below explain the gene-level statistics and their summaries; genes and repeated runs are not independent patient observations.}
\label{tab:s3}
\centering\small
\begin{tabularx}{\textwidth}{>{\RaggedRight\arraybackslash}p{0.23\textwidth}Y}\toprule
Quantity & Definition and interpretation\\\midrule
Pearson correlation & Correlation between predicted and measured values across scored locations, calculated separately for each gene. Constant targets or predictions yield undefined correlations.\\
Spearman correlation & Correlation between the ranks of predicted and measured values across scored locations, calculated separately for each gene. Numerical changes in tied ranks do not establish a biological improvement.\\
Mean squared error (MSE) & Mean squared prediction error across scored locations for each gene, evaluated in the common development-standardized log1p-count units.\\
Mean absolute error (MAE) & Mean absolute prediction error across scored locations for each gene, evaluated in the same target units.\\
Panel correlation & Median of defined gene correlations. Undefined correlations are excluded from this median and do not count as positive-correlation successes.\\
Panel error & Arithmetic mean of the gene-level MSEs or MAEs. All genes are retained in error summaries.\\
Patient and cohort summaries & Mask seeds are averaged within geometry and geometries within patient. Headline cohort values are medians of patient summaries; paired effects are calculated within patients before summarizing them.\\
\bottomrule\end{tabularx}
\end{table}
\begin{table}[htbp]
\caption{\textbf{Negative-control families.} Control definitions are experiment-specific and are given in the corresponding figure legends. Training-row controls differ from inference-time controls that permute permitted sparse inputs. Permuting anchor-expression assignments and anchor-position assignments disrupts related associations and does not constitute independent biological evidence. Exported control results are supplied in \protect\path{source_data/SuppTable4_negative_controls.csv}.}
\label{tab:s4}
\centering\small
\begin{tabularx}{\textwidth}{>{\RaggedRight\arraybackslash}p{0.28\textwidth}Y}\toprule
Control family & Interpretation\\\midrule
Training-target or training-embedding row permutation & Disrupts local morphology--target pairing during development while preserving the validation and evaluation data.\\
Anchor-expression assignment permutation & Reassigns observed expression vectors among anchors, disrupting their correspondence with spatial position.\\
Anchor-position assignment permutation & Disrupts the spatial assignment of observed anchors while retaining the permitted molecular information.\\
Image-location or image-prediction permutation & Tests dependence on the image input while retaining the allowed molecular measurements.\\
\bottomrule\end{tabularx}
\end{table}\clearpage
\begin{table}[htbp]
\caption{\textbf{Recorded integrity and replay checks.} Counts are taken from the original Wu and Bassiouni technical audits; these audits were not rerun during manuscript preparation. Method predictions and metric arrays are distinct audit units. For each prediction, hidden labels were independently replaced by NaNs and random values to test that unobserved expression could not influence model inputs. Digital records: \protect\path{source_data/SuppTable5_integrity.csv}.}
\label{tab:s5}
\centering\small
\begin{tabular}{lrr}\toprule
Audit quantity & Wu & Bassiouni\\\midrule
Mask trajectories & 132 & 968\\
Clean run bundles & 396 & 2,904\\
Control run bundles & 18 & 18\\
Method prediction replays & 1,656 & 8,766\\
Metric array replays & 6,624 & 35,064\\
Hidden-label corruption checks & 3,312 & 17,532\\
\bottomrule\end{tabular}
\end{table}
\begin{table}[htbp]
\caption{\textbf{Observation masks and sampling protocols.} Coverage is the proportion of retained locations with observed complete expression vectors, not the proportion of tissue area assayed. All unobserved labels are excluded from model inputs. The three regional geometries and ten random-mask seeds are computational replicates. The accompanying \protect\path{source_data/SuppTable6_masks.csv} records the exported mask identities, patients, regimes, geometries and sampling trajectories.}
\label{tab:s6}
\centering\small
\begin{tabularx}{\textwidth}{>{\RaggedRight\arraybackslash}p{0.28\textwidth}Y}\toprule
Protocol component & Definition\\\midrule
Measurement budgets & For $N$ retained locations and $b\in\{0.10, 0.20, 0.40\}$, the observed count is $\lfloor bN+0.5\rfloor$. Random anchor sets use nested prefixes of the same trajectory as coverage increases.\\
Distributed interpolation (V1) & Anchors may be sampled throughout the retained tissue; every remaining location is scored. Predictions use up to 16 nearest observed anchors with normalized inverse-distance-squared weights.\\
Contiguous regional completion (V2) & A prespecified contiguous region contains the 40\% of locations nearest to its center. Anchors are sampled outside that region, and only locations within it are scored. Predictions use the mean of up to 16 nearest observed anchors.\\
Regional geometry & Three centers are selected at the 25th, 50th and 75th percentile of first coordinate principal-component rank. The three geometries are averaged equally within patient.\\
Secondary spatial sampling & Label-free farthest-point sampling starts from the eligible location farthest from the eligible-coordinate centroid. It is not optimized using external expression outcomes.\\
Development masks & Earlier block and capture-window masks differ from the external regional-completion protocol and are interpreted separately.\\
\bottomrule\end{tabularx}
\end{table}\clearpage
\begin{table}[htbp]
\caption{\textbf{Distance-resolved evaluation and patient support.} Distance is measured from each scored hidden location to its nearest observed anchor. Empty bins are omitted. Bassiouni binwise Pearson correlations require at least 20 scored locations in the corresponding run; Wu distance bins report MSE and MAE only. Error summaries may include smaller occupied bins. The patient counts below summarize the Bassiouni cohort at 20\% coverage and do not define a universal distance cutoff. Exported bin-level metrics and support counts are supplied in \protect\path{source_data/SuppTable7_distance_bins.csv}.}
\label{tab:s7}
\centering\small
\begin{tabularx}{\textwidth}{>{\RaggedRight\arraybackslash}p{0.34\textwidth}Y}\toprule
Evaluation component & Definition\\\midrule
Distance bins (\um) & $[0,150)$, $[150,300)$, $[300,600)$, $[600,1200)$, $[1200,2400)$ and $[2400,\infty)$.\\
Bin support & Number of scored hidden locations contributing to the patient, method, mask and distance-bin summary; occupied-bin support may differ across patients and protocols.\\
Nearest-to-farthest comparison & Compares MSE in the nearest and farthest occupied bins within each patient.\\
Monotonicity & Assesses whether MSE strictly increases across the occupied distance bins.\\
\bottomrule\end{tabularx}
\vspace{12pt}
\begin{tabularx}{\textwidth}{Yrr}\toprule
Bassiouni patient summary & V1 & V2\\\midrule
Higher MSE in farthest than nearest occupied bin & 15/22 & 20/22\\
Strictly increasing occupied-bin MSE & 6/22 & 13/22\\
\bottomrule\end{tabularx}
\end{table}
\clearpage
\begin{table}[p]
\caption{\textbf{Selected supporting predictor architectures for the 50-gene panel.} Dimensions denote input $\rightarrow$ hidden $\rightarrow$ output, and every listed MLP has GELU, dropout 0.1 and a linear output layer. Encoder parameters remain frozen; counts include trainable head biases. All MLPs use the shared optimizer and stopping rule above. The initial local predictor and the representation-comparison predictor are distinct fitted heads.}
\label{tab:supporting_architectures}
\centering\small
\begin{tabularx}{\textwidth}{@{}>{\RaggedRight\arraybackslash}p{0.25\textwidth}>{\RaggedRight\arraybackslash}p{0.22\textwidth}Y@{}}
\toprule
Predictor & Selected dimensions & Fitting settings or fixed aggregation\\
\midrule
Initial ResNet18 ridge & $512\rightarrow50$ & Ridge penalty 100; intercept; 25,650 coefficients.\\
Initial ResNet18 MLP & $512\rightarrow128\rightarrow50$ & Lowest validation MSE; 72,114 head parameters.\\
ResNet18 mean pooling / GCN & $512\rightarrow128\rightarrow50$ & $PX$ or $QX$ on capped Delaunay adjacency; 72,114 head parameters.\\
ResNet18 mean-GraphSAGE & $1024\rightarrow67\rightarrow50$ & $[X,RX]$ on capped Delaunay adjacency; 72,075 head parameters.\\
ResNet18 representation-comparison MLP & $512\rightarrow256\rightarrow50$ & Highest validation median gene Pearson; 144,178 head parameters.\\
Phikon local MLP & $768\rightarrow256\rightarrow50$ & Same candidate grid and selection criterion; 209,714 head parameters.\\
Phikon spatial MLPs & $768\rightarrow128\rightarrow50$; $1536\rightarrow67\rightarrow50$ & Mean pooling and mean-GraphSAGE, respectively; inherited graph and head widths.\\
Sparse residual correction & 34 covariates plus intercept per gene & Gene-specific ridge correction of fixed fusion; penalty 10,000; 1,750 coefficients for 50 genes.\\
\bottomrule
\end{tabularx}
\end{table}

\clearpage
\begin{table}[!htbp]
\caption{\textbf{Coverage response and contiguous-region completion in the two original evaluation studies.} All rows use the same frozen 50-gene panel and spatial interpolation. Values are medians of patient summaries, with $n=3$ for Wu and $n=22$ for Bassiouni. V1 is distributed interpolation with inverse-distance-squared weighting; V2 is completion of a predefined contiguous region using the mean of nearby anchors. Both the scoring support and interpolation rule change between V1 and V2. The comparison therefore describes protocol-level deterioration rather than an isolated causal geometry effect. In both studies, every evaluation patient improved in Pearson correlation and MSE from 10\% to 40\% V1 coverage; every patient deteriorated in Pearson, MSE and MAE under V2 relative to V1 at 20\%.}
\label{tab:coverage}
\centering\small
\begin{tabularx}{\textwidth}{@{}Yrrrrr@{}}
\toprule
Cohort & Coverage & V1 Pearson $r$ & V1 MSE & V2 Pearson $r$ & V2 MSE \\
\midrule
Wu, original evaluation & 10\% & 0.313 & 0.1080 & 0.111 & 0.1356 \\
 & 20\% & 0.359 & 0.0987 & 0.081 & 0.1366 \\
 & 40\% & 0.393 & 0.0899 & 0.088 & 0.1385 \\
\addlinespace[5pt]
Bassiouni, original evaluation & 10\% & 0.380 & 0.1001 & 0.121 & 0.1520 \\
 & 20\% & 0.416 & 0.0923 & 0.171 & 0.1496 \\
 & 40\% & 0.465 & 0.0852 & 0.181 & 0.1518 \\
\addlinespace[5pt]
\bottomrule
\end{tabularx}
\end{table}

\clearpage
\section{Additional controls and source-data guide}
The matched molecular and neural comparisons use previously evaluated cohorts. Complete source tables are under \path{source_data/journal/} and \path{source_data/extension/}; protocols, input identities, fitted-state records and numerical audits are documented in the repository. Exact Bassiouni inputs are available and verified. The complete additional per-gene exports and model states remain author-local with release-ready manifests, not publicly deposited assets. Publication regeneration from exported tables is distinct from replay of the three missing historical primary image heads.

\begin{table}[!htbp]
\caption{\textbf{All 20 matched-information estimators in Wu at 20\% V1 coverage.} Each patient value averages ten original random trajectories. P1=1160920F, P2=CID4290, P3=CID44971. Relative MSE is candidate minus fixed IDW k16/p2, divided by that reference; negative values improve. No external winner was selected. All 10/20/40\% results, MAE and individual paired effects accompany the CSV exports. Gaussian bandwidth uses observed-anchor geometry only.}
\label{tab:journal_baselines}
\centering\footnotesize
\begin{tabular}{@{}lrrrrrr@{}}
\toprule
Estimator & $r$ P1 & $r$ P2 & $r$ P3 & MSE\% P1 & MSE\% P2 & MSE\% P3 \\
\midrule
nearest & 0.223 & 0.228 & 0.368 & +68.3 & +61.9 & +63.7 \\
uniform\_k4 & 0.328 & 0.293 & 0.474 & +8.0 & +8.2 & +7.5 \\
uniform\_k8 & 0.354 & 0.294 & 0.475 & +0.7 & +4.6 & +4.0 \\
uniform\_k16 & 0.359 & 0.265 & 0.442 & -0.8 & +7.5 & +6.8 \\
uniform\_k32 & 0.348 & 0.220 & 0.396 & +0.8 & +13.7 & +11.7 \\
idw\_k4\_p1 & 0.327 & 0.305 & 0.483 & +8.5 & +6.6 & +6.4 \\
idw\_k4\_p2 & 0.313 & 0.299 & 0.475 & +13.1 & +9.7 & +10.2 \\
idw\_k4\_p3 & 0.297 & 0.289 & 0.460 & +19.8 & +15.3 & +16.3 \\
idw\_k8\_p1 & 0.360 & 0.321 & 0.499 & -0.0 & +0.8 & +0.3 \\
idw\_k8\_p2 & 0.341 & 0.317 & 0.495 & +4.4 & +2.8 & +3.0 \\
idw\_k8\_p3 & 0.315 & 0.301 & 0.474 & +12.7 & +9.4 & +10.4 \\
idw\_k16\_p1 & 0.378 & 0.318 & 0.497 & -3.1 & +0.7 & +0.0 \\
idw\_k16\_p2 & 0.359 & 0.325 & 0.502 & +0.0 & +0.0 & +0.0 \\
idw\_k16\_p3 & 0.324 & 0.305 & 0.479 & +9.1 & +6.6 & +7.6 \\
idw\_k32\_p1 & 0.387 & 0.307 & 0.480 & -3.3 & +3.8 & +2.1 \\
idw\_k32\_p2 & 0.371 & 0.328 & 0.503 & -2.2 & -0.8 & -1.2 \\
idw\_k32\_p3 & 0.329 & 0.308 & 0.481 & +7.2 & +5.2 & +6.3 \\
gaussian\_k8 & 0.360 & 0.308 & 0.489 & -0.1 & +2.6 & +2.0 \\
gaussian\_k16 & 0.374 & 0.286 & 0.461 & -1.9 & +4.8 & +4.4 \\
gaussian\_k32 & 0.358 & 0.249 & 0.417 & -0.5 & +10.3 & +9.5 \\
\bottomrule
\end{tabular}
\end{table}
\FloatBarrier

\begin{table}[!htbp]
\caption{\textbf{Operator by historical protocol at 20\% coverage.} IDW and uniform operators share each original mask and scored set. V2 averages geometries equally after averaging seeds. The V1/V2 contrast still changes spatial support and is not a causal geometry estimate. All budgets are supplied in source data.}
\label{tab:journal_factorial}
\centering\footnotesize
\begin{tabular}{@{}lrrrrr@{}}
\toprule
Patient & Protocol & IDW $r$ & IDW MSE & Uniform $r$ & Uniform MSE \\
\midrule
1160920F & V1 & 0.359 & 0.0817 & 0.359 & 0.0811 \\
1160920F & V2 & 0.179 & 0.1029 & 0.176 & 0.1028 \\
CID4290 & V1 & 0.325 & 0.0987 & 0.265 & 0.1061 \\
CID4290 & V2 & 0.050 & 0.1367 & 0.036 & 0.1366 \\
CID44971 & V1 & 0.502 & 0.1150 & 0.442 & 0.1228 \\
CID44971 & V2 & 0.113 & 0.1781 & 0.081 & 0.1767 \\
\bottomrule
\end{tabular}
\end{table}
\FloatBarrier

\begin{table}[!htbp]
\caption{\textbf{Identical-target anchor-placement effects at 20\% coverage in Wu.} Region-exclusion minus distributed anchors, with the same H and total anchor count. Each row averages ten seeds. The holdout H is fixed across conditions and budgets. All three patients deteriorate in Pearson, MSE and MAE with both estimators. This central-region result is exploratory robustness evidence in previously inspected patients, not new blind validation.}
\label{tab:journal_identical}
\centering\footnotesize
\begin{tabular}{@{}lrrrr@{}}
\toprule
Patient & Operator & $\Delta r$ & Relative MSE\% & $\Delta$ MAE \\
\midrule
1160920F & IDW & -0.1056 & +20.1 & +0.0214 \\
1160920F & Uniform & -0.0928 & +21.7 & +0.0196 \\
CID4290 & IDW & -0.2452 & +45.4 & +0.0464 \\
CID4290 & Uniform & -0.1875 & +36.0 & +0.0361 \\
CID44971 & IDW & -0.1274 & +37.3 & +0.0386 \\
CID44971 & Uniform & -0.1605 & +38.9 & +0.0387 \\
\bottomrule
\end{tabular}
\end{table}
\FloatBarrier

\begin{table}[!htbp]
\caption{\textbf{All new controlled neural fits at 20\% TENX99 coverage.} Each value averages three identical original masks. Models fit NCBI783/NCBI785 and use TENX95 alone for stopping. Seeds are training variability, not patients. Residual is the exploratory image plus interpolated observed-residual predictor. All budgets, fixed 20\% fusion and anchor-only values are in CSVs. No architecture or seed was selected using TENX99. Nominal parameter counts are 72,114 (local/mean/GCN) and 72,075 (SAGE); SAGE self-only duplicates self inputs and has a smaller effective function class.}
\label{tab:journal_neural}
\centering\footnotesize
\begin{tabular}{@{}lrrrrr@{}}
\toprule
Configuration & Seed & Image $r$ & Image MSE & Residual $r$ & Residual MSE \\
\midrule
local local & 0 & 0.666 & 0.9142 & 0.818 & 0.2576 \\
local local & 1 & 0.686 & 0.9008 & 0.824 & 0.2569 \\
local local & 2 & 0.695 & 0.9191 & 0.821 & 0.2555 \\
MEAN spatial & 0 & 0.651 & 0.8825 & 0.812 & 0.2458 \\
MEAN spatial & 1 & 0.647 & 0.9095 & 0.813 & 0.2497 \\
MEAN spatial & 2 & 0.658 & 0.9206 & 0.813 & 0.2446 \\
MEAN self & 0 & 0.666 & 0.9142 & 0.818 & 0.2576 \\
MEAN self & 1 & 0.686 & 0.9008 & 0.824 & 0.2569 \\
MEAN self & 2 & 0.695 & 0.9191 & 0.821 & 0.2555 \\
MEAN shuffled & 0 & 0.117 & 1.2311 & 0.787 & 0.2747 \\
MEAN shuffled & 1 & 0.227 & 1.1425 & 0.771 & 0.2865 \\
MEAN shuffled & 2 & 0.213 & 1.1819 & 0.778 & 0.2889 \\
GCN spatial & 0 & 0.652 & 0.8871 & 0.809 & 0.2508 \\
GCN spatial & 1 & 0.659 & 0.8885 & 0.812 & 0.2518 \\
GCN spatial & 2 & 0.653 & 0.9240 & 0.809 & 0.2503 \\
GCN self & 0 & 0.666 & 0.9142 & 0.818 & 0.2576 \\
GCN self & 1 & 0.686 & 0.9008 & 0.824 & 0.2569 \\
GCN self & 2 & 0.695 & 0.9191 & 0.821 & 0.2555 \\
GCN shuffled & 0 & 0.124 & 1.2297 & 0.788 & 0.2743 \\
GCN shuffled & 1 & 0.230 & 1.1264 & 0.767 & 0.2909 \\
GCN shuffled & 2 & 0.228 & 1.1830 & 0.777 & 0.2908 \\
SAGE spatial & 0 & 0.667 & 0.8611 & 0.823 & 0.2340 \\
SAGE spatial & 1 & 0.688 & 0.8792 & 0.831 & 0.2346 \\
SAGE spatial & 2 & 0.679 & 0.9173 & 0.822 & 0.2329 \\
SAGE self & 0 & 0.667 & 0.9364 & 0.815 & 0.2574 \\
SAGE self & 1 & 0.656 & 0.9010 & 0.812 & 0.2590 \\
SAGE self & 2 & 0.672 & 0.9599 & 0.810 & 0.2582 \\
SAGE shuffled & 0 & 0.686 & 0.8615 & 0.812 & 0.2511 \\
SAGE shuffled & 1 & 0.695 & 0.8655 & 0.816 & 0.2564 \\
SAGE shuffled & 2 & 0.684 & 0.9124 & 0.818 & 0.2403 \\
\bottomrule
\end{tabular}
\end{table}
\FloatBarrier

\clearpage
\begin{figure}[!htbp]
\centering\includegraphics[width=\textwidth]{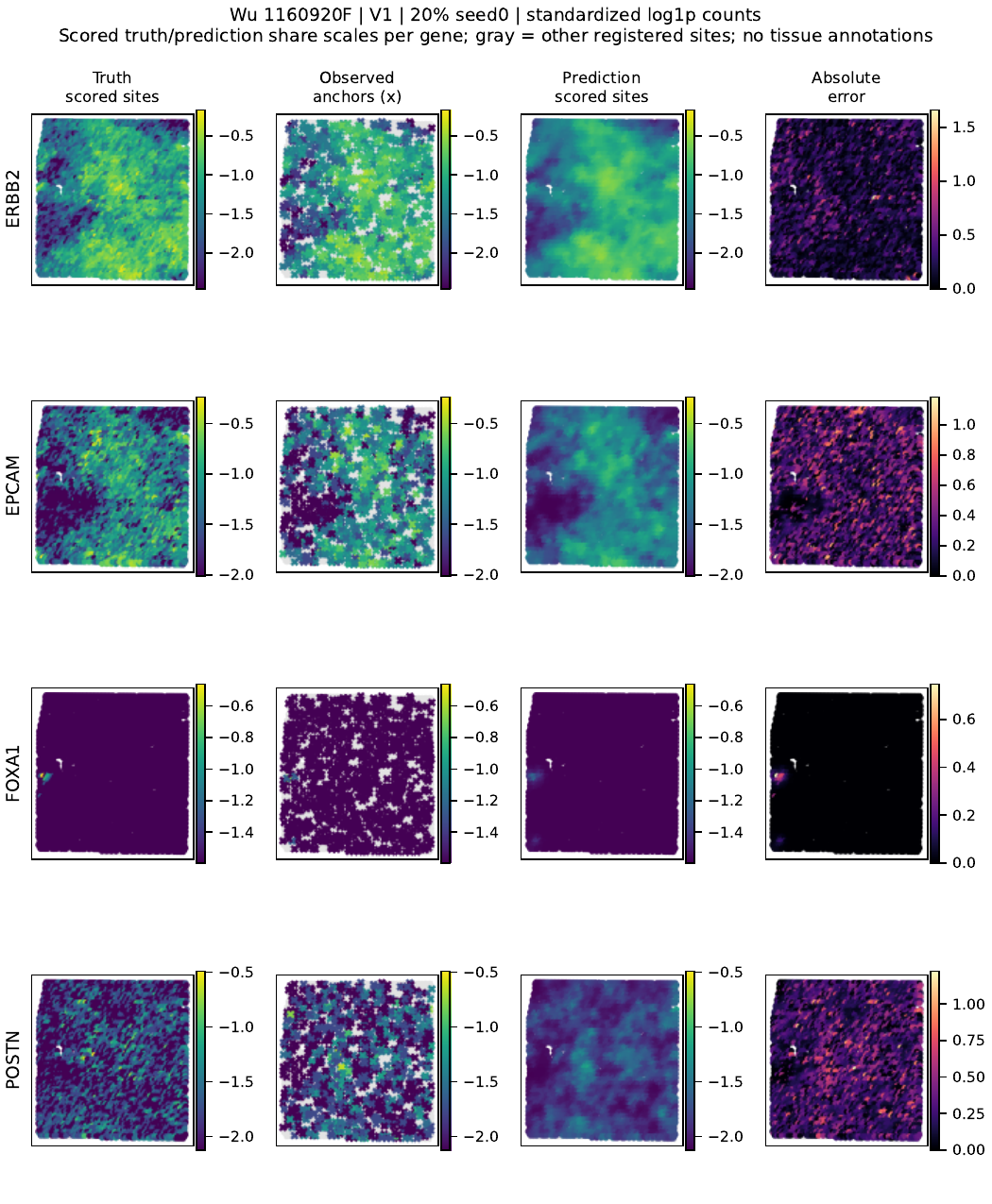}
\caption{\textbf{Predetermined real Wu V1 gene maps.} Patient 1160920F, 20\% random seed0, original mask, fixed IDW k16/p2. Genes ERBB2, EPCAM, FOXA1 and POSTN were designated before comparisons. Truth/prediction use shared per-gene scales over scored sites; errors have separate scales. Anchors are crosses and gray sites are other registered locations. Standardized log1p counts are not absolute transcript abundance. Constant scored fields are labelled rather than omitted or given misleading contrast. Source: Wu et al., Zenodo 4739739, CC-BY4.0; no tumor/cell annotations are inferred.}
\label{fig:journal_map_v1}
\end{figure}
\FloatBarrier
\clearpage
\begin{figure}[!htbp]
\centering\includegraphics[width=\textwidth]{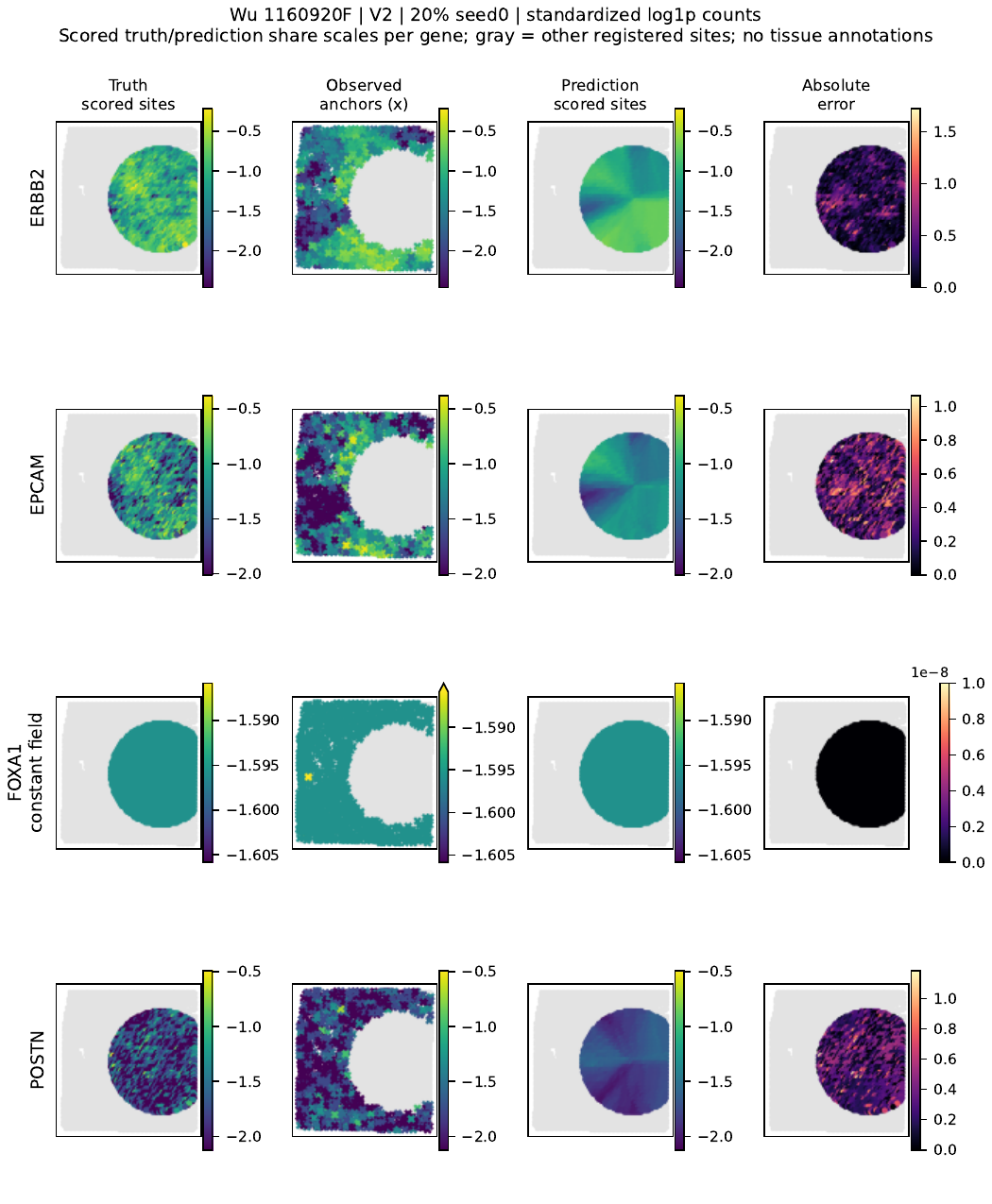}
\caption{\textbf{Predetermined real Wu V2 gene maps.} Same patient, genes, 20\% budget and seed0, central original V2 geometry 1, historical uniform k16 operator. Colored truth/prediction/error points are the original forbidden scoring region; anchor crosses lie outside. Scales and source terms match the preceding figure. FOXA1 has a constant scored field; numerical residuals do not establish biological recovery. These maps retain unsuccessful/constant examples and are not independent validation. The corresponding Bassiouni maps are provided in Supplementary Figs.~S26 and S27.}
\label{fig:journal_map_v2}
\end{figure}
\FloatBarrier

\clearpage\section{Graph comparisons and computational diagnostics}

\subsection{Complete cohort and prediction-route matrices}
All matrices use 20\% distributed coverage and the fixed 50-gene panel. Sixteen full-section configurations and ten frozen minibatch configuration families are shown, each retaining three training seeds, alongside fixed IDW and uniform molecular estimators. Metrics are averaged over the same three mask seeds and then training seeds within patient. Dots are patient summaries, and black ticks are cohort medians; their spread is not a confidence interval. The image-only route uses frozen image features and permitted coordinates; fixed fusion and observed-residual interpolation additionally use the same measured anchors as the molecular references. Thus image-only versus molecular rows have different information access. Architecture names identify configurations, not outcome-selected winners.

\begin{figure}[!htbp]
\centering\includegraphics[width=\textwidth]{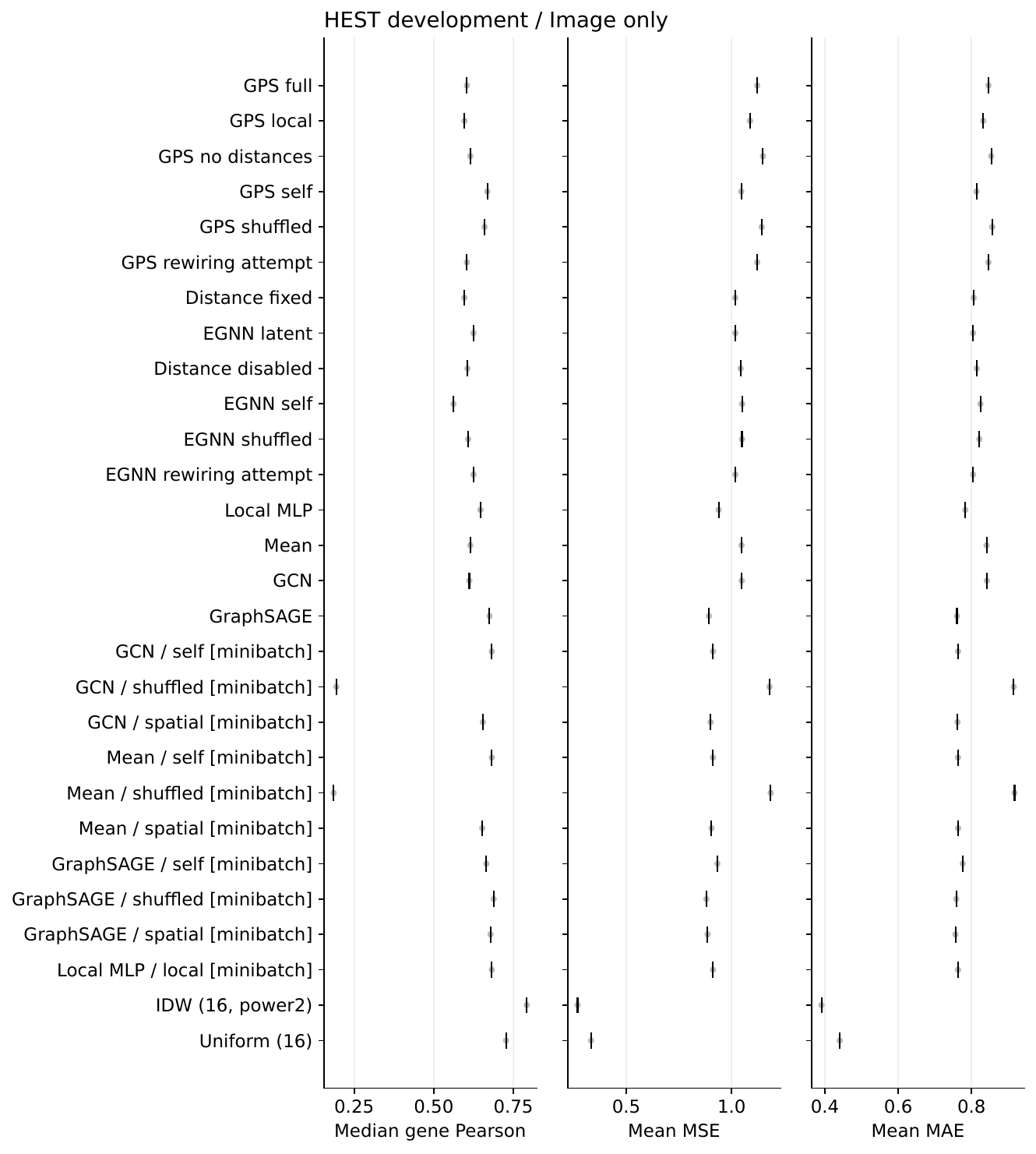}
\caption{\textbf{HEST development: image-only at 20\% distributed coverage.} Results for one previously evaluated patient, TENX99. Pearson, MSE and MAE are shown after averaging three mask and three training seeds within patient. Neural rows use image features and permitted geometry; the two molecular rows receive measured anchors. The dot represents one patient; no population interval is inferred.}
\label{fig:ExtensionMatrix_Development_image}
\end{figure}
\FloatBarrier

\begin{figure}[!htbp]
\centering\includegraphics[width=\textwidth]{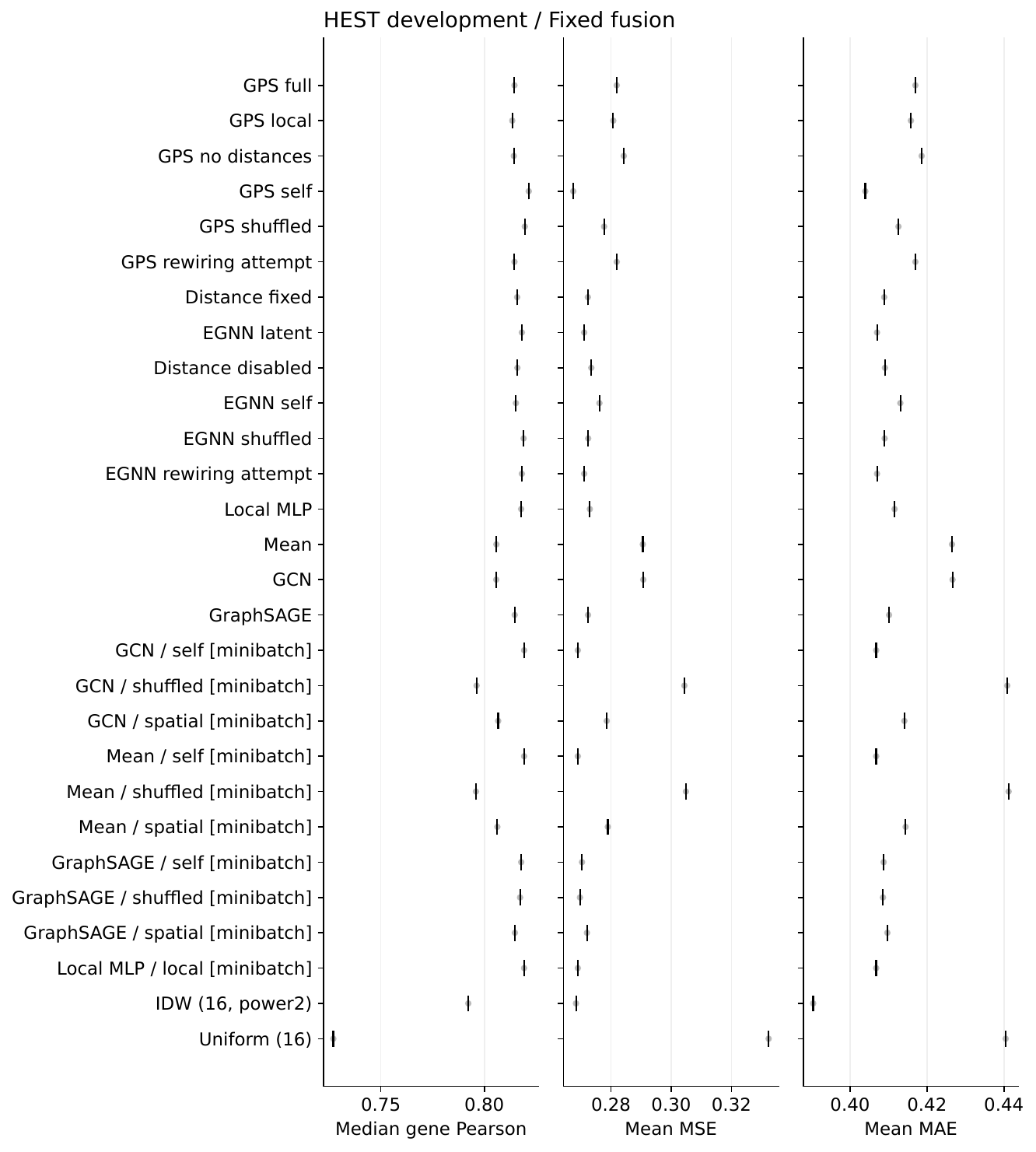}
\caption{\textbf{HEST development: fixed fusion at 20\% distributed coverage.} Results for one previously evaluated patient, TENX99. Pearson, MSE and MAE are shown after averaging three mask and three training seeds within patient. Neural rows combine 0.2 calibrated image prediction with 0.8 IDW; molecular rows use the same anchor vectors without images. The dot represents one patient; no population interval is inferred.}
\label{fig:ExtensionMatrix_Development_fixed_20pct_image}
\end{figure}
\FloatBarrier

\begin{figure}[!htbp]
\centering\includegraphics[width=\textwidth]{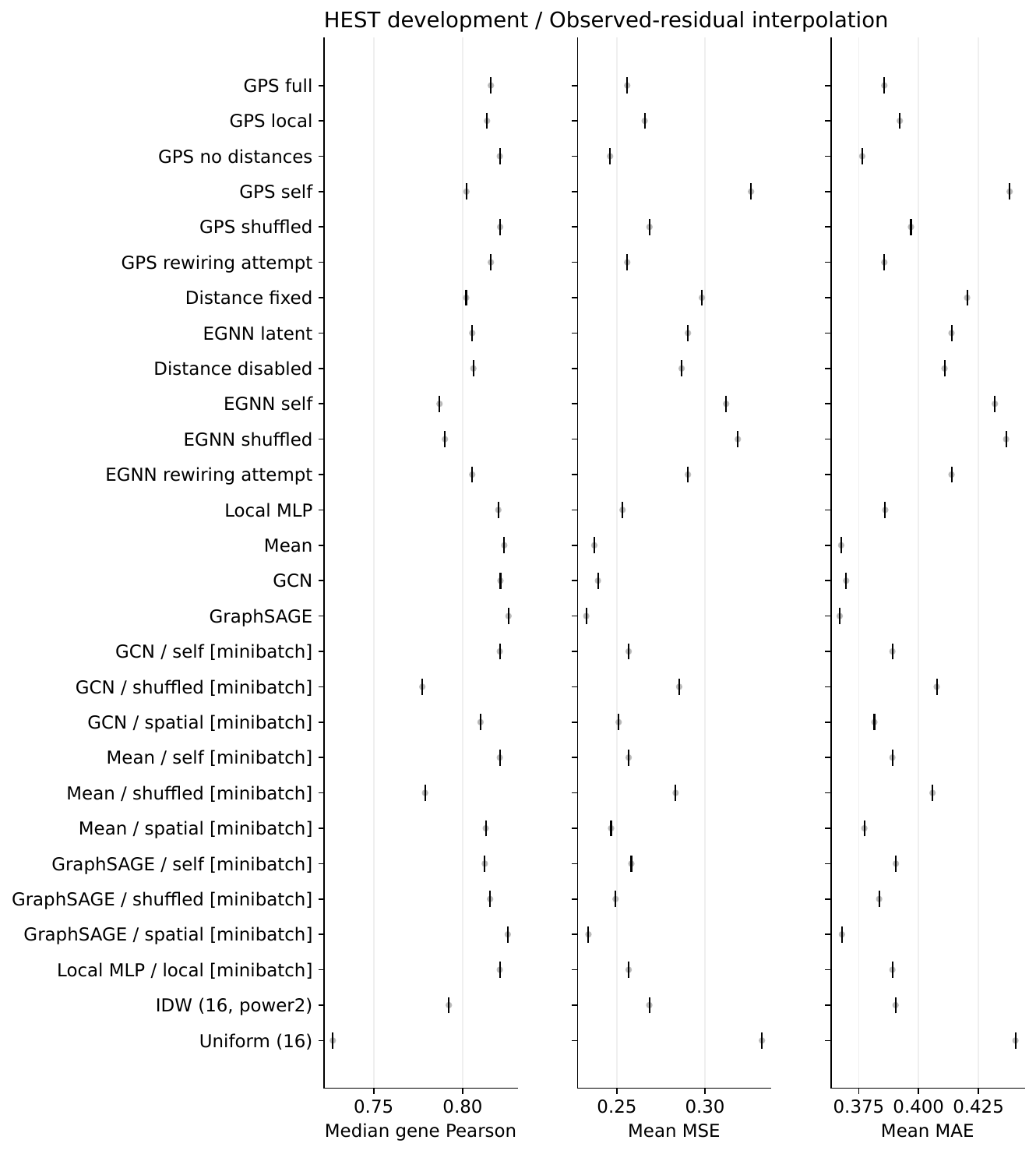}
\caption{\textbf{HEST development: observed-residual interpolation at 20\% distributed coverage.} Results for one previously evaluated patient, TENX99. Pearson, MSE and MAE are shown after averaging three mask and three training seeds within patient. Neural rows add interpolation of observed residuals to image prediction; molecular rows use the same anchors without images. The dot represents one patient; no population interval is inferred.}
\label{fig:ExtensionMatrix_Development_exploratory_residual_anchor}
\end{figure}
\FloatBarrier

\begin{figure}[!htbp]
\centering\includegraphics[width=\textwidth]{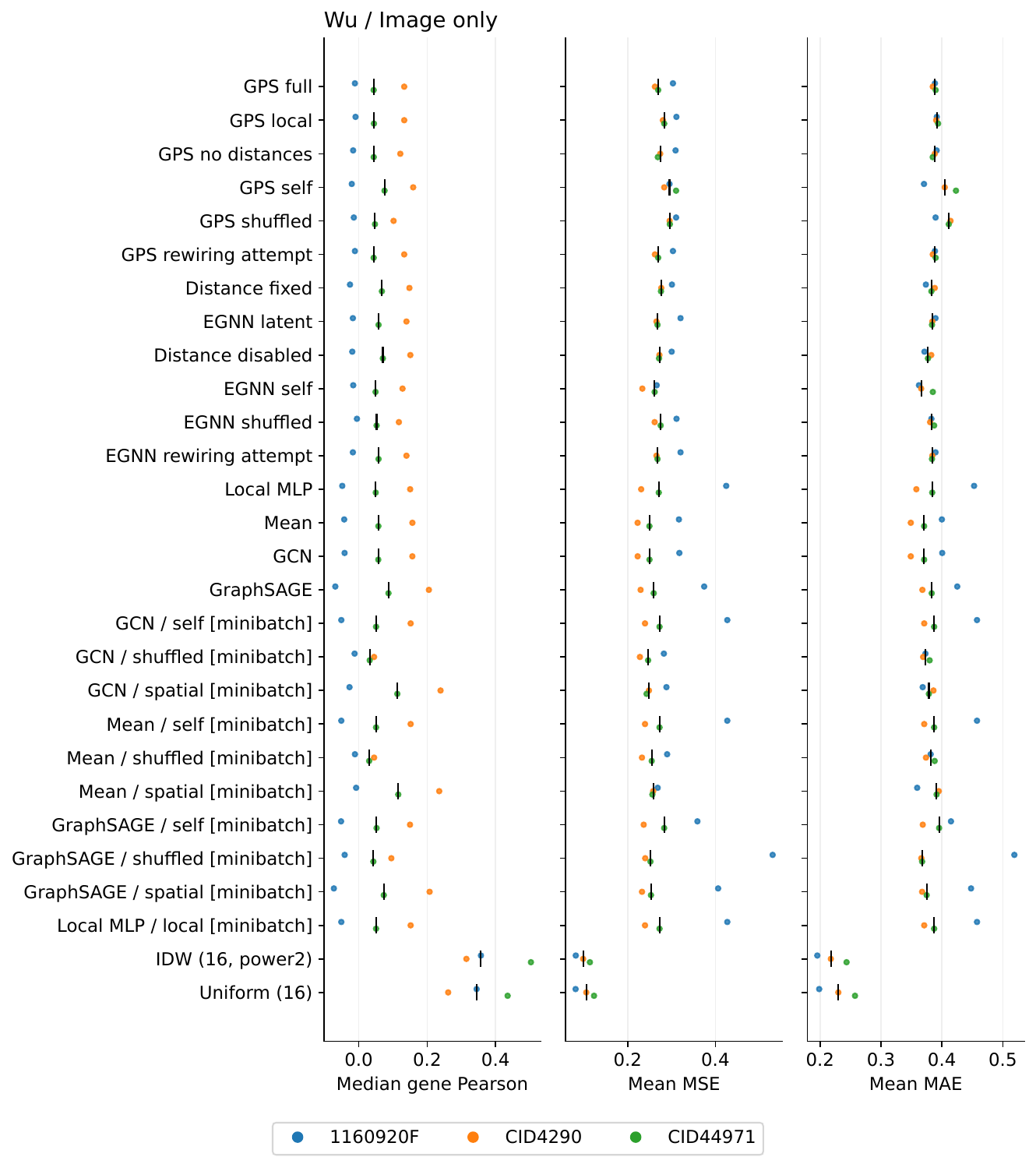}
\caption{\textbf{Wu: image-only at 20\% distributed coverage.} Results for three previously evaluated patients. Pearson, MSE and MAE are shown after averaging three mask and three training seeds within patient. Neural rows use image features and permitted geometry; the two molecular rows receive measured anchors. Colors identify 1160920F, CID4290 and CID44971 individually.}
\label{fig:ExtensionMatrix_Wu_image}
\end{figure}
\FloatBarrier

\begin{figure}[!htbp]
\centering\includegraphics[width=\textwidth]{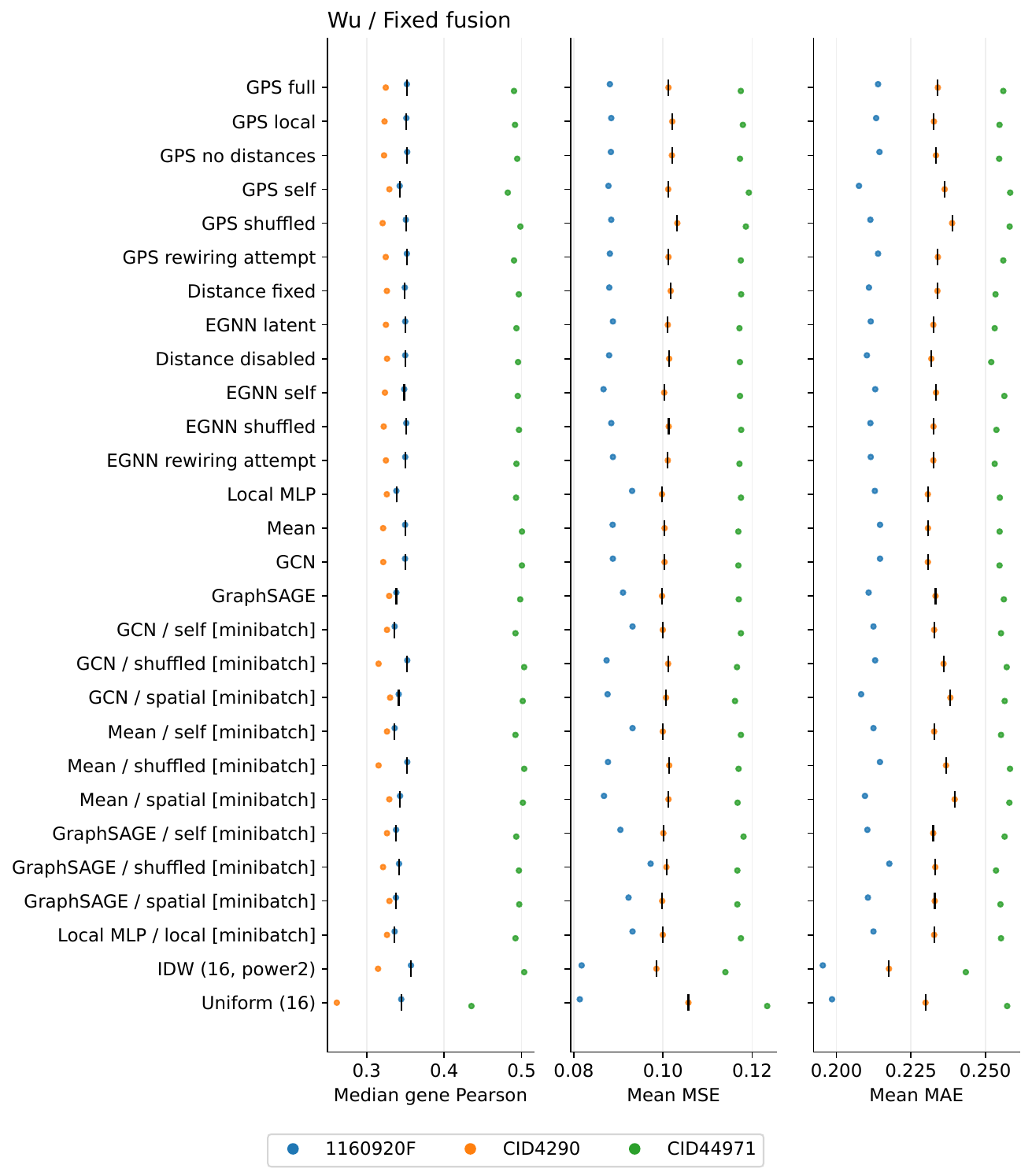}
\caption{\textbf{Wu: fixed fusion at 20\% distributed coverage.} Results for three previously evaluated patients. Pearson, MSE and MAE are shown after averaging three mask and three training seeds within patient. Neural rows combine 0.2 calibrated image prediction with 0.8 IDW; molecular rows use the same anchor vectors without images. Colors identify 1160920F, CID4290 and CID44971 individually.}
\label{fig:ExtensionMatrix_Wu_fixed_20pct_image}
\end{figure}
\FloatBarrier

\begin{figure}[!htbp]
\centering\includegraphics[width=\textwidth]{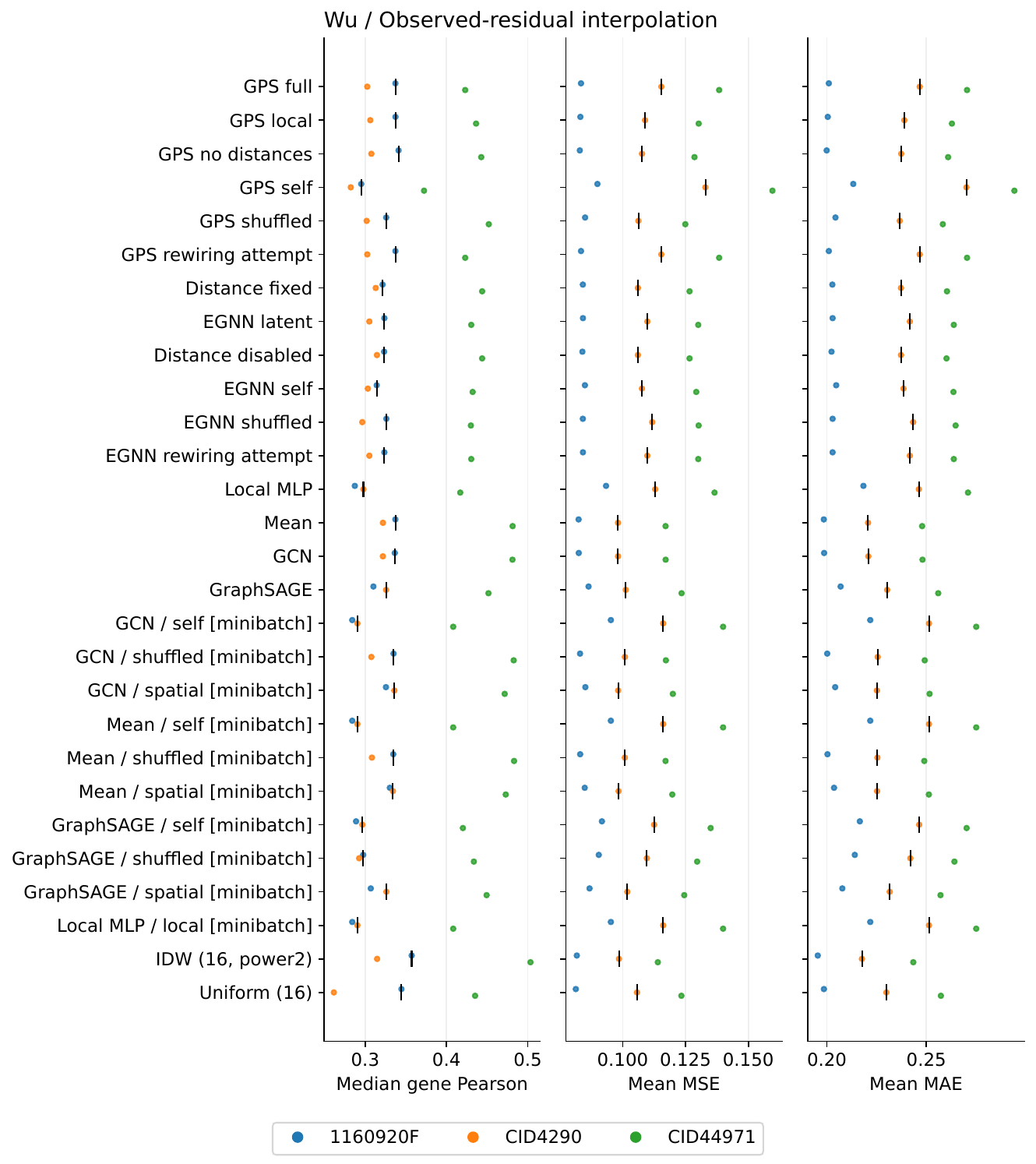}
\caption{\textbf{Wu: observed-residual interpolation at 20\% distributed coverage.} Results for three previously evaluated patients. Pearson, MSE and MAE are shown after averaging three mask and three training seeds within patient. Neural rows add interpolation of observed residuals to image prediction; molecular rows use the same anchors without images. Colors identify 1160920F, CID4290 and CID44971 individually.}
\label{fig:ExtensionMatrix_Wu_exploratory_residual_anchor}
\end{figure}
\FloatBarrier

\begin{figure}[!htbp]
\centering\includegraphics[width=\textwidth]{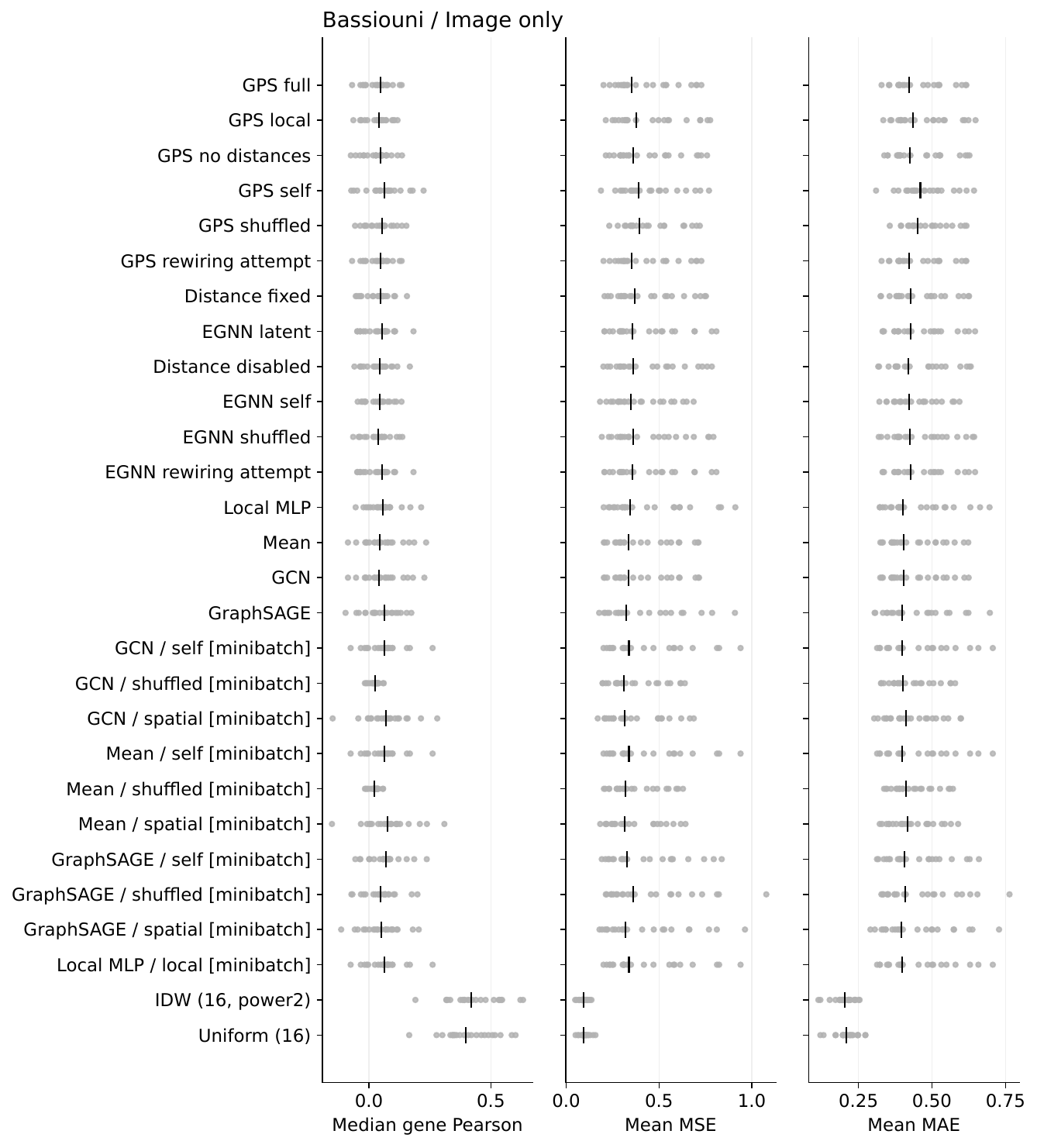}
\caption{\textbf{Bassiouni: image-only at 20\% distributed coverage.} Results for 22 previously evaluated patients. Pearson, MSE and MAE are shown after averaging three mask and three training seeds within patient. Neural rows use image features and permitted geometry; the two molecular rows receive measured anchors. Gray dots identify individual patients; black ticks are medians of their summaries.}
\label{fig:ExtensionMatrix_Bassiouni_image}
\end{figure}
\FloatBarrier

\begin{figure}[!htbp]
\centering\includegraphics[width=\textwidth]{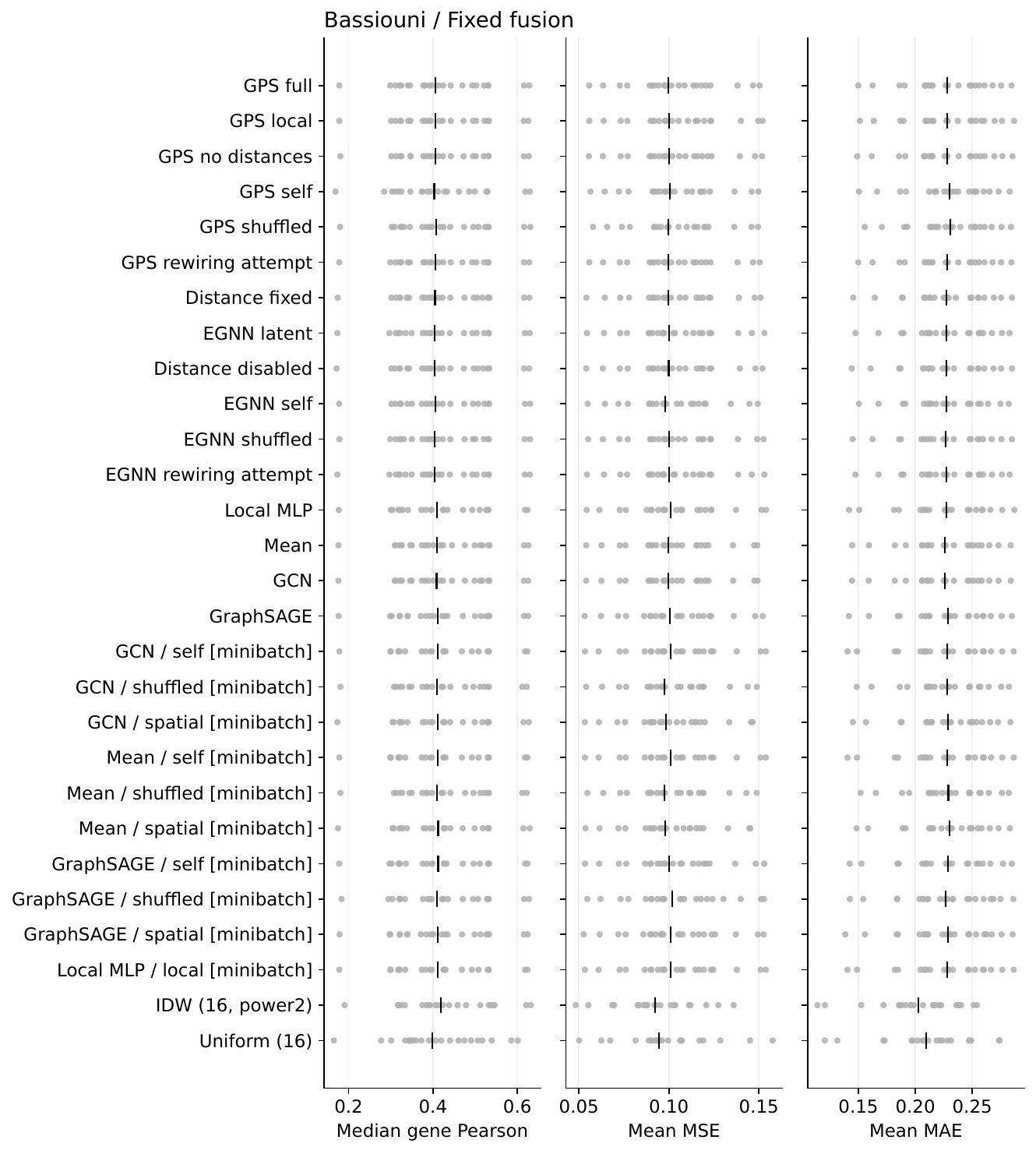}
\caption{\textbf{Bassiouni: fixed fusion at 20\% distributed coverage.} Results for 22 previously evaluated patients. Pearson, MSE and MAE are shown after averaging three mask and three training seeds within patient. Neural rows combine 0.2 calibrated image prediction with 0.8 IDW; molecular rows use the same anchor vectors without images. Gray dots identify individual patients; black ticks are medians of their summaries.}
\label{fig:ExtensionMatrix_Bassiouni_fixed_20pct_image}
\end{figure}
\FloatBarrier

\begin{figure}[!htbp]
\centering\includegraphics[width=\textwidth]{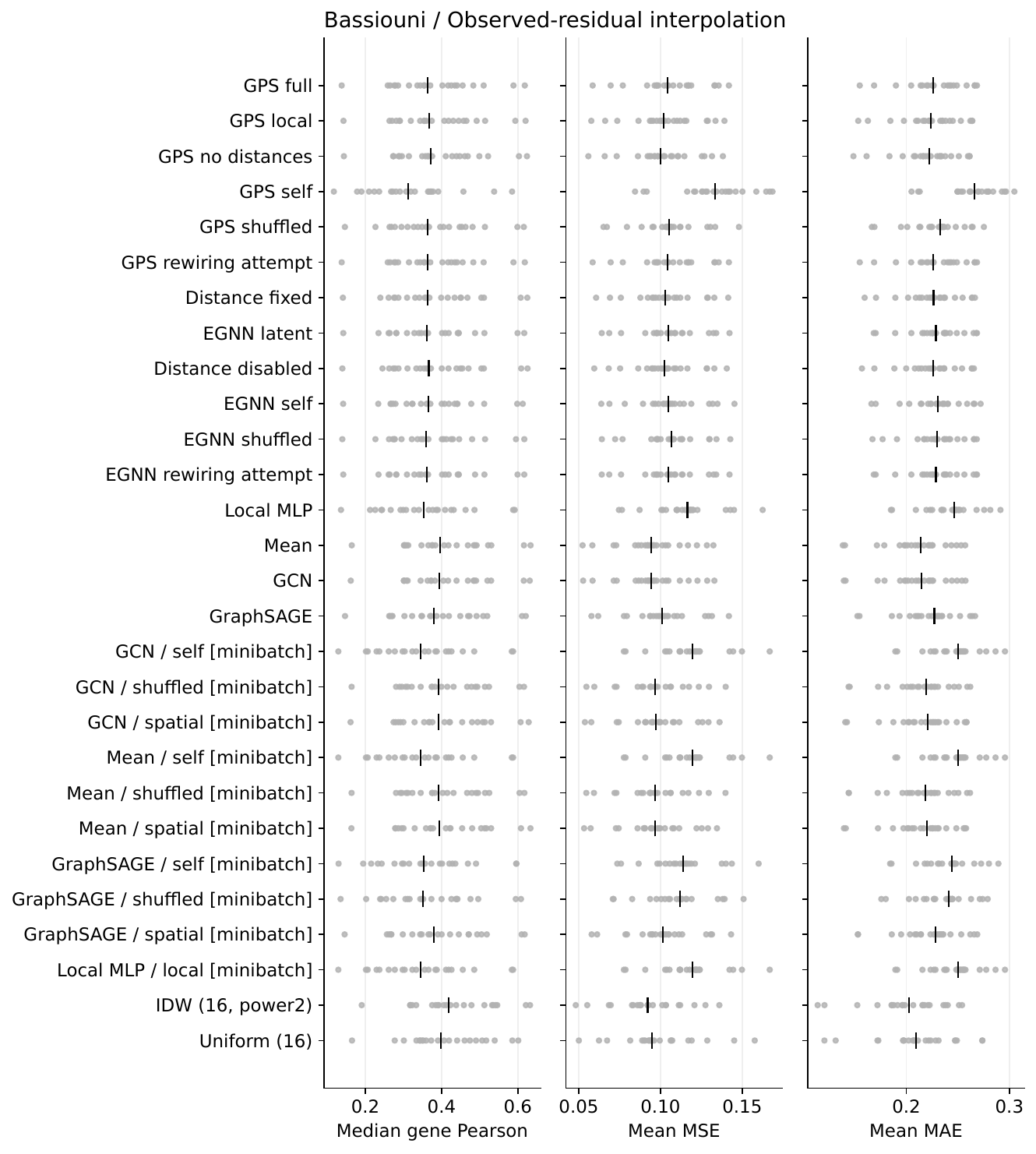}
\caption{\textbf{Bassiouni: observed-residual interpolation at 20\% distributed coverage.} Results for 22 previously evaluated patients. Pearson, MSE and MAE are shown after averaging three mask and three training seeds within patient. Neural rows add interpolation of observed residuals to image prediction; molecular rows use the same anchors without images. Gray dots identify individual patients; black ticks are medians of their summaries.}
\label{fig:ExtensionMatrix_Bassiouni_exploratory_residual_anchor}
\end{figure}
\FloatBarrier

\begin{longtable}{@{}llrrr@{}}\caption{\textbf{All 384 saved-checkpoint conditional geometric checks.} Maxima across all three seeds and recorded transforms. Physical latent errors are in micrometers. Tests are conditional on fixed image features, not the complete CNN pipeline. Baseline preaggregation remains float 32 even when the head is tested in float 64.}\label{tab:extension_geometric}\\\toprule Model & Precision & Max scalar error & Max latent error & Checks \\\midrule\endfirsthead\toprule Model & Precision & Max scalar error & Max latent error & Checks \\\midrule\endhead
Distance disabled & float 64 & 0 & $1.14\times 10^{-13}$ & 12 \\
Distance disabled & float 32 & $7.75\times 10^{-7}$ & $2.40\times 10^{-5}$ & 12 \\
Distance fixed & float 64 & $1.11\times 10^{-15}$ & $1.14\times 10^{-13}$ & 12 \\
Distance fixed & float 32 & $8.42\times 10^{-7}$ & $2.40\times 10^{-5}$ & 12 \\
EGNN rewired & float 64 & $8.88\times 10^{-16}$ & $1.14\times 10^{-13}$ & 12 \\
EGNN rewired & float 32 & $7.15\times 10^{-7}$ & $3.13\times 10^{-5}$ & 12 \\
EGNN latent & float 64 & $1.33\times 10^{-15}$ & $1.14\times 10^{-13}$ & 12 \\
EGNN latent & float 32 & $7.45\times 10^{-7}$ & $2.86\times 10^{-5}$ & 12 \\
EGNN self & float 64 & 0 & $1.14\times 10^{-13}$ & 12 \\
EGNN self & float 32 & $7.15\times 10^{-7}$ & $2.40\times 10^{-5}$ & 12 \\
EGNN shuffled & float 64 & $1.33\times 10^{-15}$ & $1.14\times 10^{-13}$ & 12 \\
EGNN shuffled & float 32 & $7.45\times 10^{-7}$ & $2.70\times 10^{-5}$ & 12 \\
GPS rewired & float 64 & $1.33\times 10^{-15}$ & $1.14\times 10^{-13}$ & 12 \\
GPS rewired & float 32 & $9.54\times 10^{-7}$ & $2.40\times 10^{-5}$ & 12 \\
GPS full & float 64 & $1.33\times 10^{-15}$ & $1.14\times 10^{-13}$ & 12 \\
GPS full & float 32 & $9.54\times 10^{-7}$ & $2.40\times 10^{-5}$ & 12 \\
GPS local & float 64 & $1.11\times 10^{-15}$ & $1.14\times 10^{-13}$ & 12 \\
GPS local & float 32 & $7.75\times 10^{-7}$ & $2.40\times 10^{-5}$ & 12 \\
GPS no distances & float 64 & 0 & $1.14\times 10^{-13}$ & 12 \\
GPS no distances & float 32 & $8.34\times 10^{-7}$ & $2.40\times 10^{-5}$ & 12 \\
GPS self & float 64 & 0 & $1.14\times 10^{-13}$ & 12 \\
GPS self & float 32 & $7.15\times 10^{-7}$ & $2.40\times 10^{-5}$ & 12 \\
GPS shuffled & float 64 & $1.55\times 10^{-15}$ & $1.14\times 10^{-13}$ & 12 \\
GPS shuffled & float 32 & $9.54\times 10^{-7}$ & $2.40\times 10^{-5}$ & 12 \\
GCN & float 64 & 0 & $1.14\times 10^{-13}$ & 12 \\
GCN & float 32 & $8.94\times 10^{-8}$ & $2.40\times 10^{-5}$ & 12 \\
Local MLP & float 64 & 0 & $1.14\times 10^{-13}$ & 12 \\
Local MLP & float 32 & $3.58\times 10^{-7}$ & $2.40\times 10^{-5}$ & 12 \\
Mean & float 64 & $3.49\times 10^{-9}$ & $1.14\times 10^{-13}$ & 12 \\
Mean & float 32 & $1.19\times 10^{-7}$ & $2.40\times 10^{-5}$ & 12 \\
GraphSAGE & float 64 & 0 & $1.14\times 10^{-13}$ & 12 \\
GraphSAGE & float 32 & $1.19\times 10^{-7}$ & $2.40\times 10^{-5}$ & 12 \\\bottomrule\end{longtable}

\begin{table}[!htbp]
\caption{\textbf{Resource accounting for every new configuration.} All three seeds are included. Training wall times were measured under the documented concurrent workload and are not a controlled throughput comparison. Standardized inference uses the same 7,760-node TENX95 section, four CPU threads, two warmups and median of five forwards, including model-specific feature aggregation/Tensor conversion but excluding frozen CNN and graph construction. Peak process RSS includes interpreter and data, not only model activations. No GPU or paid external service was used. Full per-seed values, graph construction times and environments are exported.}
\label{tab:extension_resources}
\centering\footnotesize
\begin{tabular}{@{}lrrrrr@{}}
\toprule
Model & Parameters & Best epoch range & Fit s & Forward ms & RSS MiB \\
\midrule
GPS full & 121018 & 5--60 & 5.67 & 78.4 & 546 \\
GPS local & 121018 & 5--56 & 7.04 & 102.3 & 568 \\
GPS no distances & 121018 & 5--42 & 5.45 & 69.6 & 568 \\
GPS self & 121018 & 47--53 & 15.76 & 34.6 & 568 \\
GPS shuffled & 121018 & 6--76 & 22.15 & 76.5 & 568 \\
GPS rewired & 121018 & 5--60 & 3.78 & 81.5 & 546 \\
Distance fixed & 87154 & 7--53 & 12.47 & 56.5 & 533 \\
EGNN latent & 87282 & 7--57 & 13.95 & 54.2 & 538 \\
Distance disabled & 87154 & 7--74 & 12.10 & 54.1 & 537 \\
EGNN self & 87282 & 8--35 & 1.91 & 11.7 & 534 \\
EGNN shuffled & 87282 & 7--40 & 14.76 & 54.6 & 545 \\
EGNN rewired & 87282 & 7--57 & 13.95 & 62.2 & 544 \\
Local MLP & 72114 & 8--77 & 5.30 & 9.9 & 520 \\
Mean & 72114 & 8--9 & 1.24 & 36.2 & 550 \\
GCN & 72114 & 8--9 & 1.26 & 75.8 & 581 \\
GraphSAGE & 72075 & 39--60 & 3.77 & 58.4 & 550 \\
\bottomrule
\end{tabular}
\end{table}
\FloatBarrier

\begin{longtable}{@{}lrrrrr@{}}\caption{\textbf{Every new fit, without seed selection.} Histories, optimizer states, input hashes and complete checkpoint SHA 256 values are available in the fit JSON files and inventory CSV; weights are distributed separately from Git.}\label{tab:extension_fits}\\\toprule Model & Seed & Best epoch & Ran & Seconds & RSS MiB \\\midrule\endfirsthead\toprule Model & Seed & Best epoch & Ran & Seconds & RSS MiB \\\midrule\endhead
Distance disabled & 0 & 74 & 80 & 35.82 & 632 \\
Distance disabled & 1 & 30 & 40 & 12.10 & 643 \\
Distance disabled & 2 & 7 & 17 & 5.66 & 650 \\
Distance fixed & 0 & 53 & 63 & 28.79 & 613 \\
Distance fixed & 1 & 30 & 40 & 12.47 & 638 \\
Distance fixed & 2 & 7 & 17 & 4.35 & 632 \\
EGNN rewired & 0 & 23 & 33 & 13.95 & 642 \\
EGNN rewired & 1 & 57 & 67 & 20.21 & 651 \\
EGNN rewired & 2 & 7 & 17 & 5.02 & 643 \\
EGNN latent & 0 & 23 & 33 & 13.95 & 640 \\
EGNN latent & 1 & 57 & 67 & 19.52 & 639 \\
EGNN latent & 2 & 7 & 17 & 5.50 & 648 \\
EGNN self & 0 & 35 & 45 & 4.64 & 506 \\
EGNN self & 1 & 8 & 18 & 1.88 & 504 \\
EGNN self & 2 & 8 & 18 & 1.91 & 502 \\
EGNN shuffled & 0 & 24 & 34 & 18.71 & 634 \\
EGNN shuffled & 1 & 40 & 50 & 14.76 & 635 \\
EGNN shuffled & 2 & 7 & 17 & 4.77 & 643 \\
GPS rewired & 0 & 60 & 70 & 35.00 & 631 \\
GPS rewired & 1 & 5 & 15 & 3.70 & 633 \\
GPS rewired & 2 & 6 & 16 & 3.78 & 624 \\
GPS full & 0 & 60 & 70 & 15.56 & 626 \\
GPS full & 1 & 5 & 15 & 5.67 & 629 \\
GPS full & 2 & 6 & 16 & 3.82 & 629 \\
GPS local & 0 & 56 & 66 & 30.11 & 692 \\
GPS local & 1 & 5 & 15 & 5.24 & 679 \\
GPS local & 2 & 5 & 15 & 7.04 & 674 \\
GPS no distances & 0 & 42 & 52 & 31.17 & 600 \\
GPS no distances & 1 & 5 & 15 & 3.67 & 599 \\
GPS no distances & 2 & 6 & 16 & 5.45 & 596 \\
GPS self & 0 & 51 & 61 & 20.15 & 543 \\
GPS self & 1 & 47 & 57 & 15.76 & 546 \\
GPS self & 2 & 53 & 63 & 13.54 & 546 \\
GPS shuffled & 0 & 76 & 80 & 47.42 & 631 \\
GPS shuffled & 1 & 46 & 56 & 22.15 & 630 \\
GPS shuffled & 2 & 6 & 16 & 6.00 & 634 \\
GCN & 0 & 8 & 18 & 1.26 & 524 \\
GCN & 1 & 8 & 18 & 0.98 & 522 \\
GCN & 2 & 9 & 19 & 1.36 & 522 \\
Local MLP & 0 & 8 & 18 & 2.20 & 497 \\
Local MLP & 1 & 77 & 80 & 5.30 & 497 \\
Local MLP & 2 & 74 & 80 & 6.21 & 497 \\
Mean & 0 & 8 & 18 & 1.27 & 521 \\
Mean & 1 & 8 & 18 & 1.01 & 522 \\
Mean & 2 & 9 & 19 & 1.24 & 522 \\
GraphSAGE & 0 & 48 & 58 & 6.51 & 570 \\
GraphSAGE & 1 & 60 & 70 & 3.77 & 570 \\
GraphSAGE & 2 & 39 & 49 & 2.81 & 570 \\\bottomrule\end{longtable}

\paragraph{Diagnostic interpretation.} Undefined correlations remain undefined, with valid-gene counts in the exports. Raw Spearman is secondary because near-tied ranks are floating-point sensitive. Distance curves retain occupied-bin support; different bins need not contain the same locations or patients. Conditional geometric checks and remote-feature sensitivity do not establish biological association. Degree/length controls with zero accepted swaps do not provide an independent rewired comparison. Optional anchor-conditioned learning and full-image equivariance tests were not performed.

\begin{figure}[!htbp]
\centering\includegraphics[width=\textwidth]{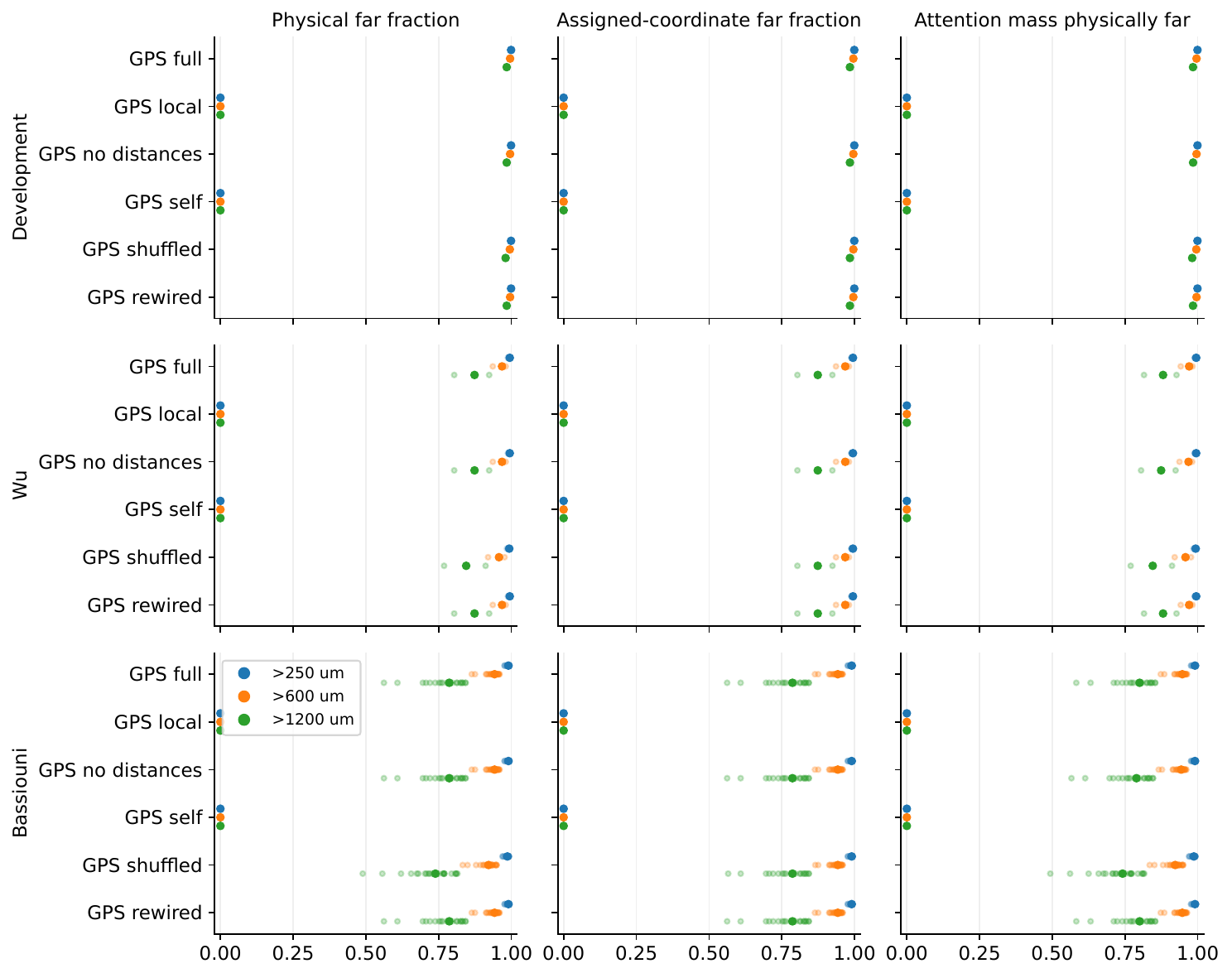}
\caption{\textbf{Conditional attention reach.} Dots are individual patient summaries after equal seed/head/layer means; large dots are cohort medians. Colors identify 250/600/1200 micrometer thresholds. This is a software/connectivity diagnostic, not evidence of molecular correlation. All GPS variants, seeds and patients are exported.}
\label{fig:ExtensionAttentionReach}
\end{figure}
\FloatBarrier

\begin{figure}[!htbp]
\centering\includegraphics[width=\textwidth]{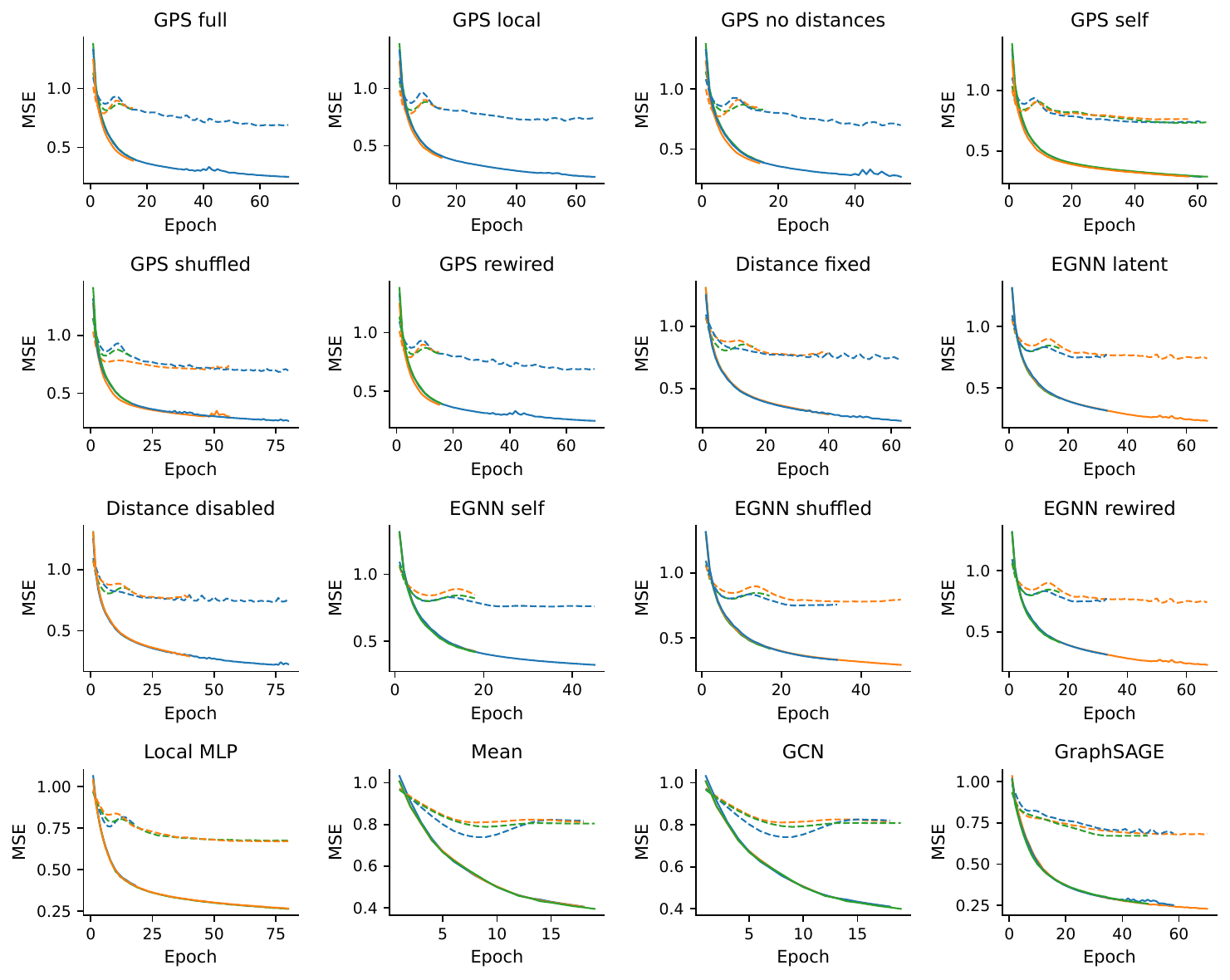}
\caption{\textbf{All 48 source-only training histories.} Every one of 16 configurations and all three seeds are retained. Solid curves show training MSE, dashed curves validation MSE. Blue, orange and green identify seeds 0, 1 and 2, respectively. Only TENX95 validation MSE selected epochs; evaluation outcomes did not select architecture or seeds.}
\label{fig:ExtensionLearningCurves}
\end{figure}
\FloatBarrier

\begin{figure}[!htbp]
\centering\includegraphics[width=\textwidth]{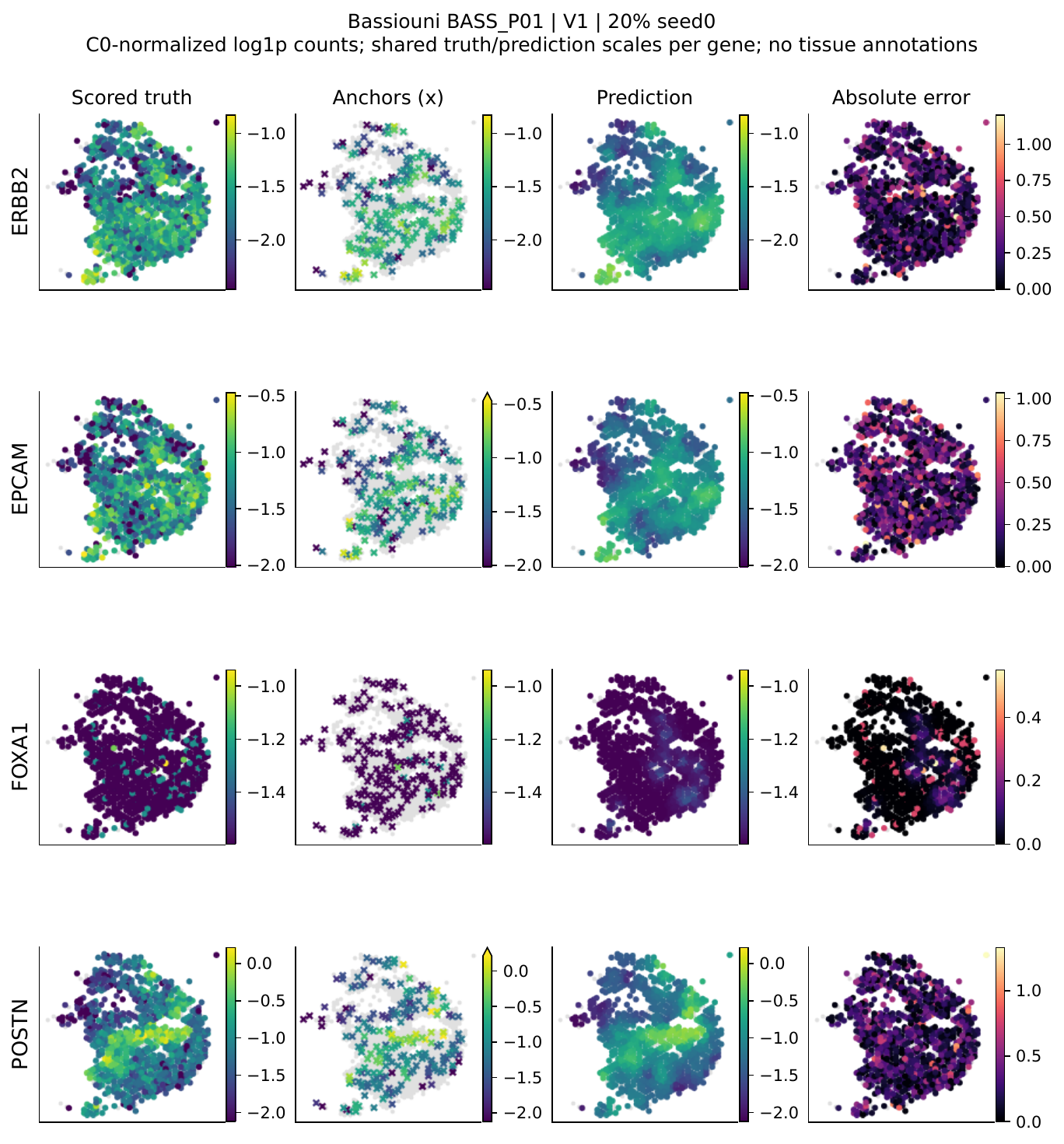}
\caption{\textbf{Predetermined Bassiouni distributed map.} BASS P01 at 20percent seed0, the four prespecified genes, original IDW k16/p2. Truth and prediction share each gene scale; anchor colorbar extensions mark clipped anchor values. Gray sites are unscored context. Constant fields are retained and explicitly marked; no tissue annotations are inferred.}
\label{fig:ExtensionBassiouniMap_V1}
\end{figure}
\FloatBarrier

\begin{figure}[!htbp]
\centering\includegraphics[width=\textwidth]{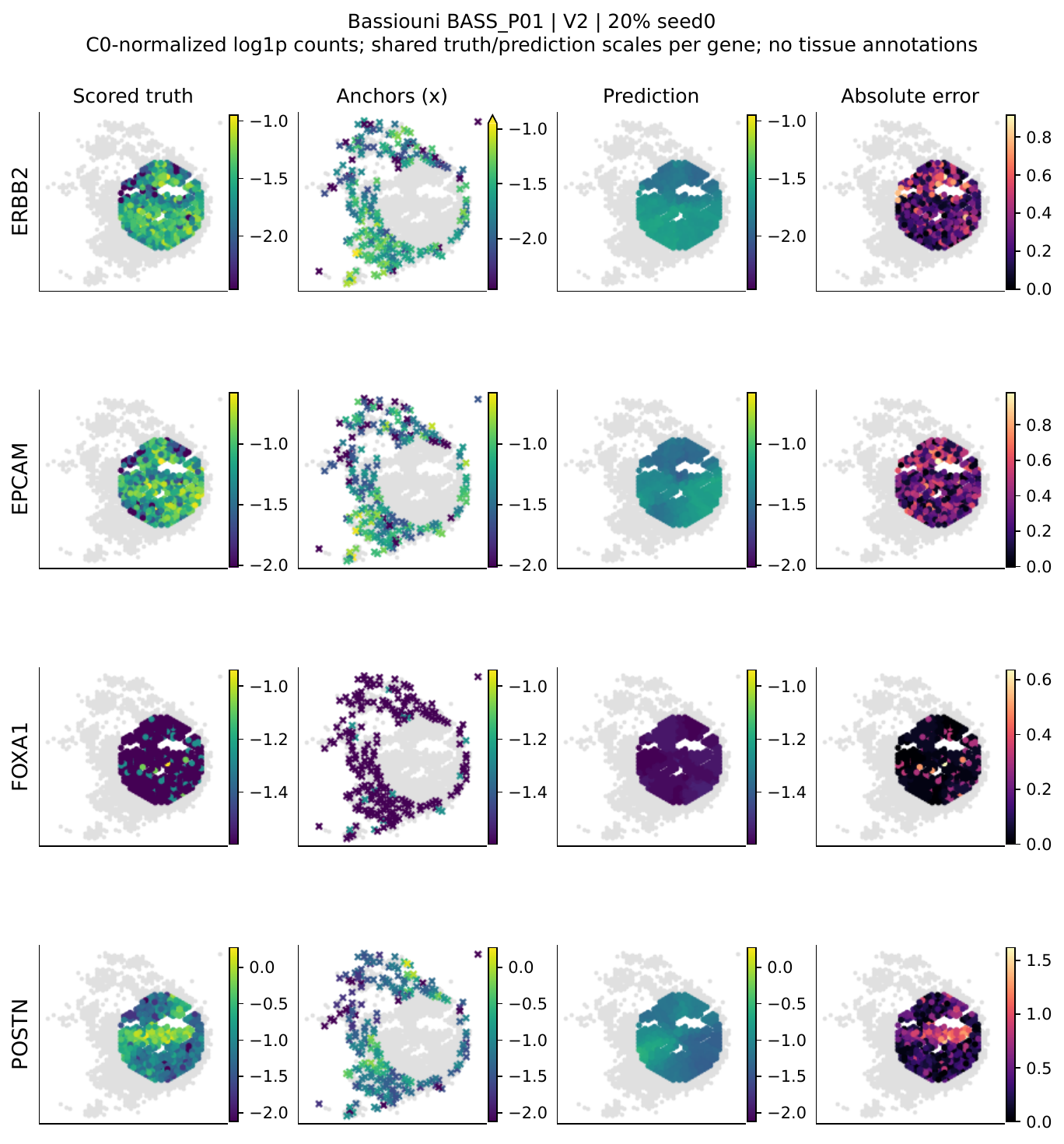}
\caption{\textbf{Predetermined Bassiouni regional map.} BASS P01 central V2 at 20percent seed0, uniform k16 and the same four predetermined genes. Historical V2 has different scored support and estimator from V1; this map comparison does not isolate geometry. Identical-H effects are evaluated separately. Truth/prediction scales are shared within each gene, errors use a separate nonnegative scale.}
\label{fig:ExtensionBassiouniMap_V2}
\end{figure}
\FloatBarrier

\begin{figure}[!htbp]
\centering\includegraphics[width=\textwidth]{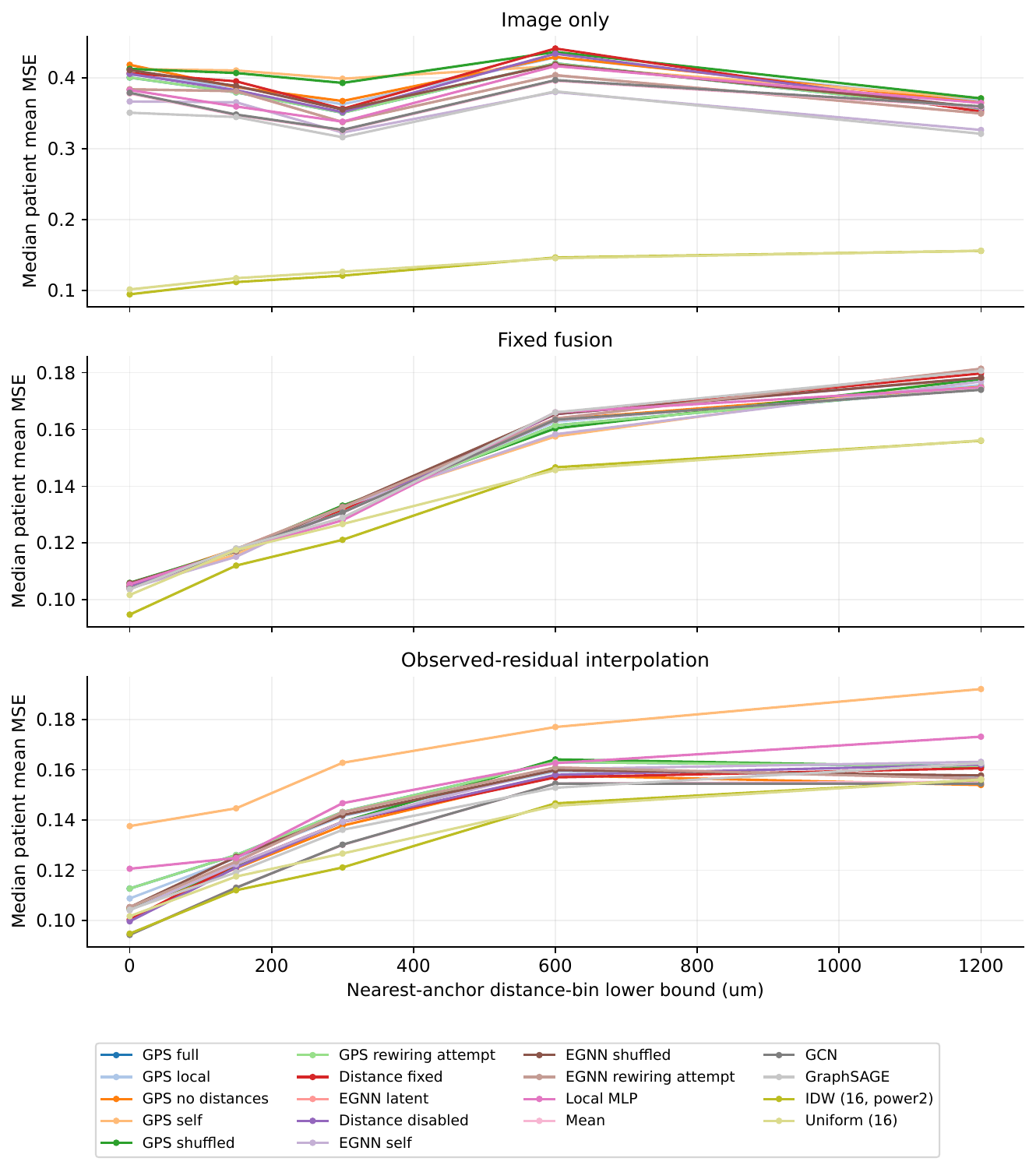}
\caption{\textbf{Distance-resolved Bassiouni reconstruction.} All16 new models and both molecular baselines,20percent region-exclusion B2. Curves are medians of per-patient conditional-bin mean errors after equal mask and training-seed averages. Occupied bins only; empty bins are omitted. Source tables retain number of patients and min/median/max mean locations per mask in each bin. Different bins need not contain the same patients or locations and do not identify a biological correlation length.}
\label{fig:ExtensionDistanceBins}
\end{figure}
\FloatBarrier

\begin{longtable}{@{}lrrrrrrr@{}}\caption{\textbf{All 48 full-section development-test fits, with every training seed retained.} TENX99 at 20\% coverage, averaging the same three original masks within each seed. Left metrics are image-only; right metrics use exploratory residual anchors. This is one previously inspected patient, not 48 biological replicates. No confidence interval or seed selection is inferred. Full precision, fixed-fusion, all budgets, external patient metrics and older frozen heads accompany the CSV exports.}\label{tab:extension_test_seeds}\\\toprule Model & Seed & Image $r$ & MSE & MAE & Residual $r$ & MSE & MAE \\\midrule\endfirsthead\toprule Model & Seed & Image $r$ & MSE & MAE & Residual $r$ & MSE & MAE \\\midrule\endhead
GPS full & 0 & 0.714 & 0.9751 & 0.7814 & 0.834 & 0.2349 & 0.3694 \\
GPS full & 1 & 0.529 & 1.1646 & 0.8594 & 0.801 & 0.2693 & 0.3967 \\
GPS full & 2 & 0.569 & 1.2240 & 0.8969 & 0.813 & 0.2634 & 0.3909 \\
GPS local & 0 & 0.708 & 0.9202 & 0.7560 & 0.826 & 0.2662 & 0.3921 \\
GPS local & 1 & 0.573 & 1.1372 & 0.8470 & 0.804 & 0.2694 & 0.3958 \\
GPS local & 2 & 0.508 & 1.2035 & 0.8928 & 0.811 & 0.2617 & 0.3885 \\
GPS no distances & 0 & 0.711 & 1.0426 & 0.8070 & 0.835 & 0.2242 & 0.3589 \\
GPS no distances & 1 & 0.556 & 1.1672 & 0.8560 & 0.809 & 0.2596 & 0.3878 \\
GPS no distances & 2 & 0.577 & 1.2357 & 0.8978 & 0.819 & 0.2545 & 0.3827 \\
GPS self & 0 & 0.664 & 1.0495 & 0.8149 & 0.803 & 0.3344 & 0.4438 \\
GPS self & 1 & 0.669 & 1.0613 & 0.8177 & 0.795 & 0.3441 & 0.4500 \\
GPS self & 2 & 0.671 & 1.0259 & 0.8074 & 0.808 & 0.3001 & 0.4204 \\
GPS shuffled & 0 & 0.717 & 0.9751 & 0.7806 & 0.822 & 0.2865 & 0.4111 \\
GPS shuffled & 1 & 0.710 & 1.2113 & 0.8818 & 0.829 & 0.2456 & 0.3793 \\
GPS shuffled & 2 & 0.552 & 1.2417 & 0.9061 & 0.812 & 0.2733 & 0.4001 \\
GPS rewired & 0 & 0.714 & 0.9751 & 0.7814 & 0.834 & 0.2349 & 0.3694 \\
GPS rewired & 1 & 0.529 & 1.1646 & 0.8594 & 0.801 & 0.2693 & 0.3967 \\
GPS rewired & 2 & 0.569 & 1.2240 & 0.8969 & 0.813 & 0.2634 & 0.3909 \\
Distance fixed & 0 & 0.680 & 0.9648 & 0.7772 & 0.820 & 0.2945 & 0.4164 \\
Distance fixed & 1 & 0.637 & 0.9801 & 0.7883 & 0.793 & 0.3136 & 0.4327 \\
Distance fixed & 2 & 0.471 & 1.1048 & 0.8520 & 0.793 & 0.2867 & 0.4118 \\
EGNN latent & 0 & 0.628 & 1.0248 & 0.8072 & 0.799 & 0.3033 & 0.4259 \\
EGNN latent & 1 & 0.711 & 0.9576 & 0.7706 & 0.825 & 0.2646 & 0.3915 \\
EGNN latent & 2 & 0.536 & 1.0697 & 0.8344 & 0.792 & 0.3030 & 0.4239 \\
Distance disabled & 0 & 0.706 & 1.0438 & 0.8028 & 0.831 & 0.2609 & 0.3887 \\
Distance disabled & 1 & 0.640 & 0.9790 & 0.7877 & 0.794 & 0.3128 & 0.4320 \\
Distance disabled & 2 & 0.471 & 1.1047 & 0.8519 & 0.793 & 0.2866 & 0.4118 \\
EGNN self & 0 & 0.595 & 0.9884 & 0.7966 & 0.778 & 0.3307 & 0.4476 \\
EGNN self & 1 & 0.546 & 1.1053 & 0.8471 & 0.791 & 0.3008 & 0.4227 \\
EGNN self & 2 & 0.545 & 1.0548 & 0.8283 & 0.793 & 0.3046 & 0.4253 \\
EGNN shuffled & 0 & 0.627 & 1.0708 & 0.8276 & 0.794 & 0.3082 & 0.4303 \\
EGNN shuffled & 1 & 0.665 & 1.0025 & 0.7999 & 0.785 & 0.3441 & 0.4550 \\
EGNN shuffled & 2 & 0.532 & 1.0732 & 0.8359 & 0.791 & 0.3038 & 0.4245 \\
EGNN rewired & 0 & 0.628 & 1.0248 & 0.8072 & 0.799 & 0.3033 & 0.4259 \\
EGNN rewired & 1 & 0.711 & 0.9576 & 0.7706 & 0.825 & 0.2646 & 0.3915 \\
EGNN rewired & 2 & 0.536 & 1.0697 & 0.8344 & 0.792 & 0.3030 & 0.4239 \\
Local MLP & 0 & 0.604 & 0.9930 & 0.8144 & 0.829 & 0.2433 & 0.3775 \\
Local MLP & 1 & 0.669 & 0.9033 & 0.7628 & 0.820 & 0.2591 & 0.3910 \\
Local MLP & 2 & 0.667 & 0.9185 & 0.7714 & 0.811 & 0.2571 & 0.3897 \\
Mean & 0 & 0.624 & 1.0157 & 0.8231 & 0.827 & 0.2361 & 0.3667 \\
Mean & 1 & 0.593 & 1.0522 & 0.8471 & 0.819 & 0.2387 & 0.3688 \\
Mean & 2 & 0.627 & 1.0723 & 0.8525 & 0.824 & 0.2366 & 0.3674 \\
GCN & 0 & 0.620 & 1.0175 & 0.8241 & 0.825 & 0.2386 & 0.3692 \\
GCN & 1 & 0.591 & 1.0536 & 0.8478 & 0.818 & 0.2406 & 0.3707 \\
GCN & 2 & 0.623 & 1.0727 & 0.8529 & 0.821 & 0.2385 & 0.3694 \\
GraphSAGE & 0 & 0.671 & 0.8701 & 0.7488 & 0.827 & 0.2333 & 0.3675 \\
GraphSAGE & 1 & 0.688 & 0.8818 & 0.7530 & 0.829 & 0.2337 & 0.3683 \\
GraphSAGE & 2 & 0.667 & 0.9245 & 0.7785 & 0.821 & 0.2307 & 0.3653 \\\bottomrule\end{longtable}
\clearpage
\section{Detailed development methods}
\label{sec:detailed_development}
These specifications distinguish the original development comparators from the additional minibatch and full-section comparisons. They retain the model-selection and calibration details needed to reproduce the corresponding figures.
\subsection{Development head searches and conditional patient holdouts}
The original development MLPs used one hidden linear layer, GELU, dropout and a linear gene-output layer. Features were standardized on fitting patients before graph aggregation; a separate fitting-only head-target scaler was inverted before scoring. Population scales below $10^{-8}$ were replaced by 1. AdamW used learning rate $10^{-3}$, weight decay $10^{-4}$, minibatches of 512 with shuffled rows and retained final partial batches, at most 80 epochs, patience 10 and minimum validation-MSE improvement $10^{-4}$. Best qualifying parameters were restored, with dropout disabled at prediction. Three training seeds 0/1/2 were retained; hyperparameter candidates used seed 0. Computation used deterministic CPU float32 tensors and four threads, without a learning-rate schedule or auxiliary loss. These settings describe the original development comparisons below, not the later full-section regime.

The initial local comparison in Supplementary Fig.~S3 selected hidden width from $\{128,256\}$ and dropout from $\{0, 0.1\}$ by the lowest validation MSE. For 50 genes, the selected architecture was $512\rightarrow128\rightarrow50$ with dropout 0.1. The ridge baseline used an intercept and a Cholesky solver, with penalty selected from $\{10^{-4},10^{-3},10^{-2},10^{-1},1,10,100\}$ by the highest validation median gene Pearson correlation, breaking ties by lower MSE and then smaller penalty; the selected 50-gene penalty was 100. Training-target and training-embedding permutations disrupted pairing within the fitting data; validation and evaluation data were unchanged.

The representation comparison in Supplementary Fig.~S6 used the same four width/dropout candidates but selected the highest validation median gene Pearson correlation, with lower MSE as the first tie-breaker. Both ResNet18 and Phikon selected width 256 and dropout 0.1 for 50 genes. Thus the initial local predictor in Supplementary Fig.~S3 and the ResNet18 predictor used in the representation comparison and original sparse pipeline have different selected widths. The latter predictor is described in the main Methods. The representation comparison also evaluated ridge penalties $\{0.1,1,10,100,1000\}$; both encoders selected 1000 for 50 genes. Equal width and search budget did not imply equal parameter count because the input dimensions differed.

Conditional leave-one-patient-out sensitivity fits used the remaining three development patients, refitted feature and head-target scalers on those patients, retained the primary-selected hyperparameters and used fixed epoch counts without early stopping on the held-out patient. For the 50-gene heads, these counts were 14 for the initial MLP, 14 for mean pooling and GCN, 7 for GraphSAGE, 12 for the representation-comparison ResNet18 head and 1 for the Phikon head. These were rounded medians of the corresponding primary fits' best epochs. Since the configurations and underlying target definition had already been selected during development, these fits are conditional sensitivity analyses.

\subsection{Original graph controls and pathology representation}
The matched controls in Supplementary Fig.~S5 retained the GraphSAGE head and training settings. The self-only control used $[X,X]$. Vertex permutation preserved the graph's degree distribution, while double-edge rewiring preserved each node's degree. The additional distance-constrained rewiring preserved the numbers of physical edge lengths in $[0,175)$, $[175,250)$ and $[250,\infty)$~\um{}. Rewiring targeted five accepted swaps per edge, with limits of 100 attempts per edge for degree-only rewiring and 1,000 for distance-constrained rewiring; the latter used candidate endpoints within 275~\um{}. Five graph randomization seeds were crossed with the three model seeds. The Phikon spatial sensitivity comparison reused the selected graph, hidden widths and dropout: mean pooling used $768\rightarrow128\rightarrow50$, and GraphSAGE used $1536\rightarrow67\rightarrow50$, with true and degree-plus-length-matched graphs. These inherited widths do not constitute exact parameter matching between encoders.

The historical Phikon representation is defined in the main Methods subsection ``Registration and frozen image features''. Its C4 extraction used float32 evaluation-mode inference, batches of 16 and eight CPU threads (Transformers 4.57.3 and PyTorch scaled dot-product attention). It used the same archived RGB patches without recropping. Historical head selection and target scaling followed the development procedures above; these differ from the direct-C0-target, validation-MSE contract of the matched-backbone follow-up. Neither analysis fine-tuned the encoder.

\subsection{Development ridge residual correction}
The learned-propagation model in Supplementary Fig.~S7 is a gene-specific ridge residual correction using fixed graph summaries. Its image prediction $H$ is the equal average of the three frozen local MLP predictions from the representation comparison. Let $o$ be the observed-location indicator and $Z$ the expression matrix containing permitted observed vectors and zeros elsewhere. A self-inclusive row-normalized operator $P$ was constructed from the symmetrized six-nearest-neighbor graph capped at 250~\um{}. The observed masses and molecular messages were
\[
 m^{(1)}=Po,\quad m^{(2)}=P^2o,\qquad
 V^{(1)}_{ig}=\frac{(PZ)_{ig}}{\max(m^{(1)}_i,10^{-6})},\quad
 V^{(2)}_{ig}=\frac{(P^2Z)_{ig}}{\max(m^{(2)}_i,10^{-6})}.
\]
Image features were standardized on fitting patients and reduced to eight whitened principal components with randomized singular-value decomposition and random seed 0. Denote these components by $X_{\rm PCA}$. The 29 shared covariates at each location were $X_{\rm PCA}$, $PX_{\rm PCA}$, $P^2X_{\rm PCA}$, $o$, $m^{(1)}$, $m^{(2)}$, $\log(1+d_1/100)$, and the fraction of the nearest 16 observed anchors lying within 300~\um{}, where $d_1$ is nearest-anchor distance in \um{}. Five further covariates for gene $g$ were the image prediction $H_g$, the fixed fusion prediction $F_g$, $Z_g$, $V^{(1)}_g$ and $V^{(2)}_g$. The resulting design has 34 covariates and an intercept for each gene.

Predictions were $\widehat y_g=F_g+\Phi_g\theta_g$, where the last column of $\Phi_g$ is one. Coefficients were fitted by
\[
 \underset{\theta_g}{\operatorname{argmin}}\;
 \left\|(y_g-F_g)-\Phi_g\theta_g\right\|_2^2
 +\alpha\left\|\theta_{g,1:34}\right\|_2^2.
\]
The intercept was unpenalized. Fitting sampled up to 256 hidden locations without replacement per training mask using that mask's seed; their expression entered only the supervised response. Masks covered 5, 10, 20, 40, 60 and 80\% of locations under random, block and capture-window sampling, with three replicates per condition. The penalty grid was $\{1,100,10000\}$, selected by mean validation MSE across masks with equal mask weights; both encoders selected $\alpha=10000$ for 50 genes. No additional expression scaler or auxiliary loss was introduced.

The underlying development fusion was $F=\lambda H+(1-\lambda)M$. Nearest-anchor mean interpolation selected $k$ from $\{1,4,8,16\}$; inverse-distance interpolation used up to 16 anchors and selected power from $\{1,2,3\}$. The stronger family and $\lambda\in\{0, 0.25, 0.5, 0.75,1\}$ were chosen by mean validation MSE separately by panel and sampling geometry. For 50 genes, random sampling selected inverse-distance-squared interpolation and $\lambda=0.25$; block sampling selected the 16-anchor mean and $\lambda=0.5$; capture-window sampling selected inverse-distance-cubed interpolation and $\lambda=0.5$. These development settings differ from the final external fusion settings. Supplementary Fig.~S7 evaluates three permutation seeds on one random 20\% observation mask, with the fitted models unchanged. The plotted B1, B3 and B4 labels denote the frozen image predictor, fixed fusion and ridge residual correction, respectively.

\subsection{Patient-disjoint HER2ST calibration and fusion selection}
Supplementary Figs.~S1 and S10 use the same eight HER2ST patients and 40 shared genes. Four outer folds held out the pairs A1/B1, C1/D1, E1/F1 and G1/H1, respectively. Within each remaining six-patient set, target-mapping candidates were evaluated by leave-one-patient-out records: the mapping for an inner patient's predictions was estimated from the other five patients. The fixed source moments came from NCBI783 and NCBI785. The candidates were the unchanged target, per-gene mean/standard-deviation mapping and per-gene median/interquartile-range mapping. Each patient had equal total weight in moment or quantile estimates. An interquartile range below $10^{-8}$ fell back to the corresponding standard deviation, and a standard deviation below $10^{-8}$ was replaced by one. Source and destination moments were estimated separately. Image predictions were expressed in source-standardized coordinates, anchors in destination-standardized coordinates, and final errors were computed after mapping predictions back to common target units.

The target-mapping comparison retained fixed image weights 0.25 for V1 and 0.5 for V2. Selection used inner random-mask seeds 0--2 and four equally weighted conditions: V1 coverage 10, 20 and 40\%, and V2 coverage 20\%. Seeds and geometries were first averaged within patient and condition, then conditions within patient, then the six patients. Candidates within 0.005 of the highest mean patient median gene Pearson correlation were eligible; among these, the candidate with lowest common-unit MSE was selected. The mean/standard-deviation transform was selected in every fold; the corresponding outer mapping used moments estimated from all six outer-training patients.

The selected-fusion analysis then used the corresponding inner records to fit and select among four families: molecular interpolation alone; one global image weight from $\{0, 0.1,\ldots,1\}$; 35 nondecreasing three-bin weight rules with weights from $\{0, 0.1, 0.25, 0.5,1\}$ and nearest-anchor boundaries 300 and 600~\um{}; and a five-parameter sigmoid gate. All weights were shared across genes, regimes and budgets within an outer fold. The sigmoid gate was
\[
 w_i=\operatorname{sigmoid}(\beta_0+\widetilde q_i^\top\beta),\qquad
 \widehat y_i=M_i+w_i(H_i-M_i),
\]
where the four features before fitting-data standardization were $\log(1+d_1/200)$, $\log(1+\overline d_{16}/200)$, $\log(1+n_{600})$ and $\log(1+v)$: nearest-anchor distance, mean distance to up to 16 nearest anchors, the number of observed anchors within 600~\um{}, and mean across-gene variance among the neighboring observed vectors. The gate minimized common-unit squared error with equal patient/condition/run/location contributions, divided by each run's interpolation MSE floored at $10^{-8}$, plus $0.01\|\beta\|_2^2+10^{-6}\beta_0^2$. The same weights defined feature means and standard deviations. L-BFGS-B used initial coefficients $(-2,0,0,0,0)$, bounds $[-12,12]$, maximum 300 iterations, function tolerance $10^{-12}$ and gradient tolerance $10^{-7}$. No outer-patient outcomes entered fitting or candidate selection.

The same correlation-then-MSE rule selected the fusion family. The first three patient-pair folds selected global image weight 0.2; the G1/H1 fold selected the sigmoid gate. Outer evaluation used mask seeds 0--9. Supplementary Fig.~S10 reports these fold-selected rules, whereas Supplementary Fig.~S1 shows the fixed-fusion target-mapping comparison. Both remain calibration-development analyses of previously examined HER2ST patients.

The full-precision primary image feature scaler, target normalization and source-to-Wu calibration constants are supplied in \path{methods_data/}. The primary head settings and recorded software versions are provided there in machine-readable form. These constants specify the published transformations without refitting them.

\subsection{Identity controls and numerical audit scope}
Local, mean-self and GCN-self inputs in the 30-head comparison were bitwise identical. Seeds 1/2 predictions also agreed bitwise; seed 0 weights differed by at most $2.98\times10^{-8}$ and predictions by $9.54\times10^{-7}$, within the stated float32 guard. These are computational identity checks, not independent biological support. At 20\% coverage, matched-seed GCN spatial-minus-self correlation changes ranged from $-0.0425$ to $-0.0149$; GraphSAGE changes ranged from $+0.0002$ to $+0.0318$. Shuffled GraphSAGE seed-mean image prediction exceeded the spatial version in TENX99. These development findings do not establish a general topology advantage.

The original Bassiouni evaluation produced 2,904 clean and 18 control bundles (8,766 method predictions); all 17,532 hidden-label substitutions were invariant. The recorded audit reproduced 8,766 predictions and 35,064 per-gene metric arrays and reconstructed 968 label-free mask trajectories. Additional Wu molecular comparisons comprised 2,700 runs and 5,400 exact substitutions; all 360 original random references and 54,000 primary gene values passed $10^{-10}$ tolerances. Additional Bassiouni comparisons comprised 19,800 runs and 39,600 exact substitutions, with 2,640 original random references and 396,000 primary per-gene values passing the same tolerances. The neural matrix retained 2,028 model--patient bundles and 289,224 exact hidden-label checks; all 384 conditional geometric checks passed. Raw Spearman is retained but excluded from exact-replay claims because of near ties. Hashes, protocols, complete results and environment records provide the detailed provenance.

\clearpage
\section{Matched-backbone records and reproducibility}
\label{sec:matched_records}
The matched comparison distinguishes graph context relative to a local visual head from improvement beyond the same molecular anchors. Main Methods provide the authoritative definitions of image preprocessing, graph operators, training, calibration, masks, reconstruction and patient pairing. The machine-readable comparison report separates these two questions by backbone, route and cohort, retaining the single development patient and all three individual Wu effects. The original 22-patient placement and 20-estimator studies remain separate sources of evidence.

The initial fixed-rate ResNet18 analysis reuses verified states, predictions and per-run/per-gene exports. Its results are preserved as a distinct sensitivity analysis. The two-rate refinement followed source/validation learning-curve inspection before any Phikon evaluation outcomes; it gave each graph and non-graph configuration identical search and selection opportunities. All candidate losses, seeds, histories and state hashes remain in the candidate manifest, while the selected manifest was committed before selected-model scoring. These source/validation diagnostics are not external performance evidence.

The primary condition uses three historical masks at 20\% distributed coverage, not the ten masks underlying the original molecular headline correlations. Every paired predictor uses identical observation and scoring supports, recorded with barcode, coordinate, gene, target and index hashes. Supporting files retain 10/40\% coverage, identical-target placement and the historical central regional mask. Historical regional completion and identical-target placement remain distinct designs.

Per-run/per-gene exports retain undefined Pearson values and raw Spearman scores. The common-valid-gene sensitivity is explicitly separate from the primary defined-gene metric. Mask and training seeds represent computational variation; biological pairing and Bassiouni resampling use patients. New selected-model scoring repeats NaN/random corruption of all unobserved labels, including unscored locations, with exact unchanged hybrid predictions, and verifies image-only anchor independence on fixed hidden support. Reused checks are identified as verified prior evidence rather than newly executed tests.

Registered RGB patch bytes are identical across matched heads. Across encoders, pretraining, embedding dimension and effective physical field differ, as described in the main Methods. No interpolation-based enlargement restores native image detail. Frozen feature extraction is distinct from head fitting and from encoder fine-tuning. New derived states and complete numerical bundles remain author-local; manifests do not constitute public deposition. The three original primary image-head binaries remain missing.

\begin{table}[p]
\centering
\caption{\textbf{Matched-backbone fitting and validation selection.} Each configuration used the same two candidate learning rates (0.001 and 0.0001) with seeds 0/1/2. One rate was selected per backbone/configuration using mean best qualifying TENX95 MSE across all three seeds; no evaluation outcome selected it. Epoch triplets retain seed order 0/1/2. Updates count all minibatches, including the final partial batch. Times sum the three selected fits; peak RSS is their maximum and includes loaded data. These observed execution costs are not a controlled speed comparison across encoders or historical execution sessions. All 72 candidate histories, validation losses, state/configuration/input hashes and selected flags are machine-readable.}
\label{tab:backbone_resources}
\scriptsize\setlength{\tabcolsep}{3pt}
\begin{tabular}{llrrrrrr}\toprule
Backbone & Head/control & Parameters & Rate & Best epochs & Run epochs & Updates & Seconds / MiB\\\midrule
ResNet18 & local & 72,114 & 0.0010 & 12/14/17 & 22/24/27 & 1022 & 11.5 / 650\\
ResNet18 & mean & 72,114 & 0.0010 & 12/16/14 & 22/26/24 & 1008 & 9.9 / 584\\
ResNet18 & GCN & 72,114 & 0.0010 & 12/14/14 & 22/24/24 & 980 & 8.6 / 637\\
ResNet18 & GraphSAGE & 72,075 & 0.0010 & 7/10/7 & 17/20/17 & 756 & 7.9 / 657\\
ResNet18 & GCN shuffled & 72,114 & 0.0010 & 1/9/15 & 11/19/25 & 770 & 7.0 / 644\\
ResNet18 & GraphSAGE shuffled & 72,075 & 0.0010 & 9/18/7 & 19/28/17 & 896 & 10.8 / 687\\
Phikon & local & 104,882 & 0.0001 & 5/7/6 & 15/17/16 & 672 & 7.5 / 681\\
Phikon & mean & 104,882 & 0.0001 & 5/7/6 & 15/17/16 & 672 & 9.1 / 681\\
Phikon & GCN & 104,882 & 0.0001 & 5/7/6 & 15/17/16 & 672 & 9.3 / 681\\
Phikon & GraphSAGE & 104,792 & 0.0010 & 1/1/1 & 11/11/11 & 462 & 4.5 / 730\\
Phikon & GCN shuffled & 104,882 & 0.0001 & 7/7/7 & 17/17/17 & 714 & 11.6 / 704\\
Phikon & GraphSAGE shuffled & 104,792 & 0.0001 & 11/12/12 & 21/22/22 & 910 & 12.6 / 768\\
\bottomrule\end{tabular}
\end{table}

\FloatBarrier

\end{document}